\documentclass[acmlarge]{acmart}
\setcopyright{none}

\hypersetup{draft}

\usepackage{xspace}

\usepackage{wrapfig}
\usepackage{todonotes}
\usepackage{xspace}
\usepackage{paralist}
\usepackage{amsmath}
\usepackage{booktabs} 
\usepackage{algorithm}
\usepackage[noend]{algpseudocode}
\usepackage{flushend}
\usepackage{multirow}
\usepackage{listings}
\usepackage{xcolor}
\usepackage{enumitem}
\usepackage{comment}
\usepackage{footmisc}
\usepackage{longtable}
\usepackage{wrapfig}
\usepackage{lipsum}
\setitemize{noitemsep,topsep=0pt,parsep=0pt,partopsep=0pt}
\usepackage{array,booktabs,arydshln,xcolor}

\newcommand{\ew}{\ensuremath{E_w}}
\newcommand{\pw}{\ensuremath{P_w}}
\newcommand{\udp}{\textit{udp}}
\newcommand{\ty}{\ensuremath{\rho}}

\newcommand{\rs}{\!\!}
\newcommand{\FPD}{{\sffamily\scshape FPDetect}\xspace}

\begin{document}

\title{\FPD{}: Efficient Reasoning About Stencil Programs Using Selective Direct Evaluation}

\author{Arnab Das}
\affiliation{%
  \institution{University of Utah}
}
\email{arnab.das@utah.edu}

\author{Sriram Krishnamoorthy}
\affiliation{%
  \institution{Pacific Northwest National Laboratory}
 }
\email{sriram@pnnl.gov}

\author{Ian Briggs}
\affiliation{%
  \institution{University of Utah}
}
\email{ianbriggsutah@gmail.com}

\author{Ganesh Gopalakrishnan}
\affiliation{%
  \institution{University of Utah}
}
\email{ganesh@cs.utah.edu}

\author{Ramakrishna Tipireddy}
\affiliation{%
  \institution{Pacific Northwest National Laboratory}
 }
\email{Ramakrishna.Tipireddy@pnnl.gov}

\renewcommand{\shortauthors}{Das and Krishnamoorthy, et al.}



\newcommand{\etc}{etc.\@\xspace}
\newcommand{\ie}{i.e.\@\xspace}
\newcommand{\eg}{e.g.\@\xspace}
\newcommand\ignore[1]{}
\newcommand\ggcmt[1]{\todo[inline, size=\small, color=green!40]{GG: #1}}
\newcommand\ggcmtside[1]{\todo[size=\scriptsize, color=green!40]{GG: #1}}

\newcommand\adcmt[1]{\todo[inline, size=\small, color=yellow!60]{AD: #1}}
\newcommand\adcmtside[1]{\todo[size=\scriptsize, color=yellow!40]{AD: #1}}

\newcommand\ibcmt[1]{\todo[inline, size=\small, color=blue!60]{IB: #1}}
\newcommand\ibcmtside[1]{\todo[size=\scriptsize, color=blue!40]{IB: #1}}

\newcommand\skcmt[1]{\todo[inline, size=\small, color=orange!60]{SK: #1}}
\newcommand\skcmtside[1]{\todo[size=\scriptsize, color=orange!40]{SK: #1}}


\begin{abstract}

We present \FPD{}, a low overhead
approach for detecting logical errors and soft errors affecting
stencil computations without generating false positives. 
We develop an offline analysis that tightly estimates the number of floating-point bits preserved across stencil applications. 
This estimate rigorously bounds the values expected in the data space
of the computation.
Violations of this bound can be attributed with certainty to errors.
\FPD{} helps synthesize error detectors customized for user-specified levels of accuracy and coverage.
\FPD{} also enables overhead reduction techniques based on deploying these detectors
coarsely in space and time. Experimental evaluations demonstrate the practicality of our approach.


\end{abstract}

\maketitle

\section{Introduction}
\label{sec:intro}
Approaching the end of Moore's law, chip manufacturers are seeking
smaller lithographies, higher densities often achieved by
emerging 3D stacking
methods, newer memory/interconnect technologies, and increasing use
of GPUs and other accelerators.
These trends are increasing
the likelihood of soft
errors~\cite{DBLP:journals/ijhpca/SnirWAABBBBCCCCDDEEFGGJKLLMMSSH14,baumann,borkar,Quinn,EDA-018} already noticed
to be high in GPU-based systems~\cite{Tiwari:2015:RLL:2807591.2807666}.
The increased pressure toward specialization~\cite{Hennessy:2019:NGA:3310134.3282307}
may reduce the economies of scale achieved through general purpose parts,
and put cost-reduction pressures on verification methods, which already are
stressed~\cite{Jayaraman:2017:SPF:3135069.3107615}.
This can increase residual logical bugs in chips.
The increasing complexity of compiler optimizations will also
increase the likelihood of introducing
logical bugs~\cite{DBLP:conf/popl/BaoKPRS16}.
Software-level error
detectors can serve as uniform and application-aware ways
of trapping both soft errors~\cite{rivers2011ErrTolSerClaPro}
and logical bugs~\cite{DBLP:conf/popl/BaoKPRS16}, and
these detectors are needed more than ever before.

Unfortunately, system resilience research has,
so far, gone nowhere in terms of adoptions
into practice.
The main underlying reason is that in today's systems,
these resilience solutions geared toward soft errors will largely
cause a slowdown, with no apparent benefits from the point of view of
soft errors.
It may even aggravate the problem due to the high false positive rates in today's
resilience solutions.
Last but not least, today's resilience solutions have not been demonstrated to
have the ability to reliably flag logical errors.

Our main contribution in this paper is to demonstrate that
for a narrow class of
applications---specifically
stencil-based---we 
can indeed develop solutions that benefit {\em both} logical error protection and
soft error protection, owing to our solutions having two key attributes: (1)~no false
positives, and (2)~acceptable overheads.
With this combination, we strongly believe that resilience solutions will
be welcomed more readily, at least for their immediate impact on logical bugs.


Stencils belong to an important class of iterative solvers operating on dense D-dimensional
arrays modeling partial differential equations (PDEs) with applications belonging
to fluid dynamics, cosmology, combustion, etc.\footnote{While our approach applies to sparse stencils, we focus on stencils operating on dense matrices.}
They are part of the essential building blocks for solving larger more composite problems involving
multiple PDEs.
%

The ability to write good assertions that capture a stencil computation's
evolving profile of data is tricky and error-prone.
Existing work on system resilience that address data integrity
include time-series data analysis methods~\cite{DBLP:journals/tpds/DiC16} 
and data outlier detection methods constructed using machine
learning
(e.g. ~\cite{DBLP:conf/cluster/SubasiDBULCKC17,DBLP:conf/cluster/SubasiK17}).
Unfortunately, these approaches have high
overheads, and can only help loosely characterize
the expected data value ranges.
They additionally bring in training bias into the models constructed.
Given that they both overestimate and underestimate the
data ranges, detectors based on them
generate both false positives and false negatives.
With soft error detection,
false positives are virtually unacceptable, given that the natural
soft error rates themselves are quite low
(false positive needlessly
engage checkpoint/recovery systems).

\FPD is the first approach that takes the novel approach
of basing data-space protection on {\em accurate
floating-point round-off error estimation}.
Like any calculation carried out by numerical code,
stencil-based calculations are also subject to floating-point rounding error.
However, given that stencils are more structured, we show that
one can develop a rigorous floating-point round-off error
estimation approach for them, guaranteeing a certain number of
mantissa bits after every stencil application.
Consequently, in the \FPD approach,
if ``round-off'' appears suddenly exaggerated, we can attribute
it reliably---without any false positives---to a logical bug or a soft error.

Contemporary floating-point error estimation
approaches using interval analysis~\cite{boldo,DBLP:conf/arith/Kramer97}
tend to give excessively
conservative estimates(due to loss of correlation)
of round-off error---especially for large programs
that are iterated over time.
These conservative estimates with large error bounds
implies guarantees for only the higher order mantissa bits,
i.e., only those errors that
cause very high magnitude changes can be
trapped using them.\footnote{That is, should these methods be used
to synthesize error detectors, which hasn't been done.}
FPTaylor~\cite{FanChiang} builds symbolic taylor forms
for the error expressions resulting in tight guarantees
but do not scale to be usable beyond a few hundred operators.
\FPD on the other hand provide tighter guarantees
on the error for large stencil programs, 
thus helping protect more mantissa bits,
and also allows a {\em tunable} approach
to trading off overhead against detection precision,
as we show in this paper.

Classical approaches like dual modular redundancy (DMR) can
rigorously detect silent data corruptions by utilizing a duplicated execution
thread and checking for result matches.
This unfortunately can double the overall computation time.
Clover~\cite{clover} exercises a selective instruction duplication strategy 
to bring down the SDC detection cost to around 26\% using tail-DMR.
However, this approach is meant only for soft errors; it cannot be
used to trap 
compiler bugs because they impact {\em both} the execution threads.

In this paper, we show that by 
focusing on a narrow (but important) class
of applications---in our case stencils---we can arrive at
a {\em unified solution} for detecting compiler bugs and soft errors  while
also offering rigorous guarantees.
%
We offer our tool
\FPD{}, a {\em software based error detector}, 
whose core approach is based on rigorous floating-point round-off error analysis
for an iterated application of the stencil across $T$ computational steps.
%
%
To the best of our knowledge, this is the first work
that encompasses two important correctness
checks around the single
central idea of offline round-off analysis.
\noindent{\sc Roadmap:\/}
\S\ref{sec:overview} provides an overview of \FPD
including error analysis, and optimizations.
\S\ref{sec:application} is an overview of how \FPD
helps detect software bugs and soft errors.
To reduce overheads, we
establish a crucial result
that if we leapfrog stencil applications and the detectors
$T$ steps ahead do not detect a soft error, then
we can establish a {\em certified baseline}
that allows earlier time steps to be forgotten
based on the fact that uncaught
soft errors in earlier steps will not affect future detection.
This allows \FPD
detectors to be deployed sparsely in space
and time (e.g., once per $64$ time steps)
(\S\ref{sec:strategies} and~\S\ref{sec:placement}).
We perform offline detector synthesis to
create lookup tables, and instantiate specific
detectors just before execution
(\S\ref{sec:offline-online}).
Our evaluation (\S\ref{sec:evaluation}) includes
results on software bug detection and resilience
(both simulated through fault injection).
Additional related work (\S\ref{sec:rel-work})
and conclusions (\S\ref{sec:conclusions}) follow.




\section{Overview}
\label{sec:overview}
\label{sec:error-analysis}

\vspace{.5ex}
\noindent{\bf Background on Floating-point error analysis:\/}
Floating-point error analysis
is a vast area, and we provide only a brief overview
(see~\cite{Muller:2009:HFA:1823389,goldberg} for details).
A floating point number system, $\mathbb{F}$,
in radix, $\beta$, is a subset 
of the set of real numbers,
and can be expressed as a 5 tuple $(\beta, s,m,e,p)$.
We use binary ($\beta=2$) double precision with number of {\em precision bits}, $p=53$.
$s \in \{-1,1\}$ is the sign bit,
$e$ is the exponent in the range $-1022 \leq e \leq 1023$,
and $m$ is the mantissa or the significand, and represents
the magnitude $s\cdot m \cdot 2^e$.
If $x \in \mathbb{R}$,
then $x_f \in \mathbb{F}$ denotes an element in
$\mathbb{F}$ closest to $x$ obtained
by applying the rounding operator $(\circ)$ to $x$. We use the bound consistent 
for all IEE754~\cite{ieee754} rounding modes.

In floating point arithmetic, the {\em absolute error} is
the magnitude difference between the values yielded by
computations done in the space
of real numbers (``true answer'') and those done in floating point.
The relative error is the ratio of the absolute error and the true
answer.
Every real number $x$ lying in the range of $\mathbb{F}$
can be approximated by an element  $x_f \in \mathbb{F}$
with a relative error no larger than the
{\em unit round-off} $\boldsymbol{u} = 0.5\times\beta^{1-p}$. 
Here $\beta^{1-p}$ corresponds to the
{\em unit of least precision(ulp)} for exponent value of 1.
We use $\mu$ to denote ulp(1),
such that $\mu = 2\boldsymbol{u}$.
In our case $\boldsymbol{u} = 2^{-53}$,
and hence $\mu = 2^{-52}$. 
Hence, for all rounding modes($\circ$), 
\( \circ(x) = x(1 + \delta),\quad |\delta| \leq \boldsymbol{u} = \mu/2.\)
Given two exactly represented floating point numbers
$x_f$ and $y_f$,
arithmetic operators $\diamond \in \{ +,-,\times\}$
have the following guarantees across
any rounding modes:
\begin{equation}
\label{eq:float-basic-op}
x_f \diamond_f y_f = (x_f \diamond y_f)(1 + \delta) = (x_f \diamond y_f) + (x_f \diamond y_f)\delta,
\quad |\delta| \leq 2\boldsymbol{u} = \mu
\end{equation}

In our work, we employ affine arithmetic~\cite{stolfi} for error estimation.
In an affine representation, each input variable
$\hat{x}$ is represented by
$\hat{x} = x_0 + \sum_{i=1}^n x_i \epsilon_i $,
with
$x_0$ denoting the central value of the corresponding interval,
and coefficients $x_i$ being finite floating
point numbers. The $\epsilon_i$ are
{\em formal noise variables}
which are unknown until concretized but assumed to
lie in the interval [-1,+1]. 
The $x_i$ model the noise coefficients such that the round-off
error in equation$~\eqref{eq:float-basic-op}$ can be expressed as
$x_i = (x_f \diamond y_f)\times 2$, with a noise variable $\epsilon_i \in [-1,1]$.

We define two operators, 
$\sigma$ and $\gamma$ that allows us to retrieve
the associated noise variable and coefficient information
for each affine variable.

\begin{compactitem}
\item {\em $\sigma$:} Defines the mapping from an affine variable
					  to its set of noise variables. Thus
							$\sigma(\hat{x}) = \{\epsilon_i\}_{i=1}^n$.
\item {\em $\gamma$:} Defines the mapping from a 2-tuple of 
					  {\em (affine variable, noise variable)} to its 
					  corresponding noise coefficient. Thus $\gamma(\hat{x},\epsilon_i)$ return $x_i$ if $\epsilon_i \in \sigma(\hat{x})$ else return 0.
\end{compactitem}

Using these operators, at every operator site, a fresh noise variable is introduced
and their collective impact in the forward path is tracked using affine analysis. For codes
implemented with the  round to nearest model (as in most cases), one can 
configure ${\bf u} = \mu$, to obtain stricter bounds.\footnote{A
full derivation of error bound using affine arithmetic are presented in the
separate {\bf supplementary material} for the interested reader in {\bf Appendix A}}.

\begin{figure*}[h]
\centering 
\begin{minipage}{0.32\textwidth}
\includegraphics[width=1.0\columnwidth]{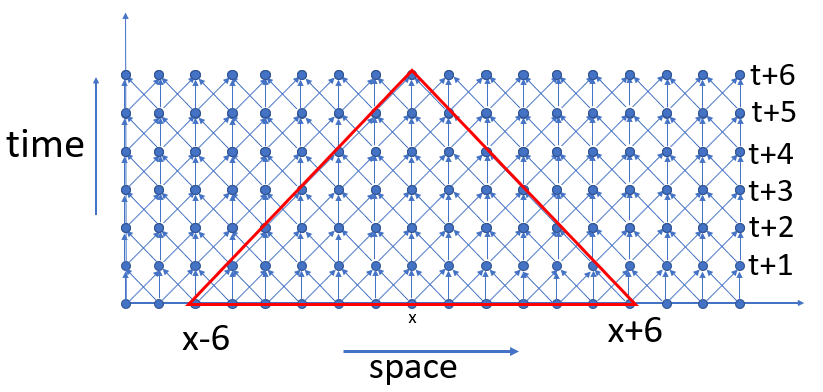}
\caption{\label{fig:iteratedStencil} Simplified 1D stencil over 6 time steps}
\end{minipage}\hfill
\begin{minipage}{0.32\textwidth}
\includegraphics[width=1.0\columnwidth]{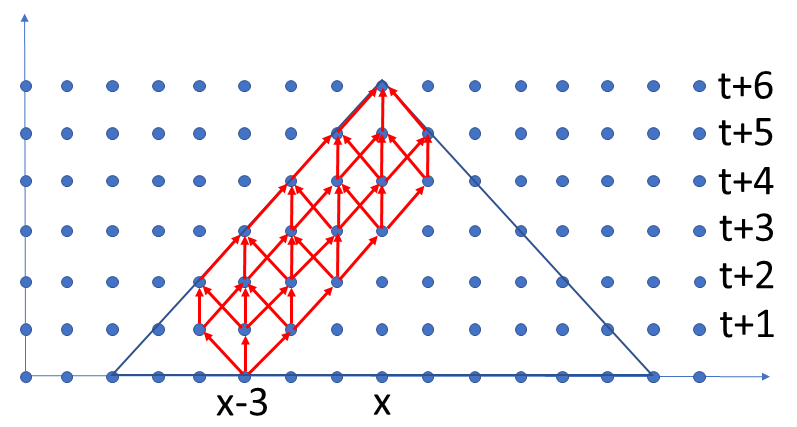}
\caption{\label{fig:PathDominance} Illustration of path dominance}
\end{minipage}\hfill
\begin{minipage}{0.32\textwidth}
\centering
\includegraphics[width=1.0\columnwidth]{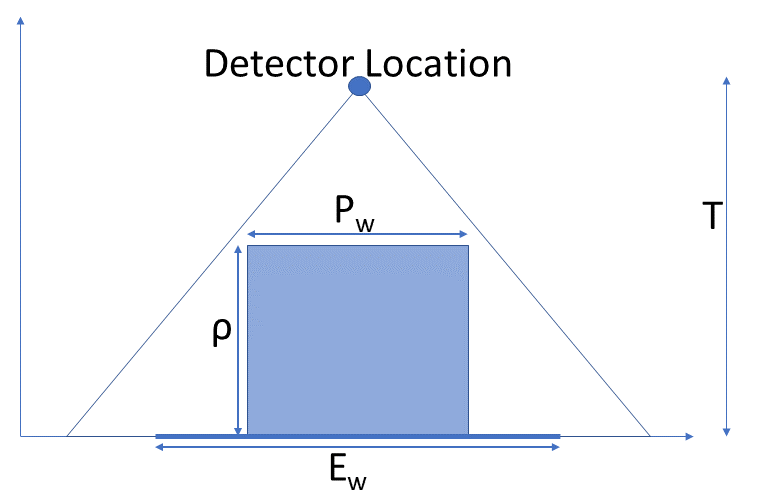}
\caption{\label{fig:ew_and_pw} Essential (\ew{}) and protected widths (\pw{})}
\end{minipage}
\end{figure*}

\vspace{.5ex}
\noindent{\bf Illustrative example:\/}
We now illustrate
the concepts and practical details behind \FPD{}
using the simplified 1D stencil of 3 inputs leapfrogged (iterated) over 6 time steps
(Figure~\ref{fig:iteratedStencil}).
In this example, the stencil coefficients are $0.25,0.5,0.25$
(say, modeling discretized heat flow) working on problem array $A$.
The value $A[x,t+1]$ can be calculated
to be $0.25*A[x-1,t] + 0.5*A[x,t] + 0.25*A[x+1,t]$.
We call this approach of obtaining the values at the next
time step the {\em iterated} evaluation scheme.
One can also imagine obtaining the value at $A[x,t+6]$ through
{\em direct} evaluation that leapfrogs 6 steps, provided one calculates
the {\em effective path coefficients}.
Table~\ref{table:unroll-table} illustrates how these coefficients
may be calculated by taking
the sum of the product of the path weights
($Set_2$, for instance, is $0.375 = 0.25*0.25+0.25*0.25+0.5*0.5$).

Now, given a specific unroll (leapfrog distance) of $T$,
we can then generate a similar table for that unroll, obtaining,
for a $k$ point stencil, $(k-1)T+1$ effective coefficients. For example
for a 3-point stencil over $T=5$ steps, we obtain $11$ separate
coefficients for contribution from each of the corresponding unrolled
points.
Then, given $(k-1)T+1$ runtime values of these points, the direct evaluation scheme
can obtain the value $T$ steps ahead, and is the basis of
realizing \FPD detectors.
We in fact implement this dot product of input values
with the effective coefficients being in higher precision, and also
employ Kahan's~\cite{kahan1996ieee}
summation algorithm, as well as vectorization, thus
minimizing the {\em relative error}
of this answer.\footnote{We also employ vectorization
  to reduce overheads in the \FPD detector.}
Thus we obtain an estimate for the relative error with minimal precision loss; call it 
$R_d$ for {\em relative error under
  direct evaluation}.

While we know $R_d$, we cannot give any bounds on the number
of mantissa bits preserved unless we also know the
iterated evaluation's relative error, say $R_s$ ($s$ for ``serial''
evaluation).
A key contribution we make is to tightly estimate
$R_s$ analytically.

\begin{small}
\begin{table}[t]
  \caption{Evolution of Coefficient values as we unroll the stencil} 
  \centering
\begin{tabular}{cccccccc}
  \toprule
  \bfseries Set Number & \multicolumn{7}{c}{Coefficients at Offsets} \\\cmidrule(lr){2-8}
	  offset index $\rightarrow$ & -3 & -2 & -1 & 0 & 1 & 2 & 3 \\
  \midrule
    $Set_1$ & 0 & 0 & 0.25 & 0.5 & 0.25 & 0 & 0 \\ 
    $Set_2$ & 0 & 0.0625 & 0.25 & 0.375 & 0.25 & 0.0625 & 0 \\ 
    $Set_3$ & 0.015625 & 0.09375 & 0.234375 & 0.3125 & 0.234375 & 0.09375 & 0.015625 \\
    \bottomrule
  \end{tabular}
  \label{table:unroll-table}
\end{table}
\end{small}

\vspace{.5ex}
\noindent{\bf Iterated Evaluation:\/}
That there are many reconvergent computational paths to be
taken into account during iterated evaluation
for generating the output at $[x,t+6]$
from the inputs at $t$. 
In such situations, unless we keep the errors on various
re-convergent paths correlated, we will obtain uselessly
exaggerated error bounds.
Fortunately, affine arithmetic
is well known for being able to handle
error analysis in such situations.
As an example, $(x-x)$ yields $0$ under affine arithmetic
(whereas interval arithmetic will yield an interval
with the error in $x$ doubled~\cite{Muller:2009:HFA:1823389}).

\begin{figure}[h]
    \centering
    \begin{minipage}{0.45\textwidth}
\includegraphics[width=0.85\columnwidth]{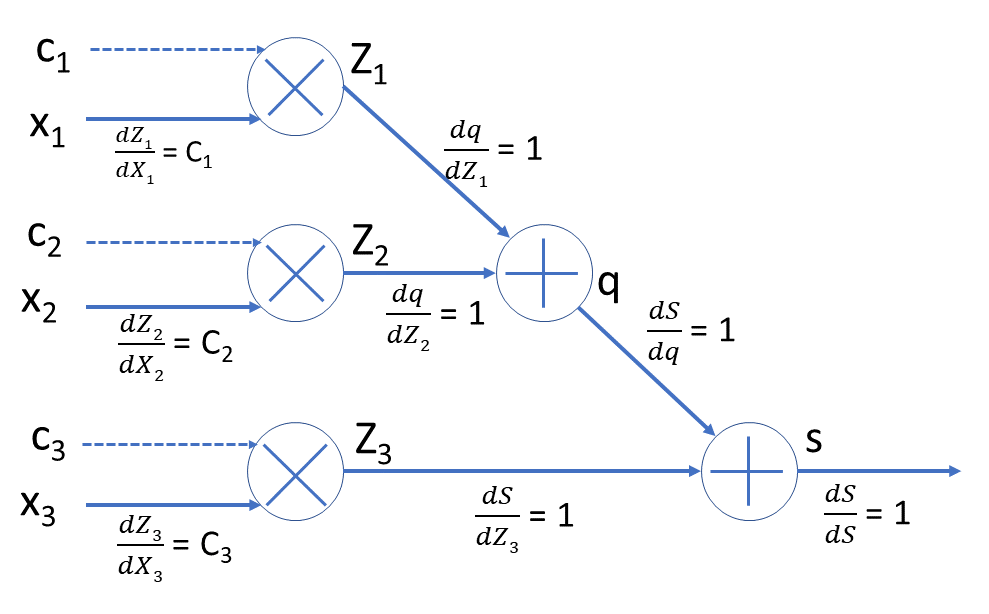}\\
\caption{\label{fig:cfg}
Computational graph with highlighted derivatives for 1 step of a 1-d 3-point stencil
}
    \end{minipage}
    \hfill
    \begin{minipage}{0.45\textwidth}
\includegraphics[width=\textwidth, height=0.5\columnwidth]{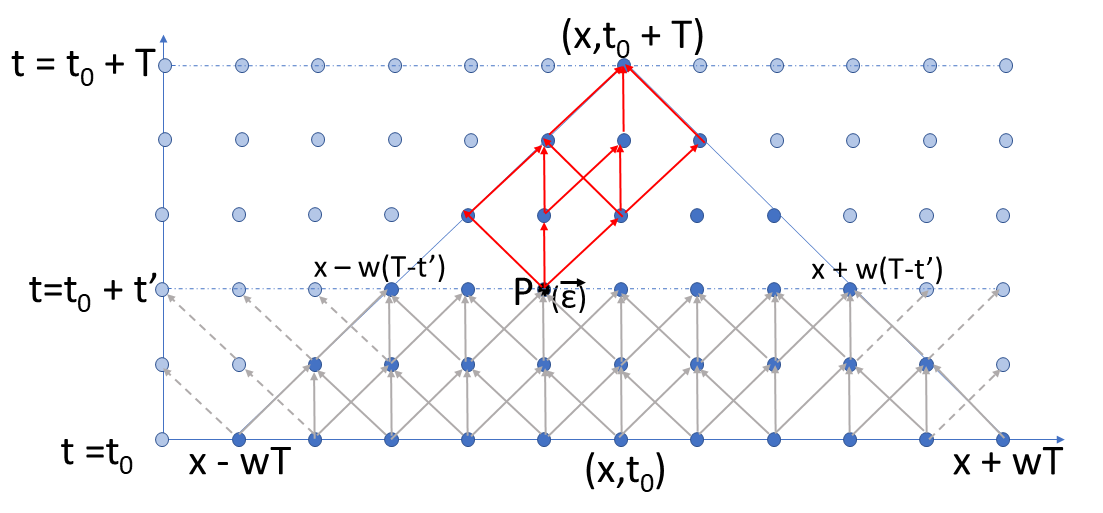}
\caption{\label{fig:error-propagation}
Error propagation from point (x-1,t0+t') to (x,t0+T) (1-d stencil)
}
\vspace*{-1em}
    \end{minipage}
\end{figure}

\begin{figure}[t]
\includegraphics[width=.95\columnwidth]{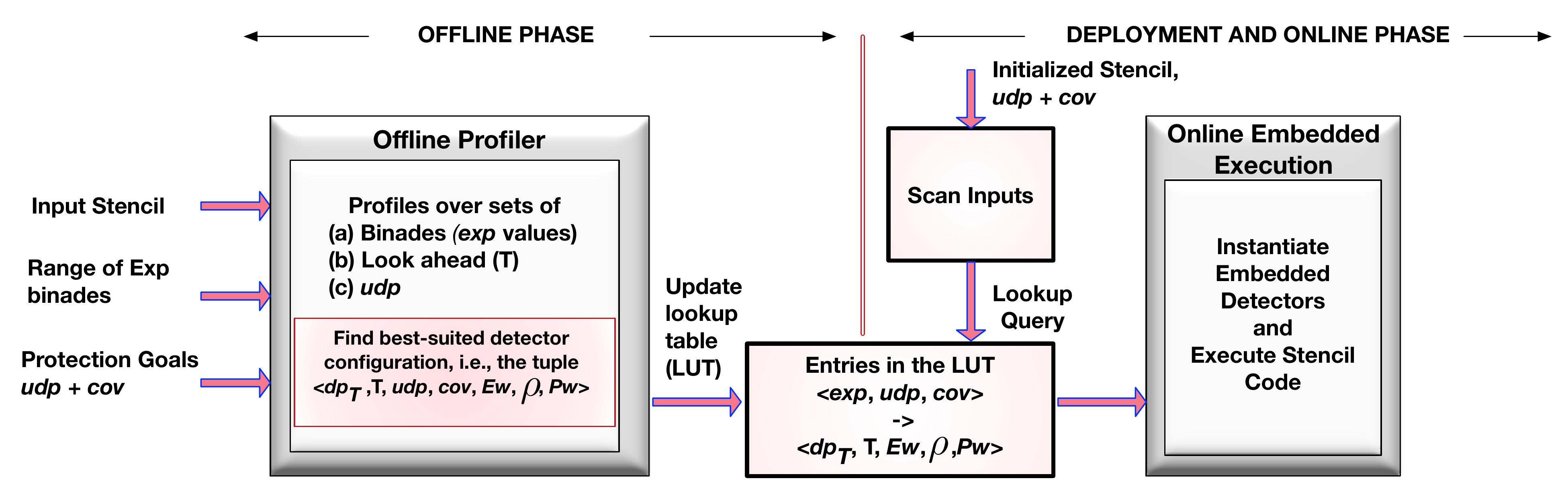}
\caption{Overview of \FPD{}}
\label{fig:trishul-overview}
\end{figure}

\FPD{} comprises of an offline phase which includes
static error analysis of the stencil combined with 
conservative profiling of the stencil (detailed in~\ref{sec:offline-online}) 
for a set of protection goals as shown in Figure~\ref{fig:trishul-overview}.
%
%
Our error analysis method uses affine arithmetic building upon
the works of~\cite{higham,boldo,DBLP:conf/popl/DarulovaK14} but are
applied to much larger expressions.
%
%
As a simple ilustration, consider a single step iteration of a 1-d 3-point
stencil with stencil coefficients
$\{c_1, c_2, c_3 \}$ centered around $x_2$ to evaluate $x_2$ for the next time step,
denoted here as S in Figure$~\ref{fig:cfg}$.
Thus the computational graph evaluates $S=((c_1\times x_1)+(c_2\times x_2))+(c_3\times x_3)$.
If $\epsilon^\prime$
is a noise variable belonging to $\sigma(x_1)$,
then the error contribution's propagation at $S$ will be
  \begin{equation}
  \label{eq:chain-rule}
  \begin{split}
  	\gamma(x_1, \epsilon^\prime)\cdot K_{x_1} &=  \gamma(x_1, \epsilon^\prime)\cdot \bigg| \dfrac{dS}{dx_1}\bigg| 
  	                                     =  \gamma(x_1, \epsilon^\prime)\cdot \bigg| \dfrac{dS}{dS}\cdot
 										 	\dfrac{dS}{dq}\cdot
 											\dfrac{dq}{dz_1}\cdot
 											\dfrac{dz_1}{dx_1}\bigg| 
 										= \gamma(x_1, \epsilon^\prime)\cdot 1 \cdot 1 \cdot 1 \cdot c_1 \\
                = \gamma(x_1, \epsilon^\prime) \cdot c_1
  \end{split}
  \end{equation}
 Thus for all such $\epsilon^\prime \in \sigma(x_1)$, the $\gamma(x_1, \epsilon^\prime)$ gets propagated by $c_1$
 (and
  similarly for $x_2$, $x_3$, $z_1$, $z_2$, $z_3$ and $q$).
  Effectively, every incoming node or an internal
  compute node has
  a locally generated error term, and a propagation factor that propagates this
  error to the output node.

We can
  iteratively evaluate the stencil and arrive at this
  output value  by
(conceptually) picking all intermediate points,
    such as $P$ that is highlighted
    in Figure~\ref{fig:error-propagation},
and first finding the error flowing into $P$.
Then find out how this error
   is {\em modified} by all the path coefficients
   from $P$ to the output.
Do this for each point in the iteration space that
iteratively contribute to the output at the detector
location.
Accumulating these error terms and normalizing them gives us
the relative error $R_s$ under iterated evaluation.
%
    While iterated evaluation mimics the stencil evaluation itself,
    {\em we are conservatively estimating this error during the offline phase}
    of \FPD.
   Our error analysis
 suitably combines the error for all such conceptual
 points $P$ using affine arithmetic.

\vspace{.5ex}
\noindent{\bf Detector precision ($dp_T$):\/}
We now can define
a precision  threshold to be checked for 
correctness at the detector locations.\footnote{A full derivation of
  $dp_T$ including affine arithmetic and error analysis 
  details are presented in the separate {\bf supplementary material} for the interested reader in 
  {\bf Appendix A}.}
Called {\em detector
  precision} $dp_T$ (where the subscript $T$ indicates it is a function of
the number of lookahead Tsteps)
it is defined as
\(	\hbox{bits preserved}_T  = (dp)_T = p - \log_2(\max(\boldsymbol{R_s}, \boldsymbol{R_d})) \)
which says ``remove the maximum of the uncertainties between
the direct and iterated evaluations.''
Given that we minimize $R_d$,
the key factor impacting the number of bits preserved
will be the main stencil's iterated evaluations.
Soft errors or logical bugs that throw the actual error beyond
the $dp_T$ limit can then be trapped with the guarantee of
{\em no false positives}.
Furthermore, all events whose impact falls within the guaranteed 
precision will be trapped.
The only case  of omission (which is outside of our model) is
when two faults simultaneously land within 
the scope of two adjacent detectors, and both these fault-flows
cancel out before reaching their respective detectors.
%

\vspace{.5ex}
\noindent{\bf Offline/Online Split:\/}
Our approach consists of an offline phase
and an online one (Figure~\ref{fig:trishul-overview}).
The user presents the uninitialized input stencil and 
the range of intervals
(in binades---meaning exponent spans) expected
for each input.
In the offline phase we explore a rich search space.
It is composed of the exponent binade ranges, and 
protection goals defined by required precision
accuracy and protection coverage. The goal is to
obtain a best fit detector configuration for a pair 
of $(precision, coverage)$ for each 
exponent binade and populate a look-up table
for each application. Each application is identified 
by its unique coefficient set.
Note that any change of coefficients 
for a given stencil 
results in re-synthesis as a new application,
creating a new entry in the lookup table.
%
%
In online phase, we scan the input to obtain
an interval that maps to a
canonical exponent binade range. This allows
\FPD to query the LUT for any unseen data
at runtime
with user defined precision and coverage goals
to obtain a best fit detector configuration.
Given these inputs, we can
calculate {\em the worst-case binade
  differences} present at the input.
The binade difference influences $R_s$ and
$R_d$: the higher the worst-case  binade differences
at the input, the worse 
$R_s$ and $R_d$ themselves are, and the lower
the value of $dp_T$.
The goal of the offline phase is
to produces a {\em look-up table} (LUT) which has 
the guarantee of $dp_T$ bits at the detector output,
provided the runtime data lives within the binade
differences upon which the stencil
is ``profiled.'' 
The worst case binade difference is extracted only at the beginning
for the initialized input set in our experiments.
Our rigorous guarantees are with respect to these binade
differences
(a fact that can be optionally revalidated
at runtime).


\vspace{.5ex}
\noindent{\bf Protected Width:\/}
Our runtime goal is to not deploy the detectors 
at every time step, but only at every $\rho$ time 
steps (Figure~\ref{fig:ew_and_pw}, where x-axis corresponds to the
grid points in the stencil evaluation) by certifying the correctness
of points within the $\rho$ detection latency.
Moreover,
our goal is also to spatially distribute the detectors
sparsely: the $\rho \times P_w$
boxes shown in this figure capture this goal,
and these boxes are stacked horizontally.
To arrive at these ``stacking'' optimizations,
the user also provides the protection goals
({\em user defined precision} or $udp$
defined in \S\ref{sec:strategies}).
Then during the online phase, we can provide
$udp$ number of bits to be preserved in the computational
volume presented in each $\rho \times P_w$
box, provided $dp_T$ bits are guaranteed by
the detector.

The actual stencil computation and detector
deployment in a realistic
problem are as presented in Table~\ref{tbl:stencil-params}.
In particular, \FPD can handle multiple computational arrays
that iteratively update each other over time.
With respect to the parameters specified in Table~\ref{tbl:stencil-params},
\FPD automatically instantiates and configures the requisite detectors.
The pseudo-code in Figure~\ref{fig:stencil-spec} also presents the automatically inserted
{\sc detect\_errors()} calls corresponding to
where the error is detected.




\begin{figure}[h]
\begin{minipage}{0.52\columnwidth}
\begin{lstlisting}
//D-dimensional array type
using MultiDimArray = ...;
//stencil routine on N multi-dimensional arrays
auto stencil(MultiDimArray A[N]) {
 for(int t=0; ; t +=[@$\rho$@]) {
  for(int tt=t; tt<t + [@$\rho$@]; tt++) { //in one time tile
   if(converged) {
    DETECT_ERRORS(); //<--(Detector-2) 
    return ;
   }
   for(int x=0; x<N; x++) { //every array 
    for([@$\vec{i} \mid \vec{l} \le \vec{i} \le \vec{h}$@]) //every valid index
     for(int y=0; y<N; y++) //every RHS array
      for([@$ \vec{s} \in S_{x,y}$ @]) //the stencil
       TMP[[@$\vec{i}$@]] += [@$c_{x,y}(\vec{s})$@] * A[y][[@$\vec{i}+\vec{s}$@]]; //update
    for([@$\vec{i} \mid \vec{l} \le \vec{i} \le \vec{h}$@]) //every valid index
     A[x][[@$\vec{i}$@]] = TMP[[@$\vec{i}$@]];
    } //for 'x'
   } // for 'tt'
   DETECT_ERRORS(); //<--(Detector-1)
 }  // for 't'
}
\end{lstlisting}
\caption{\label{fig:stencil-spec} Stencil computation pseudo-code. A compiler can transform the code in lines 11--19 into any equivalent form. Detector-1 is used throughout the computation at the time tile boundaries. Detector-2 is used at the end of the computation.}
\end{minipage}%
\hfill %
 \begin{minipage}{0.45\columnwidth}
 \begin{table}[H]
 \begin{tabular}{lp{1.1in}}
 \toprule
 Parameter & Remark \\
 \midrule
 $\vec{l} = \{ l_1,\dots, l_D \}$ & lower bound of array region \\
 $\vec{u} = \{ u_1,\ldots, u_D \}$ & upper bound of array region \\
 N  & number of arrays \\
 $\rho$ & time tile size \\
 $ S_{x,y}  = \{\vec{s_1},\ldots, \vec{s_n}\}$ & Stencil offsets to update each point of array $A_x$ from points in $A_y$\\
 $c_{x,y}(\vec{s})$ & stencil coefficients to update array $A_x$ using array $A_y$\\
 \bottomrule
 \end{tabular}
 \vspace{1.0ex}
 \caption{\label{tbl:stencil-params}Stencil computation parameters. $\vec{l}$, $\vec{u}$, and $\vec{s_i}$ contain integers. $N, T \in \mathbb{Z}^+$. Implicitly, $\forall \vec{s} \notin S_{x,y}: c_{x,y}(\vec{s}) = 0.$ $\forall \vec{s},x,y: c_{x,y}(\vec{s}) \in [-1,1]$. }
 \end{table}
 \end{minipage}
\end{figure}

Figure~\ref{fig:stencil-spec} presents the pseudo-code for a stencil program operating on dense D-dimensional arrays.
It involves updating the ``neighborhood'' of each point of every array in a given region, denoted by $\vec{l}$ and $\vec{u}$, with scaled values from other arrays. The neighborhood used to update array $A_x$ from array $A_y$ is defined by the stencil  $S(x,y)$. 
At each time step $t$, every valid array index in each array $x$ (i.e., bounded by $\vec{l}_x$ and $\vec{u}_x$) can be updated from the array locations in the stencil neighborhood of all other arrays $y$ using the coefficient function $c_{(x,y)}$. A distinct stencil coefficient function can be defined for each pair of arrays in an update operation. 
A temporary array $TMP$ is employed to handle
write-after-read dependencies in the array being updated.
The DETECT\_ERRORS() instatntiates the required detectors for the corresponding 
time tile. In the case of program converging and exiting before finishing  the
entire loop of the time tile, there is an extra set of trailing detectors instantiated
by DETECT\_ERRORS at the loop exit for protection of the end segment of the 
computation.

\vspace{.5ex}
\noindent{\bf Optimizations, and Overall Algorithm:\/}
As Figure~\ref{fig:PathDominance} shows,
we can determine a collection of
inputs such as $[x-4,t]$ that can be removed
from consideration {\em because they have no effect
  within $dp_T$ bits of precision}.
This calculation is described in \S\ref{sec:strategies}.

Our hope to leapfrog by $T$ steps will not be realized
unless we can ``seal off'' past computational points
from future consideration.
In \S\ref{sec:placement}, provide the analysis underlying
this strategy.
In our example stencil,
it will allow us to
shift the baseline by $\rho$ steps
as illustrated in Figure~\ref{fig:ew_and_pw}.
The new baseline is obtained at $\rho$ steps 
in the future while $T$ is the leapgrogging step.
In fact, the following is an invariant for
any detector instantiation: $\ \forall (\rho, T): \rho \leq T$.

Key highlights of our approach include: (1)~rigorous floating-point
precision analysis made possible by exploiting the structured nature of
our computations, (2)~being able to define concepts such as
essential (\ew{}) and protected (\pw{}) widths that help us
reduce the amount of computation necessary
to achieve a certain level of protection, and
(3)~cost analysis for optimal detector deployment
(explained in \S\ref{sec:placement}) for a given amount of protection.


\section{Applications}
\label{sec:application}

\FPD{}'s direct evaluation strategy has the ability to
check bounds on the values generated in the stencil's computational
space at runtime.
This can help trap software level 
bugs that may be introduced during compiler transformations.
%
%
It also helps detect
soft-errors within a thresholded precision limit providing robust and precise
coverage of the compute space for protection against silent data corruption (SDC).




\subsection{\bf{Software Bug Detection}}
\label{sec:sec:sbd}

%

Loop optimization frameworks generate optimized code for stencil
programs.  Pluto~\cite{uday08cc} performs polyhedral transformations
that incorporate a host of optimizations to generate efficient code
for stencils and other loop programs. Pochoir~\cite{tang2011pochoir}
is a compiler to generate efficient implementations of stencil
computations from a high-level specification. 
%
%

There has been a renewed interest, especially in the context of
kernels such as stencils, in the generation of optimized programs that
exploit various architectures (CPUs, GPUs, ..), instruction sets, and
customized hardware such as FPGAs. The code generated by such tools
needs to be checked to detect bugs in the toolchain early and avoid
errors creeping into production code. While one ideally desires a
fully verified toolchain, developing such toolchains is a non-trivial
task. \FPD{} can be used as part of the testing toolchain for these
tools as they are developed. Thus, \FPD{} can enable faulty
compilers being identified and corrected before being used in
production environments.
Specifically, compiler transformed programs can harbor bugs, as shown in the work on
Polycheck~\cite{DBLP:conf/popl/BaoKPRS16}.  Prior efforts have studied
the design of approaches to verify or check the correctness of code
generated by such
optimizers~\cite{DBLP:journals/toplas/VerdoolaegeJB12,DBLP:conf/popl/BaoKPRS16,DBLP:conf/isola/SchordanLQP14}.
These approaches detect bugs that impact the changes in control and
data flow, with limited support for semantic transformations.
Unlike the control or data flow based approaches, \FPD{} is not
limited by the nature of the transformations employed and carefully
captures the semantics of the stencil operations.
On the downside, the analysis is imprecise in a different fashion than
control or data flow analysis, which are not affected by the floating
point round-off errors.

%
%
We show that,
with respect to the benchmarks chosen,
commonly considered loop transformation bugs can be
trapped 
with around $2\%$ overhead,
allowing tested by unverified codes
to be shipped to third-parties who may
keep the detectors turned on till gaining sufficient trust.

\subsection{\bf{Soft-Error Detection}}

\FPD is intrinsically designed and
 optimized to help protect applications
by detecting SDCs with small enough detection latecy.
We consider a single event error model wherein a single error
at some time $t$ {\em transiently} corrupts one or more values
of one or more participating arrays in the computation
resulting in a soft-error. 
These errors are non-permanent (transient) in nature and cannot be
replicated making them extremely difficult to detect and isolate.

%

The soft error detectors generated using the
\FPD approach are, by construction,
devoid of any modeling bias unlike
those are generated through
machine learning~\cite{DBLP:conf/cluster/SubasiDBULCKC17,DBLP:conf/cluster/SubasiK17}
or time-series data analysis
methods~\cite{DBLP:journals/tpds/DiC16} which
tend to reflect modeling bias.
\FPD's detectors
faithfully represent
the evaluation of the stencil
by sampling the actual data space sparsely.
It computes the output generated by a stencil
application by faithfully accounting
for the precision contribution
of each point in the detector's dependence cone
(as noted in Section~\ref{sec:overview}).
\FPD enables the user
with a parametric knob called the \textbf{user defined precision (udp)}
using which they can define the required accuracy at
the program output.
Given a $udp$ value of, say, $20$, 
and a detector designed for T-step direct evaluation
with an error bound of $dp_{T}$,
\FPD finds the {\em minimal set of points which contribute at least 
  20 bits of precision to retain $dp_T$ bits of precision at the detector}.

We actually develop
two detection strategies
in the context of \FPD.
The first employs one direct evaluation and the
other employs two direct evaluations.
The latter is necessary
given that the last leapfrogging step may not finish at a multiple of $T$
steps; in those cases, the second direct evaluation
provides the effect of a trailing detector
at the end of the stencil computation.


\vspace{.5ex}
\noindent{\bf Single-direct detector:\/}
At time $t_0$, direct evaluation is used to
compute the estimated value of a point $A_x[t_0+T][\vec{i}]$, $T$ time
steps ahead.
When the iterative stencil computation reaches time $t_0+T$
and evaluates this point, it is compared with the previously estimated
value.
If the two values differ more than can be ascribed to
floating-point round-off error, it indicates an error impacting either the
direct evaluation or in some part of the iterative evaluation leading
to that point, triggering a soft-error detection event.
This
strategy involves a single additional dot-product.
The error is
detected after $T$ time steps, requiring that the computation executes
at least till time step $t_0+T$.

\vspace{.5ex}
\noindent{\bf Double-direct detector:\/}
Values at two time steps $t_0$ and $t_0\!+\!t'$ can be used to
estimate a value at a future time $t_0\rs+{\rs}T${\rs}\; ($T{\rs}>{\rs}t'$).
This involves two
direct evaluations, one each at time steps $t_0$ and $t_0{\!}+{\!}t'$, whose
estimates can be compared in the latter of the two time steps. This
allows flexible detection and does not require that the iterative
computation reach time $t_0+T$.
A mismatch between the two estimated
values (beyond the bound on floating-point round-off error) indicates
an error in the computation between time steps $t_0$ and $t_0+t'$.
This detection strategy requires two dot products,
potentially
doubling the overhead.
We minimize the overall overhead
by using the single-direct detector as much as
possible, and use the
double-direct detector only at the expected end of
the computation
(typically determined by monitoring convergence).

{\em A crucial efficiency consideration} built into \FPD is that
it detects soft errors by placing detectors at time-tile boundaries (line 20
of Figure~\ref{fig:stencil-spec}).
This allows a 
programmer or an optimizer to transform
the statements within each time tile into any semantically equivalent form.
For example, a polyhedral optimizer
(e.g., Pluto~\cite{uday08cc} or PolyOpt~\cite{polyopt-web})
can optimize this stencil {\em without interference from the detector.}

\section{Optimized Detector Synthesis}
\label{sec:strategies}

At a high level,
\FPD operates by comparing the iterative and direct evaluations
within $dp_T$ bits of precision.
However,
during the time interval $t{\rs}={\rs}t_0$ to
$t{\rs}={\rs}t_0+T$,
not all points in the
detector's dependence cone
contribute to $dp_T$ bits of precision; this
is because of the round-off effects
of using (finite-precision) floating-point arithmetic.
This allows us to selectively carve out
points from the input space.
Two key concepts
\footnote{These are presented in more detail as {\bf ``Precision Driven Optimizations''}
in {\bf Appendix B} of the supplementary material for the interested reader.}
introduced in this section---namely that of 
\textbf{Essential Width} ($E_w$)
and
\textbf{Protected Width} ($P_w$)---helps define the
amount of computation that can be carved out while still
offering our guarantees.
Later in \S\ref{sec:placement},
we describe an optimized detector synthesis scheme based on
these concepts.

\FPD works by synthesizing detectors for the expected ranges of binades
that the computation begins with, and is also assumed to be present
at every certified baseline.
%
A T-step detector's support comprises of 
$(2*T*\vec{w}+1)$ points centered around $\vec{x}$, that is the
points in $X_D = [\vec{x} - \vec{w}T, \vec{x} + \vec{w}T]$, where $\vec{w}$
represents the footprint of one iteration of the stencil expressed
as a vector to include higher dimensional spaces.
Each point $x_i$ in the $X_D$ belongs to the tightest encompassing interval $I$.
Upon multiplying each $x_i$ by their respective T-step coefficients, $c_i$, generates
separate intervals for each individual product term denoted as 
\[
[ \underline{y_i}, \overline{y_i}] = [\min(c_i\underline{x_i}, c_i\overline{x_i}),
						\max(c_i\underline{x_i}, c_i\overline{x_i}) ]
\]

This results in a new set of $N$ points, $Y = [y_1, y_2, \dots, y_N]$,
where $N = (2T\vec{w} +1)$.

%
%
Define summation $S_Y = \sum_{j=1}^N [\underline{y_j}, \overline{y_j}]$
which falls in the interval
$[\underline{S_Y}, \overline{S_Y}]$.
%
%
If
$[\underline{y_i}, \overline{y_i}]$ does not influence $S_Y$ in the scope of $dp_T$,
then point $y_i$ is a candidate for removal.
We do this pointwise analysis for each $y_i$.
%
%
For this we define,
$S_{Y\setminus i} = \sum_{j=1, \neq i}^N [\underline{y_j}, \overline{y_j}]$
and its bounds
as $[\underline{S_{Y\setminus i}}, \overline{S_{Y\setminus i}}]$.
Now define
$d_{\min}(y_i)$ as the distance between the exponent of the lower bound of $S_{Y\setminus i}$ and the exponent 
of the upper bound for $y_i$:
\begin{equation} \label{eq:min-distance}
	d_{\min}(y_i) = \max(0,exp(\underline{S_{Y\setminus i}}) - exp(\overline{y_i})) 
\end{equation}
This equation and the next help check whether $y_i$'s {\em maximal} contribution
manages to affect the {\em minimal contribution} from all the remaining points (if not,
due to the exponent differences, the magnitude
contributed by $y_i$ gets ``shifted out''
during the process of normalizing the exponents while doing floating-point addition):
\begin{equation} \label{eq:maxContrb}
	maxContrib_p(y_i, S_Y) = p - d_{\min}(y_i)
\end{equation}
Equation $~\eqref{eq:maxContrb}$ gives the maximal precision contribution of
point $y_i$ to the final sum $S_Y$ when $S_Y$ is computed up to $p$ precision bits.
A $y_i$ becomes part of the exclusion set, if $maxContrb_p(y_i, S_Y) < 0$.
Note that $y_i = (c_i\times x_i)$ already includes one multiplicative term addding an extra half ulp error
to the analysis.



Now, for a detector placed at $A_u[\vec{x}, t_0+T]$, we define \textbf{Essential Width} ($E_w$)
as the width of the multidimensional rectangular region around $\vec{x}$
such that the direct evaluation over $E_w$
is {\em sufficient} to guarantee $dp_T$ precision at the detector.\footnote{The
  word ``sufficient'' is important because we might, in general, have 
  non-contiguous chunks of inputs that can be discarded. In \FPD, we extend $E_w$
  to make it a contiguous region.}
Specifically, the region is defined by extents $E_{wl}$ and $E_{wr}$ such that $E_w = E_{wl}+ E_{wr} +1$.
The caveat from equation $~\eqref{eq:maxContrb}$ is that all such excluded candidates
could {\em collectively}
 affect the output within the precision limits; therefore
 we must engage in an iterative process of considering subsets of
 points to carve out such that equation $~\eqref{eq:subset-ew}$ holds.
In the general case, $E_w$ is a vector and is written $\vec{E_w}$.
Figure~\ref{fig:ew_and_pw} 
illustrates $E_w$ for evaluation of a 1-D stencil over T-steps.
\begin{lemma}
	\label{False-positive}
	 In the absence of an error, direct evaluation over the essential width ($E_w$) ensures the correctness of the computation
	to at least $dp_T$ bits of precision at the detector location.
\end{lemma}

For all points in the region excluded from $E_w$ 
$\forall \tilde{X}: \tilde{X}\subset ([\vec{x}\pm \vec{w}T] - \vec{E_w})$,
the following holds \\ [-0.7em]
\begin{equation}
	\label{eq:subset-ew}
	\forall x_i \in \tilde{X}: maxContrib_{dp_T} \left(\sum (y_i = c_i\times x_i), S_Y \right ) \leq 0
\end{equation}
%

One form of false positives involves detection of soft errors when no
error affects any iteration point on which the detector depends (aka
the detector's dependence cone). The preceding lemma guarantees that,
despite the reduced evaluation cost, in the absence of errors
affecting the dependence code, the direct evaluation is equivalent to
the iterative evaluation within $dp_T$ bits and no soft error
notification (false alarms) is triggered.


\ignore{-->
\paragraph*{Minimizing direct evaluation cost: essential width (\ew{}).}
The precision of value computed using direct evaluation need not be
any greater than that produced using iterative evaluation.  This is
referred to as detector precision, denoted by $dp$. Due to
finite-precision arithmetic, the maximum possible contribution from
some points at time $t$ to a detector at time $t+T$ might be less than
this desired precision. The contributions from such points can be
ignoring in performing the direct evaluation. We define essential
width (\ew{}) as the width of a rectangular region centered
around $\vec{i}$ that includes any point at time $t$ that can contribute
to the most significant $dp$ bits of the value
$A_x[t+T][\vec{i}]$. Performing the direct evaluation using the
essential width, rather than the all possible incoming dependencies
significantly reduces the detection overhead without impacting
detection accuracy.
--}

\ignore{ -->
in the full support might contribute insignificant quantities
to the final value.  We define essential width (\ew{}) as the maximal
set of points from the full support needed to evaluate the detector to
provide correctness within the detector precision ($dp$). While full
support is necessary to evaluate the stencil in real space (that is
with infinite precision), when we compute with finite precision we
will incur round-off error. In the previous section, we identified the
minimal precision preserved between the real and floating-point
computation. We utilize this minimal precision as the detector
precision and \ew{} identifies those essential points that influence
the detector value within the detector precision. Using \ew{}
significantly reduces the computational overhead while ensuring good
detection coverage.
--}

\vspace{.5ex}
\noindent{\bf Covering multiple points with a single detector: {\bf Protected Width} (\pw{}):\/}
	As depicted in Figure~\ref{fig:error-propagation}, each point inside the dependence cone 
	of the detector has varying contribution depending on its
	effective path contribution. To guarantee that an error affecting a bit at
    an iteration point is detected by a detector protecting it, the minimum contribution of
    that bit must impact the $dp$ bits at the detector. This will result in
    the values computed by iterative and direct evaluation being different, triggering a soft
    error notification.

    To quantify the minimal possible influence a $y_i$ has 
	on the final sum $S_Y$, we define another distance metric $d_{\max}(y_i)$ as the distance between
	the exponent of the upper bound of $S_{Y\setminus i}$ and the lower bound of $y_i$. \\ [-1em]
		\vspace*{-0.4em}
	\begin{equation} \label{eq:max-distance}
		d_{\max}(y_i) = \max(0, exp(\overline{S_{Y\setminus i}}) - exp(\underline{y_i}))
	\end{equation}
	Then, the minimal influence $y_i$ has on the final sum when computed correct to $dp_T$ bits 
	of precision is evaluated as \\ [-1em]
		\vspace*{-0.4em}
	\begin{equation}
		minContrib_{dp_T}(y_i, S_Y) = dp_T - d_{\max}(y_i)
	\end{equation}

	A detector's reach can then be configured to guarantee an user defined precision (\udp{}) by
	carving out a set $Y_{udp}$ from $X$ such that: \\ [-1em]
	\begin{equation} \label{eq:udp-set}
		Y_{udp} = [y_i : minContrib_{dp_T}(y_i, S_Y) \geq udp,\quad y_i \in Y]
	\end{equation}

	The most significant \udp{} bits of precision of each point inside $Y_{udp}$ is then protected 
	if $S_Y$ is evaluated correctly to $dp_T$ bits of precision.
	Equation $~\eqref{eq:udp-set}$ leads to our primary soft-error detection model that provides the guarantee that
	an error affecting within the most significant $udp$ bits of points inside $Y_{udp}$ is detected.
	In our stencil model, for a detector placed at $A_u[\vec{x}, t_0+T]$, we define 
	\textbf{Protected Width} ($P_w(T,\ty{},udp)$) as the width of a multidimensional rectangular region
    centered around $\vec{x}$ such that with respect to a detector placed at time $t_0+T$ and spatial position $\vec{x}$,
	for each $\vec{p} \in \vec{x} \pm  P_w(T,\ty{},udp)$ at time $t_0$+\ty{}, the above guarantee holds. 

	\begin{lemma}
		\label{PW-guarantee}
An error affecting any of the MSB \udp{} bits of a point inside the protected width is detected.
	\end{lemma}

	For a point $y_i$ in $y$, if $y_i$ belongs to $Y_{udp}$ and its minimal precision contribution is $p_i$
	at the final output, then $p_i \geq udp$. If $y_i \in [\underline{y_i}, \overline{y_i}]$, then an error,
	$err_{y_i}$ affecting within the MSB $udp$ bits will be bounded by $err_{y_i} \geq 2^{exp(\underline{y_i}) - p_i +1}$.
	Since, we are matching $dp_T$ bits at the detector, hence the threshold of detectable error is bounded by 
	$2^{exp(\overline{S_Y})-dp+1}$.

For our guarantee to hold, 
\(		\hbox{generated error} \geq \hbox{error threshold}. \)
This means                
\(	2^{exp(\underline{y_i}) - p_i +1} \geq 2^{exp(\overline{S_Y})-dp_T+1}. \)
	Taking logarithms and simplifying the above terms leads to the following relation that must hold true 
	for all points inside the $P_w$ region:
\(	exp(\overline{S_Y}) - exp(\underline{y_i}) \leq dp_T - p_i, \)
and
\(	udp \leq p_i \leq dp_T - (exp(\overline{S_Y}) - exp(\underline{y_i})). \)

    In determining the protected region,
    where a detector is used to protect iteration points in multiple
    time steps, we choose \ty{} such that the protected width chosen
    at \ty{} is also valid for all time points $t_0 \leq t \le t_0 + \ty{}$:
    \begin{equation}
      \forall\ 0\le t\le \ty{}: P_w(T,\ty{},udp) \le P_w(T,t,udp)
    \end{equation}
    This enclosed region forms the
    protected region for the given detector	location.
	Figure~\ref{fig:ew_and_pw} illustrates the protected region in terms of $P_w$ and \ty{}
    steps for a given \udp{}.


\ignore{ -->
\todo{Fix the following paragraph}
Given a single detector, say at time $T$, the set of points in which
errors affecting \udp{} will be reflected in the detector is referred
to as the protected region. The Cartesian sub-region of the protected
region consisting of points in the same time step, say at $T-\ty{}$,
is represented by the {\em protected width} \pw{}. A single detector's
coverage is defined in terms of its \pw{}. Note that \pw{} is
influence by the detector precision, distance of time step being
protected from the detector, and \udp{}.  Figure~\ref{fig:singleBase}
shows \pw{} at \ty{} away from the current baseline.
--}


\vspace{.5ex}
\noindent{\bf Probabilistic detection:\/}
While the above strategy guarantees detection, the stencil structure
and user-specified \udp{} requirements can severely constrain the
protection region, incurring high detection overheads. To further
trade-off overheads for detection capability, we consider a detector
that only provides the detection guarantee on only a fraction of the
points in the protected region associated with a detector. Given a
specified coverage requirement $cov$, we choose the protected width
\pw{} such that the fraction of points with guaranteed protection is
at least $cov$. An error impacting the MSB \udp{} bits of an iteration
point is guaranteed to detected by its enclosing detector, if it is
among the points for which the detection is guaranteed. Alternatively,
such an error is detected with probability of at least $cov$.



\ignore{-->

Using the two approaches to evaluate a point in the stencil
computation space, we develop two detection strategies, one using a
one direct evaluation and another using a two direct evaluations.


\vspace{.5ex}
\noindent{\bf Single-baseline detector:\/}
Figure~\ref{fig:singleBase} shows the single-baseline detector. A
baseline signifies a state of the stencil that has been proved to be
correct and hence used for future analysis. In this type, our current
baseline forms the required support for the direct evaluation, with
the detector points at the apex of the T-step. In particular, we use
direct evaluation to predict an expected value to be computed by the
stencil T steps ahead. This value is then compared with the actual
value produced by iterative evaluation of the stencil over $T$ steps.
This strategy involves each detector requiring a single dot product of
the effective coefficients with the support points in the current
baseline.

\begin{figure}
\centering 
\includegraphics[width=0.9\columnwidth]{detector-method2.png}
\caption{Double baseline}
\label{fig:doubleBase}
\end{figure}

\vspace{.5ex}
\noindent{\bf Double-direct detector:\/}

Figure ~\ref{fig:doubleBase} shows the double-baseline detector. In
this type, instead of waiting $T$ steps, we use another baseline (not
yet verified) at \ty{} (denoted ty in figure) to once again predict
the value at $x$ at time $t_0+T$. The two predicted values are
compared within the required detector precision to detect errors in
point below the $t_0 + t_y$ baseline correct to $udp$ bits of
precision. In general, the dual-baseline detector has higher
computational overhead, due to two detector evaluations, but allows
for a lower detection latency (\ty{}) as compared to the
single-baseline detector ($T$). In addition, this detector can be used
when the actual stencil cannot be executed, say, because of
termination.
--}




\section{Optimized Detector Placement}
\label{sec:placement}

\label{sec:det-placement}

\vspace{.5ex}
\noindent{\bf Dampening errors:\/}
We consider a single event fault model wherein a
single fault occurring at time $t$ can transiently 
corrupt one or more participating arrays. Based on properties
discussed in the preceding section, an error at time $t$
escaping detection by the nearest set of detectors must have 
impacted less significand bits. 
Suppose the coefficient set $C=\{c_i\}_{i=1}^n$ denotes the set of forward contributions
of a point in the stencil. Then Lemmas-~\ref{never-amplify} and \ref{error-dampening}
given below hold under the following assumptions:
\begin{compactitem}
	\item $\forall i, |c_i| \leq 1$ : This implies all forwards contributions
			move the error magnitude towards the least significant bits
	\item $\sum_i c_i \leq 1$ : The collective error spread from an affected 
			point to all neighboring points in the forward path is less
			than equal to itself 
	\item There is no error cancellation within the scope of a detector.
		  Multiple detectors may trigger a detection event as long as the triggering
		  errors are localized to the scope of those detectors
		  without any intervening error cancellations from its neighbors.\footnote{Such
                  cancellations, if they occur,
                  do not affect the final output.}

\end{compactitem}

	The first two conditions are driven by the stability criterions and smoothening impact
	of the stencil.

\begin{lemma}
	\label{never-amplify}
	An error may never amplify in the forward path.
\end{lemma}

\begin{lemma}
	\label{error-dampening}
	Error dampening guarantees an error missing detection does not affect future 
	computations within the required precision limits. 
\end{lemma}



%
Under the dampening error model, the detectors can be arranged to cover the iteration space as follows:

\begin{compactitem}

\item
  \textit{Certified baselines:}
  Error dampening allows us treat all computations up to a time $t$ that have passed the error detection
checks as effectively being error-free. We will use these checked points at a given time step $t$
as the new {\em certified baseline} to perform direct evaluation of the next detector.

\ignore{--
\paragraph{Dampening Errors}
We consider a single fault model wherein a
single fault occurring at time $t$ can transiently corrupt one or more
participating arrays.
Based on the properties discussed in the preceding section, an error
at time $t$ escaping detection by the nearest following detector must
have impacted less significant bits than the guaranteed $udp=k$ bits
of precision. Suppose there are multiple such error points due to a
single-event fault that escaped detection by the nearest detector. Let
$E$ denote the $\max$ of these error terms. If our detectors
guaranteed correctness withing $k$ bits of precision, that the error
escaped detection by the nearest detectors implies $|E| \le 2^{p-k+1}$,
where $p$ is the number of precision bits in the floating point
system. Considering the worst-case scenario, every point at time $t$
could carry an error of this magnitude at time step $t$. Since the
coefficients of all the points to a future time step sums to less than
equal to $1$, it means the total contribution of this error ($E$) to a
future value in one or more of the participating arrays is less than
equal to itself (strictly non-increasing), thus remaining confined in
the $(p-k)$ least significant (LSB) bits of all future
contributions. Hence
guaranteeing that $udp$ precision is protected in all computations
points below a baseline time step ensures that, throughout the
computation, all points collectively covered by detectors are
protected at least to $udp$ precision. Intuitively, in the stencil
programming model considered, errors do not {\em amplify} and corrupt
more significant bit positions.


Error dampening allows us treat all computation up to a time step $t$
that have passed the error detection checks as effectively being
error-free. We will use these checked points at a given time step $t$
as the new {\em certified baseline} to perform direct evaluation of
the next detector.



--}

\item \textit{Horizontal detector placement:}
If a set of detectors is placed horizontally,
i.e., at the same time step $t$,
$P_w$ apart in all dimensions,
then the points at time $t$, once they pass the detection checks,
will constitute a new certified
baseline. Thus for a $d-$\! dimensional stencil of size 
$N{\rs}={\rs}\{N_i\}_{i=1}^d$ along each dimension, if the detectors are placed
$P_w{\rs}={\rs}\{pw_i\}_{i=1}^d$ apart along each dimension, the number
of detectors required is $\prod_{i=1}^d ({N_i}/{pw_i})$.

\item \textit{Vertical detector placement: stacking.}
Because the detector characteristics are influenced by the input
range, at every baseline the input range needs to be
computed. Alternatively, a bound on the input range can be
predetermined based on the initial/boundary conditions and used for
the entire computation. In both scenarios, the detectors can be
vertically stacked to create new certified baselines as execution
moves forward.

\end{compactitem}
 \begin{figure}
 \centering 
 \includegraphics[width=0.6\columnwidth, height=0.3\columnwidth]{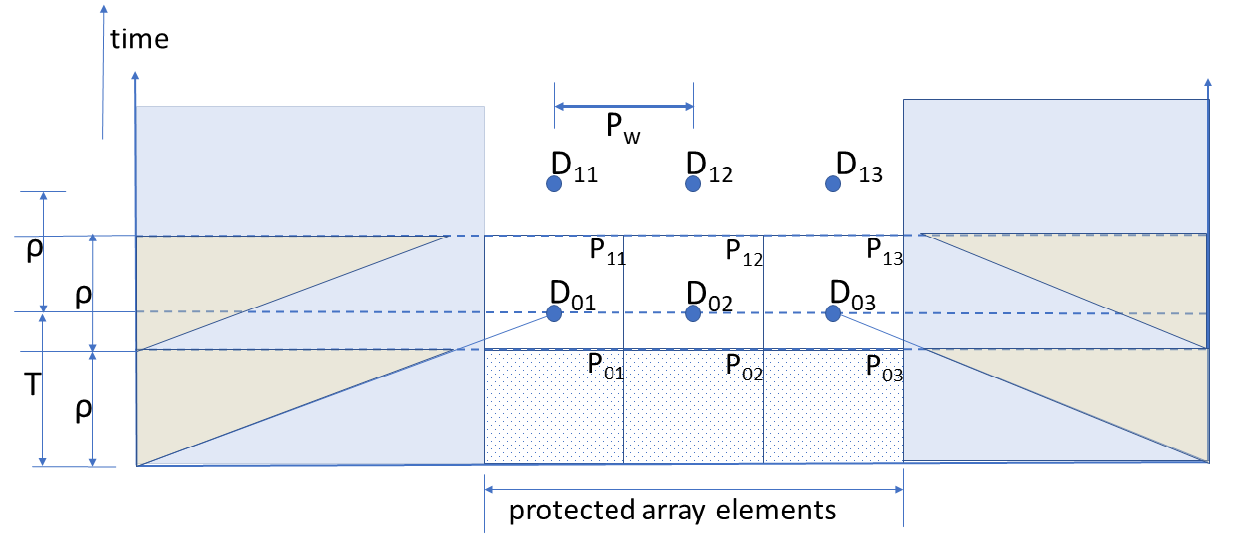}
 \caption{Detector arrangement for a 1-d stencil. Horizontal: array elements; vertical: time steps}
 \label{fig:det-stack}
 \end{figure}

\vspace{.5ex}
\noindent{\bf Illustration:\/}
Figure~\ref{fig:det-stack} illustrates the placement of detectors for
the first two certified baselines for a 1-d stencil. $D_{01}$,
$D_{02}$, etc. show the placement of the first set of detectors at
time step T. These are computed from the array values at time step
0. $D_{11}, \ldots$, the second set of detectors at time $T+\ty{}$,
computed from the certified baseline at time step \ty{}. The detectors
are vertically spaced \ty{} time steps apart, and horizontally spaced
$\pw$ apart. The protected regions are marked as squares and labeled
at the right top. For example, $P_{01}$ is the region protected by
detector $D_{01}$. The shaded region denotes the boundary region
unprotected by the detectors. Of this region, the shaded gray region
has no influence on the detectors protecting points in that time step
(outside their dependence cones). Errors in this region are not
detected. The remaining region has an influence on at least one of the
detectors. Errors in these iteration points might be detected, but are
not part of the guaranteed coverage.

	\begin{wrapfigure}{l}{0.3\textwidth}
    \begin{minipage}{0.3\textwidth}
\begin{equation} \label{eq:coveqn}
	\% cov =  \prod_{i=1}^d \left (\dfrac{N_i - 2wT}{N_i}\right ) * 100
\end{equation}
	\end{minipage}
	\end{wrapfigure}

\vspace{.5ex}
\noindent{\bf Boundaries:\/}
Although our method
is applicable to the boundary points,
the contributions from the boundary points need to computed differently from the interior,
incurring additional overhead. For cases where time dependent boundary 
conditions and Neumann boundary conditions~\cite{BVP} exists, \FPD lacks 
prior knowledge of how the boundary points are forced in the intermediate time tiles.
For fixed Dirichlet boundary conditions we can provide full coverage.
To simplify the
implementation, we ignore protection for the boundary
regions in the generalized cases. Specifically, any computation point influenced by the
boundary between certified baselines is not checked. Note that this reduces the
maximum possible coverage to be below 100\% , however still providing an accurate bound on the
guaranteed coverage per equation $~\eqref{eq:coveqn}$ akin to probabilistic coverage that
exemplifies the flexibility afforded by trade-off detection quality versus cost.

%
An error
affecting such an unprotected region can spread to the rest of the
computation space and result in erroneous output. This is akin to
probabilistic coverage, with the probability of detection reduced by
fraction of the total computation space left unprotected at the boundaries.
Let $N_i$ denote the problem size along $i'th$-dimension in a d-dimensional problem.
Equation $~\eqref{eq:coveqn}$ expresses the percentage of covered region for a detectors instantiated with $Tstep=T$.
Here, $w$ is the stencil width\footnote{The implementation uses the exact expression,
  accounting for stencil asymmetry and non-Cartesian shapes.}.
  



\vspace{.5ex}
\noindent{\bf Optimal detector configuration:\/}
%
For a given choice of detector, the essential width \ew{} is a
function of the $T$step, the input exponent range, and the
detector precision ($dp$). $P_w$ additionally depends on \udp{}
and \ty{}. The cost function is evaluated as the relative cost with
respect to the actual stencil evaluation.
%
Section~\ref{sec:offline} details the cost function and the offline analysis used to
construct optimal detector configurations.
\ignore{-->
  Because
the double-baseline detector has an additional expression to be
evaluated, it almost doubles the detection cost as compared
to the single-baseline detector. Hence, in terms of cost, single-baseline
detector remains the most effective.
--}





\section{Overall Algorithm}
\label{sec:offline-online}

\FPD performs an offline manuever to obtain a look-up table
with optimal detectors configurations for the detectors. 
For each anticipated choice of \udp{}, probability of detection, and
anticipated input range, the offline
analysis (\S\ref{sec:online}) determines the detector
configuration to be used during online execution. A detector
configuration consists of:
T, the distance at which the single-direct detector is evaluated;
\ty, the number of iterations between certified baselines; \ew{}, the essential
width to evaluate the detector; Detector coefficients for the support
in the essential width; and \pw{} the spatial separation between detectors in the same time step. 
These parameters completely specify the detector configuration at runtime.

\subsection{Offline Determination of Detector Configurations}
\label{sec:offline}

\vspace{.5ex}
\noindent{\bf Constraints on detector configuration:\/}
We use offline analysis to efficiently identify optimal
detector configurations for use during online execution.
To constrain the search space of possible configurations,
we use an upper-bound on the $Tstep$ (say,
$T_{\max}=256$ in our evaluation) that one might use for the detector
evaluation. Larger $Tstep$ values are useful and often minimize the
detector operational cost. However, this requires larger coefficient
sets, needed for the direct evaluation, to be stored in memory. For a
$d-$dimensional stencil with $N-$arrays, the space overhead encountered to
store the effective coefficient space is $O(N\cdot T_{\max}^{d+1})$.
The range of \udp{} values to be considered is bounded by the number of bits
($1-53$) and the maximum precision ($dp$) that can be preserved by
both direct and iterative stencil evaluations informed by the round-off error
analysis.
We determine the
detector configurations for a finite set of coverage choices between 0\% and 100\%,
corresponding to the fraction of points in a detector's protected region
that are protected to the desired \udp{} bits. The actual coverage including
the boundary is computed online.


\vspace{.5ex}
 \noindent{\bf Relativized input exponent ranges:\/}
  Consider two exact floating point numbers represented in triplet form as
  $a_f = (s_a, m_a, e_a)$ and $b_f = (s_b, m_b, e_b)$ where the mantissa $m_a$ and $m_b$
  are represented in $p=53$ bits. For a floating point addition (subtraction), 
  the mantissa of the number with the lower of the two
  exponents has to be shifted by $|e_a - e_b|$ to match their exponent values.
  Thus, the binary addition (subtraction) of the mantissa
  depends the relative distance
  between the exponents for shifting (not on actual exponents).

  In stencils, update rules are modeled as weighted sums involving only addition (subtraction based on sign)
  and multiplication. Multiplication by exact scalars involves binary multiplication of the mantissa,
  followed by addition of the exponent terms, maintaining linearity in the exponents. For example, let the coefficients
  associated with $a_f$ and $b_f$ be $\alpha_1 = (s_{\alpha_1}, m_{\alpha_1}, e_{\alpha_1})$ and 
  $\alpha_2 = (s_{\alpha_2}, m_{\alpha_2}, e_{\alpha_2})$, respectively. 
  The product $\alpha_1 a_f$ will have an
  exponent of $(e_{\alpha_1} + e_a)$ and $\alpha_2 b_f$ will have $(e_{\alpha_2} + e_b)$. 
  Thus, the relative distance  of exponents between these two terms will 
  $|(e_{\alpha_1}-e_{\alpha_2}) + (e_a - e_b) |$ which depends essentially on the coefficient's exponents (that 
  remain unchanged) and the relative operand exponent differences.

   This fact can be utilized to map   different interval ranges to some canonical exponent range that models the maximum relative distance
  of points inside the interval. To do this, we scan the input to find the smallest data point (in magnitude) 
  and the largest data point (in magnitude) and characterize that interval with the exponent difference
  between these two data points.
  For an input interval, if $(m_a,e_a)$ represents the smallest value (in magnitude) and
	$(m_b, e_b)$ is the largest value (in magnitude) seen in the interval, then factoring out $e_a$
	produces the following mapping\\[-0.8em]
  \begin{align*}
  [(m_a, e_a), (m_b, e_b)] &\equiv (1,e_a) [(m_a,0), (m_b, e_b-e_a)] 
  \end{align*}

  We can further factor out the corresponding mantissa and increment the mapped exponent interval width by 1 to have a larger
  bounding interval for the given input range. Since our analysis is conservative, bounds that hold for larger exponent ranges
  also hold for smaller exponent ranges.\\[-0.8em]

  \begin{equation}
  \begin{split}
 	 [(m_a, e_a), (m_b, e_b)] \equiv (m_a,e_a) [(1,0), (\dfrac{m_a}{m_b}, e_b-e_a)] 
   \equiv [(1,0), (1, e_b-e_a +1)]
  \end{split}
  \end{equation}

The operation of a stencil on an input range can be {\em relativized}
to a small number of canonical exponent ranges.  This allows us to use
the offline analysis performed on specific exponent ranges on
different set of actual exponents that belong to the same range. In
our experimental setup, we profiled exponent ranges form 0 through 20
($exp_{\max}=20$). The maximum \udp{} is set as $40$.  Given the range
of exponents and \udp{} choices, the offline algorithm construct a
lookup table that returns a 5-tuple $(T,\ty{}, dp, \pw{}, \ew{})$
corresponding to an optimal detector configuration for the given
interval, required \udp{}, and minimum detection probability. \\ [0.5em]




\vspace{.5ex}
\noindent{\textbf{Offline algorithm}:\/}\footnote{\label{suppl-algo-desc}The algorithmic listing is available in the supplementary material (Appendix C).}
We consider all T and \ty{} such that
$ T \leq T_{\max}, \ty{} < T $.
%
%
For a d-dimensional stencil, \pw{} is
	d-dimensional vector, corresponding to protected region of iteration points centered at the detector.
    \ew{} is represented by a rectangular region with a computed number of points along left and right of
    the detector along each direction. 
%
	The cost function is derived as the ratio of the total detector overhead
	to the total cost of the stencil application evaluation. Post simplification
	of the cost function for a 5-tuple
	detector configuration $(exp,udp,T,\ty{},Coeffs)$ is given as\\[-1em]
	\begin{wrapfigure}{l}{0.2\textwidth}
    \begin{minipage}{0.2\textwidth}
\begin{equation}
  Cost = \dfrac{1}{\rho}\prod_{i=1}^d \left(\dfrac{ew_i}{pw_i}\right) \label{eq:cost-function}
\end{equation}
	\end{minipage}
	\end{wrapfigure}
 derived fraction of overhead cost to total stencil compute cost.
Even though the tuple elements $exp$, $T$, $udp$, and $Coeffs$ do
	not explicitly appear in the cost function, they implicitly influence
	the cost through \ew{} and \pw{}.

The algorithm takes as input parameters for which an offline
profiles need to be determined: the maximum number of time steps
(Tmax), the set of input ranges (as exponents in exp\_set), set
of \udp{} values (as udp\_set), and probabilistic coverage values
(cov\_set). The algorithm determines the configurations one input
range choice at a time.  Using floating-point round-off analysis, the
maximum number of bits that will be preserved for each possible time
step $t$ is computed (as maxdp). For each $t$, the cost of evaluating
each detector is computed as the product of the \ew dimensions
to guarantee $maxdp[t]$ bits of precision of the direct evaluation $t$
time steps away. Then, for each candidate protected region, the
fraction of points with guaranteed coverage of $b$ bits (where $b$ is
a candidate \udp{} in the input parameter udp\_set) is computed. If
this fraction is greater than a desired coverage and if the associated
cost is lower than that of any configuration seen thus far, this
configuration is chosen. After evaluating all feasible solutions, the
algorithm returns the last chosen configurations.

\ignore{-->

The function \textsc{EvalEssentialWidth}
attempts to find the maximal set of points in the full support of the
detector and derive a reduced support set while still preserving the
required detector precision. It starts with the full support (S). Let
$I$ and $X$ denote partition of the full support (set of all points
contributed to the detector) into points included and excluded,
respectively, from \ew{}. This partitioning is valid if the maximum
contribution by all the points in $X$ cannot influence the minimum
contribution of the points in $I$ within the computation's precision
$dp$ at the detection point $T$. Otherwise, some points in $X$ are
moved to $I$ until this is true. 
The minimum and maximum
contributions are computed using the full set of coefficients and the
exponent range being considered. To simplify the generated code, we
only considered Cartesian regions to bound points in $I$, and denoted
by \ew{}. 
--}

\ignore{-->
The function \textsc{EvalProtectionWidth}
finds the largest set of points in which errors in the \udp{} most
significant bits can be detected using a detector with $dp$ precision.
For multi-array problems, we maximize \ew{} over all the arrays by
calling $\textbf{select\_best\_ew}$, such that the final \ew{}
satisfies each of the contributing arrays.  The function
\textsc{EvalProtectedWidth} profiles the stencil over a set of $udp$s,
where for each udp it tries to find the maximal set of points that
contributes at least $udp$ bits of precision to the final value of the
weighted sum. In this case, for a given support (S), for a point $p_i$
in S, find the partial sum with maximal contributions from all other
support points and minimal contribution from itself. If the minimal
contribution from $p$ is significant enough to contribute at least
$udp$ bits of precision to the final sum, we mark it as protected and
place in \pw{}, else the point is excluded from \pw{}. The routine
$select\_best\_pw$ finds the minimal \pw{} over the participating
arrays while $gen\_pw\_pyramid$ further trims the \pw{} to
ensure\\
$	\pw{}(\ty=t_1) \geq \pw{}(\ty=t_2) \implies t_1 \leq t_2$
--}

The cost of the offline procedure is dominated by dimensionality of
the space to be explored. While this exhaustive evaluation of every
feasible configuration can be
expensive, for the benchmarks considered (in the next
section), each offline procedure completes within several
minutes. Search-space exploration techniques might further lower
this cost.
%
%
%
%
%
%
%
%
%
%
%
%
%
%
%
%
%
%
%
%
%
%
%
%
%
%
%
%
%
%

\subsection{Online Detector-Embedded Execution}
\label{sec:online}
We briefly discuss the algorithm%
\footref{suppl-algo-desc}
to obtain the detector configurations and embed the
detectors within the original stencil code.
During online deployment of \FPD's detectors,
the input values are scanned to
compute the input exponent range to be handled in the
floating-point space.
To account for the unprotected boundary regions, \FPD
maps user's input coverage value to an {\em equivalent internal
coverage value} (as in\_cov) that corresponds to the detection probability
on the grid excluding the boundaries. The adjustment is done with the guarantee 
that the {\em fraction of points covered due to (in\_cov) is at least as many required
by cov over the entire computation space}. If \FPD fails to find
an equivalent mapping, 
it conveys an error message for unsupported coverage.
The input range, user-specified \udp{}, and modified detection probability (in\_cov)
are used to lookup the detector configuration.

After an initial evaluation of the detector and the
stencil for \ty{} iterations,
the execution of the stencil
is split into two segments:
iterations till the next detector evaluation (\texttt{eval\_detector})
and iterations till the next detector check
(\texttt{detector\_check}). Thus, the stencil is executed with
{\em interleaved detector evaluation and checking}.
Note that
the algorithm assumes \ty{} is greater than half the detector evaluation
T-step. This scenario requires at most two ``live'', i.e., unchecked
detectors per spatial position in the iteration space, at any time.
If not, at each spatial position,
multiple detectors need to be evaluated and retained until they are
checked.
When all iterations have been executed, the iterations
past the last certified baseline need to be checked. 
These leftover iterations are checked using the
trailing detection strategy.
%




\section{Evaluation}
\label{sec:evaluation}






\newcommand{\tilphi}{\tilde{\varphi}}
\newcommand{\opn}{\operatorname}
\newcommand{\re}{\operatorname{Re}}
\newcommand{\im}{\operatorname{Im}}
\newcommand{\diag}{\operatorname{diag}}
\newcommand{\phm}{\phantom{-}}
\newcommand{\mbb}[1]{\mathbb{#1}}
\newcommand{\mb}[1]{\mathbf{#1}}
\newcommand{\mc}[1]{\mathcal{#1}}
\newcommand{\argsr}{(\theta,\partial_\theta,r\partial_r)}
\newcommand{\argsl}{(\theta,\partial_\theta,\lambda)}
\newcommand{\rdr}{r\partial_r}
\newcommand{\jd}{\displaystyle}
\newcommand{\js}{\scriptstyle}
\newcommand{\jt}{\textstyle}
\newcommand{\jss}{\scriptscriptstyle}
\newcommand{\lami}{(\lambda-\lambda_i)}
\newcommand{\lamo}{(\lambda-\lambda_0)}
\newcommand{\der}[2]{\frac{\partial #1}{\partial #2}}
\newcommand{\half}{\frac{1}{2}}
\newcommand{\pd}[2]{\frac{\partial#1}{\partial#2}}
\newcommand{\p}[2]{\frac{\partial#1}{\partial#2}}










\newcommand{\dudt}{\ensuremath{\frac{\partial u}{\partial t}}}

\begin{figure*}
  \footnotesize
  \begin{minipage}{0.20\textwidth}
    \textbf{Parabolic (heat):} \\
    $\frac{\partial u}{\partial t} = \nabla^2 u(x,y) + f$\\
    $\alpha{}=3, \zeta{}=1.2$\\[2.5em]
  \end{minipage}%
  \begin{minipage}{0.30\textwidth}
    [Dirichlet initial/boundary condition]\\
    h1. f = $\zeta{} -2 - 2\alpha{}$, $u = 1 + x^2 + \alpha{}y^2 + \zeta{}t$\\
    h2. f = $0$                     , $u = 1 + x^2 + \alpha{}y^2 + \zeta{}t$\\
    h3. f = $2\zeta{} -2 - 2\alpha{}$, \\
        $u = 1 + x^2 + \alpha{}y^2 + \zeta{}t^2$\\[1em]
  \end{minipage}%
  \hfill
  \begin{minipage}{0.45\textwidth}
	  [Neumann initial/boundary condition] $f=0$\\
    h4. $u(x,y,t)= e^{-\pi{}^2t/2}\sin\pi{}(\frac{x+y}{2})$\\
    h5. $u(x,y,t)= 4 + e^{-\pi{}^2t/2}\cos\pi{}(\frac{x+y}{2})$\\
    h6. $u(x,y,t)= 2 + \\
    e^{-\pi{}^2t/2}\left[ \sin\pi{}(\frac{x+y}{2})+\\ \cos\pi{}(\frac{x+y}{2}) \right]$\\ [-2em]
  \end{minipage}\\
  {\centering\rule{\textwidth}{0.5pt}}\\
  \begin{minipage}{0.30\textwidth}
    \textbf{Poisson equation:} \\
    $-\frac{\partial^2 u}{\partial x^2} -\frac{\partial^2 u}{\partial y^2} = \mbox{RHS}$\\

    [Dirichlet initial/boundary condition]\\
    p1. RHS $= -6, u = 1 + x^2 +  y^2$ \\
    p2. RHS $= -6(2+x+y), u = 1+x^3+y^3$ \\
    p3. RHS $= -2-12y, u = 1+x^2+2y^3$
    \end{minipage}%
   \hfill
  \begin{minipage}{0.25\textwidth}
    [Deflection in a membrane]\\
	  $p_1 = 4 e^{-5((x - 0.6)^2 + (y - 0.6)^2)}$\\
    $p_2=2 e^{-5((x - 0.3)^2 + (y - 0.3)^2)}$\\
    $p_3=4 e^{-5((x - 0.3)^2 + (y - 0.6)^2)}$\\
	  p4. RHS = $p_1, u = 0$\\
	  p5.RHS = $p_1+p_2, u = 0$\\
	  p6. RHS = $p_1+p_2+p_3, u = 0$
  \end{minipage}%
  \begin{minipage}{0.41\textwidth}
	  [Neumann initial/boundary condition]\\
    p7. RHS = $ 10e^{-\frac{(x-0.5)^2 + (y-0.5)^2}{0.02}}$, $\frac{\partial u}{\partial n}(x, y) = - \sin(5x)$\\
    p8. RHS = 0, $ \frac{\partial u}{\partial n}(x, y) = - \sin(5x)$\\
    p9. RHS = $20e^{-\frac{(x-0.25)^2 + (y-0.25)^2}{0.01}}$, $\frac{\partial u}{\partial n}(x, y) = - \sin(5x)$\\[2em]
  \end{minipage}
  {\centering\rule{\textwidth}{0.5pt}}\\
  \begin{minipage}{0.37\textwidth}
    \textbf{Hyperbolic (Second-order wave):} \\
    $\frac{\partial^2 u}{\partial t^2}(x,y,t)= c^2\nabla^2 u(x,y,t)$\\

    [initial/boundary conditions: $c=1$, $\dudt = 0$]\\
    w1. $u(x,y,t) = \\ \cos(\sqrt{2}\pi{}t)\sin(\pi{}x)\sin(\pi{}y) +x^2 - y^2$\\
    w2. $u(x,y,t) = \\ \sin(\sqrt{2}\pi{}t)\cos(\pi{}x)\cos(\pi{}y) +x^2 - y^2$\\
    w3. $u(x,y,t) = \\ \sin({\sqrt{2}\pi{}t})\cos(\pi{}x)\cos(\pi{}y) +x^2 - y^2$\\[0.5em]
  \end{minipage}%
  \hfill
  \begin{minipage}{0.35\textwidth}
    [initial and boundary conditions: $c=0.7$, $\dudt = 0$]\\
    
    w4. $u(x,y,t) = 16 + \\ 2\sin(\frac{\pi}{4}x)\sin(\frac{\pi}{4}y)\cos(\frac{\pi}{2}c^2t)$\\
    w5. $u(x,y,t) = 16 + \\ \sin(\frac{\pi}{2}x)\sin(\frac{\pi}{2}x)\cos(\frac{\pi}{4}c^2t)$\\
    w6.$u(x,y,t) = 16 + \\ 2\sin(\frac{\pi}{2}x)\cos(\frac{\pi}{4}x)\sin(\frac{\pi}{2}c^2t)$\\[3em]
  \end{minipage}%
  \begin{minipage}{0.27\textwidth}
    \textbf{Hyperbolic (convection diffusion):} \\
    $\dudt{} + \alpha\nabla{}u(x,y) = \zeta\nabla^2u(x,y)$\\
    initial/boundary condition from\\ $u(x,y,t) = \frac{1}{4t+1}e^{\frac{-(x-\alpha{}t-0.5)^2 - (y-at-0.5)^2}{\zeta(4t+1)}}$\\
    
    c1. $\alpha = 0.8, \zeta = 0.01$\\
    c2. $\alpha = 0.4, \zeta = 0.4$\\
    c3. $\alpha = 0.1, \zeta = 0.8$
  \end{minipage}%
  \caption{\label{fig:benchmarks} Benchmark PDEs (p1--p9,h1--h6,w1--w3,c1--c3) solved
    using the stencil finite-difference method. Initial and
    boundary conditions are derived from the equations provided. All
    stencils span the $[0,1]$ spatial domain and are run 4000
    time steps.}
\end{figure*}

\vspace{.5ex}
\noindent{\bf Benchmarks:\/}
We evaluate \FPD{} on the stencil kernels shown in
Figure~\ref{fig:benchmarks}.  We choose a set of benchmark problems
from elliptic, parabolic, and hyperbolic PDEs evaluated as
stencils  using the explicit finite difference method. Benchmark
examples considered are Poisson equation with different combinations of
Dirichlet and Neumann boundary conditions, heat equation with
different initial and boundary conditions, and second-order wave and
convection-diffusion equations. Our error analysis and optimization
techniques are based on the input data interval and not the exact 
values. We test our hypothesis rigorously by exercising each benchmark
equation with multiple initial and boundary conditions.
In addition to the exponent and sign bits, fault-injection-based evaluation 
was conducted for a user-defined precision of 15 mantissa bits.
This corresponds to a total protection of 27 most significant bits.



\vspace{.5ex}
\noindent{\bf Experimental setup:\/}
All benchmarks are compiled using ICC 18.0.5 with \texttt{-qopenmp
-O3} options and executed on dual 14-core Intel Xeon CPU E5-2680v4 2.60GHz
CPUs system (total 28 processor cores) with 64GB of RAM.
The Pluto optimized code were generated with `PLUTO version 0.11.4-350-g8debc44'.


\vspace{.5ex}
\noindent{\bf Detector coverage excluding boundaries:\/} %
To reduce the complexity of the detection strategy, we did not protect
statement instances impacted by the boundary between certified
baselines. This limits the maximum coverage that can be guaranteed by
our implementation. For all benchmarks and configurations evaluated,
we observed this fraction to be greater than 90\% of the computation
space, demonstrating the simplification does not significantly
limit detection ability.

\vspace{.5ex}
\noindent{\bf Space overheads:\/}
At stencil runtime, \FPD requires space to store:
\begin{itemize}
\item the set of all possible coefficients of interest,

\item the direct evaluation configurations for specific input ranges and precision needs,  and

\item the result of the selective direct evaluation by each detector.

\end{itemize}

We bound the offline analysis with $T_{\max}\rs=\rs 256$ iterations and require the full
coefficient set for $T\rs=\rs 0$ to $T_{\max}$.
%
This allows a runtime choice of time tile size between 1 and 256. 
This incurs a space overhead of $128MB$. While incurring a large memory footprint, only a
fraction of this data--one rectangular portion the single-direct detector--is
used during online execution and the accessed portions are reused for all detectors at a time step.
In addition, the trailing detector accesses one rectangular portion determined by the number of iterations
not executed by the last tile. 
This results in good cache behavior, limiting their impact on overall performance.

The offline configuration lookup table is built by profiling
over 20 exponents, 40 $udp$ and coverage values from 0 to 100\% (in increments of 5\%).
Thus, the offline lookup table is indexed by a 3-tuple of $(exp,udp,cov)$, with each
lookup returning a 6-tuple configuration of $(T, \rho, dp, \vec{P_w}, \vec{E_{wl}}, \vec{E_{wr}})$, where
$\vec{E_{wl}}$, $\vec{E_{wr}}$, and $\vec{P_w}$ are $d-$dimensional vectors for a $d-$dimensional stencil program. Here, $\vec{E_{wl}}$ and $\vec{E_{wr}}$ corresponds to the left and right extents of $\ew$.\footnote{$\vec{E_{wl}}$ and $\vec{E_{wr}}$ are equal in the case of symmetric stencils.}
\FPD scans the data at runtime to determine the necessary exponent range.
This is combined with the user-specified precision and coverage to index the
lookup table and obtain the corresponding detector configuration.
For a 2-dimensional problem, a key-value pair in the table occupies
a space of 48 bytes. Over all the profiled configurations, the configuration lookup table requires 
a total space of $\approx$750KB per benchmark.

In addition to the coefficients and the lookup table, we require one floating-point number
per array at each detector location to store the value computed through selective direct evaluation. 
The maximum number of instantiated detectors comes to 
around 387K for the wave benchmarks. For most other benchmarks is
around 100K for a problem size of $10K\! \times\! 10K$ and $T_{\max}\rs=256$.

\begin{figure*}
\centering
\includegraphics[width=\textwidth]{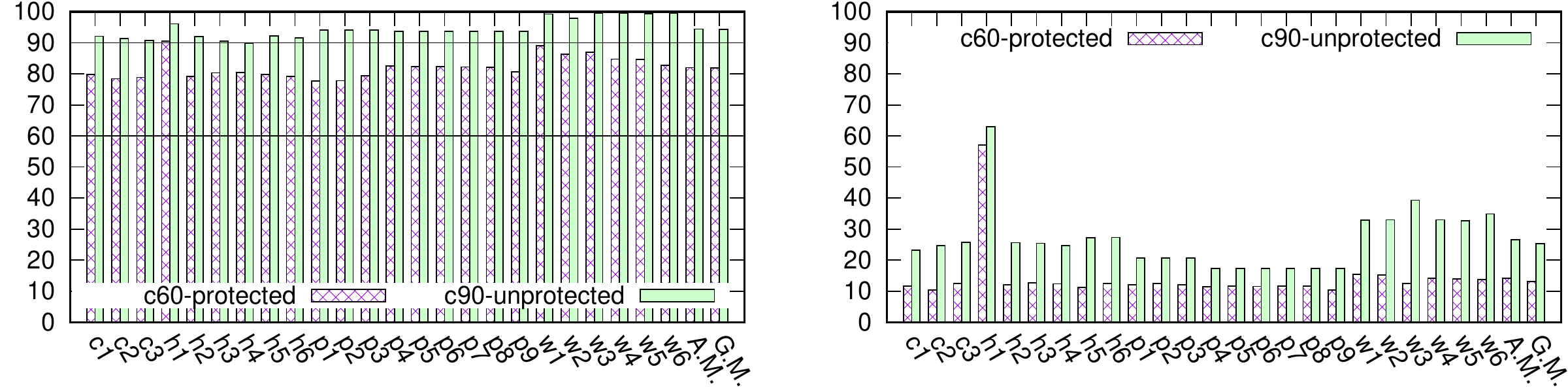}
\caption{\label{fig:detection-rate}
Detection rates of single bit-flip error injections in bits protected (left) and unprotected (right) by the detector
(A.M.: arithmetic mean; G.M.: geometric mean)
}
\end{figure*}


\subsection{Software Bug Detection}
\label{sec:sw-bug-detection}

We evaluate the effectiveness of \FPD{} in detecting
logical errors injected into the code generated for the stencil
programs in Table~\ref{fig:benchmarks} optimized by
Pluto~\cite{Bondhugula:2008:PAP:1375581.1375595}.  Specifically, we
inject three classes of bugs injected in the evaluation of
PolyCheck~\cite{DBLP:conf/popl/BaoKPRS16}, a tool designed to check
errors introduced by loop transformers: incorrect loop bounds, invalid
array accesses, and invalid loop reorderings.

For each software bug introduced, we evaluated its impact using
bitwise comparison of the result with that of the non-buggy
version. Any mismatch is treated as a bug. Some bugs might
not result in an erroneous result for the problem instances evaluated.
\FPD{} cannot aid in their detection.

The bugs introduced in the source potentially affect multiple runtime
operations, making them easier to detect that soft errors. Exploiting
this, we evaluate \FPD{}'s effectiveness in detecting software bugs
when deployed with a small \udp{} value of 4. For each benchmark, the
default deteector precision ($dp$) determined for this \udp{} is used
in the evaluation.

\vspace{.5ex}
\noindent{\bf Pluto-generated code:\/}
To generate optimized versions using Pluto, we implemented all the
stencils in the form of affine loop nests enclosed in \texttt{scop}
pragmas. This code input to Pluto has consists of median 53 source
lines of code (SLOC)\footnote{Source lines of code are measured using
sloccount tool} per benchmark. Pluto takes
these \texttt{scop}-annotated benchmarks as input to generate
optimized versions with median 1157 SLOC. The median of the number of
loops in the transformed code was 365 across all the benchmarks with a
median nesting depth of 6. In checking such large and complex
generated codes, tools such as \FPD{} are essential.
 

%
%

\vspace{.5ex}
\noindent{\bf Loop bound and array access bugs:\/}
We automated the injection of loop bound bugs.  For loop bounds, the
injected bug either offsets the lower bound by one in the positive
direction or offsets the upper bound by one in the negative direction.
%
We automated the injection of array access bugs.
Specifically, array accesses were made incorrect by either dividing or
multiplying the indexing term by two.  

\vspace{.5ex}
\noindent{\bf False positives:\/}
Some bugs do not result in any difference in our bitwise comparison of
the result with the non-buggy version. Therefore, our approach incurs
no false positives in our evaluation. This is because some
source-level bugs do not manifest at runtime.  For example, a loop
iterator might be constrained by multiple loop bounds
(e.g., \texttt{min} or \texttt{max} of multiple expressions), with the
loop bound never reaching the error-injected expressions.\footnote{An
 example can be found in the supplementary material in Appendix D. } In a
situation with multiply nested loops where the indexing for an inner
loop depends conditionally on outerloop indices, a bug impacting the
outer loop bound doesn't often affect the inner loop and hence doesn't
impact the stencil's output.

Table~\ref{table:software-bugs} summarizes \FPD{}'s bug detection
effectivness for the above two categories of bugs. The table only
lists detection percentages when the bitwise comparison with the
non-buggy version flags an error-injected version as being in error.

\begin{table}[htbp]
	\caption{Software bug detection results. }
	\centering
	\begin{tabular}{ccccccc||ccccccc}
	\toprule
	\multirow{2}{*}{} & \multicolumn{3}{c}{Loop bound} & \multicolumn{3}{c}{Array access} & \multirow{2}{*}{} & \multicolumn{3}{c}{Loop bound} & \multicolumn{3}{c}{Array access} \\
	\cmidrule(lr){2-4} \cmidrule(lr){5-7} \cmidrule(lr){9-11} \cmidrule(lr){12-14}
	& {\#SL}  & {\#RL} & {\%det}  & {\#SL}  & {\#RL} & {\%det} & & {\#SL}  & {\#RL} & {\%det}  & {\#SL}  & {\#RL} & {\%det}\\
	\midrule
	H1 & 374 & 374 & 100 & 300 & 267 & 99  & H2 & 370 & 370 & 100 & 300 & 258 & 100	\\ 
	H3 & 374 & 374 & 100 & 300 & 267 & 98  & H4 & 486 & 106 & 99  & 300 & 26  & 100	\\ 
	H5 & 486 & 106 & 100 & 300 & 26  & 100 & H6 & 486 & 106 & 100 & 300 & 26  & 100	\\ 
	P1 & 10  & 10  & 100 & 18  & 17  & 100 & P2 & 12  & 12  & 100 & 18  & 17  & 100	\\ 
	P3 & 12  & 12  & 100 & 18  & 17  & 100 & P4 & 370 & 310 & 100 & 300 & 223 & 87 	\\ 
	P5 & 370 & 310 & 96  & 300 & 223 & 82  & P6 & 370 & 310 & 96  & 300 & 223 & 82 	\\ 
	P7 & 414 & 104 & 98  & 300 & 48  & 97  & P8 & 416 & 104 & 100 & 300 & 44  & 98 	\\ 
	P9 & 416 & 104 & 98  & 300 & 44  & 98  & W1 & 360 & 280 & 100 & 300 & 215 & 55 	\\ 
	W2 & 360 & 280 & 44  & 11  & 6   & 54  & W3 & 360 & 350 & 31  & 300 & 266 & 45 	\\ 
	W4 & 260 & 280 & 45  & 300 & 215 & 60  & W5 & 360 & 280 & 53  & 300 & 215 & 62 	\\
	W6 & 360 & 350 & 100 & 300 & 266 & 100 & C1 & 360 & 360 & 100 & 300 & 262 & 100	\\ 
	C2 & 360 & 360 & 100 & 300 & 262 & 100 & C3 & 360 & 360 & 100 & 300 & 262 & 100	\\ 
    \bottomrule
\end{tabular}
  \label{table:software-bugs}
\end{table}

The ${\#SL}$ enumerate all the possible source locations for the types of bug injected.
${\#RL}$ enumerates the number of source locations that were reached at runtime. 
$\%det$ denotes the fraction of errors flagged by bitwise comparison with correct execution that was detected by \FPD{}.
We observe that \FPD{} has a high but not perfect detection rate. We
investigated the scenarios in which \FPD{} missed detection and all of
them were a result of the logical error having a low impact on the stencil's output.
Specifically, the impacted point's effect on the stencil was beyond the required detector precision ($dp$) evaluated by our conservative error analysis.

To illustrate, consider two cases where \FPD{} detect such an impact and another where it escapes \FPD{}'s detection. In the first case, the detector evaluates to the expected value to
\texttt{0x40101b539cbac780} (we present in hex values for readability),
while the bug affected stencil computes 
%
\texttt{0x40101b53a11cf49f}.
They match in the first 23 bits of the mantissa. This example, evaluated for $W1$ has $dp=30$, hence is trapped.
In the second scenario, 
detector evaluate the expected value to 
\texttt{0x4004178de3e4ab00}
while the buggy stencil evaluates 
\texttt{0x4004178de3e4aac4},
differing in only the last 9 bits of mantissa. 
Given the low detector precision of 30 for wave benchmarks, this error goes undetected.
In general,  our conservative error analysis and low \udp{} choice
predicts only upto $dp$ bits at the detector,
attributing precision beyond $dp$ bits to round-off errors.
While this leads to missed detection opportunities, we still observe high detection rates.



\vspace{.5ex}
\noindent{\bf Loop order bugs:\/}
Unlike changes to array accesses and loop bounds, changing loop orders
required non-local changes to the Pluto-generated code. Therefore, we
handcrafted the error injected versions. The handcrafted scenarios
included swapping nested loop pairs and re-ordering of non-nested loop
blocks.  The optimized codes included a maximum of 360
nested \texttt{for} loops with a maximum loop nesting depth of 11.
Testing all possible combinations of loop reorderings will be
prohibitively expensive. We limited our testing to 15 loop order bug
injections per benchmark.

Across all benchmarks, a median of 3 injected bugs per benchmark did
not result in a bitwise difference in the output. This could either be
due to the reordered version being correct due to commutativity of
loops, or the error being too small to persist under finite precision
arithmetic for the inputs chosen. These bugs were also not detected
by \FPD{}. Other than benchmarks W1 and W3, loop order bugs injected
in all other benchmarks were detected. In W1 and W3, there was one injected bug
in each that resulted in the bitwise comparison flagging an error was not
detected by \FPD{}. In both cases, the difference was within 8 least
significant bits (corresponding to $<10^{-12}$ relative error), and could not
distinguished from floating point round-off errors.



In summary, we observe that \FPD{} can detect a large class of
software bugs with a low \udp{} of as low as 4 bits. 
In all cases evaluated, we observed
negligible overheads ($<2\%$), making it a useful component in a test
suite or as a low cost online checking tool for flagging systematic
errors. As observed earlier, increasing the \udp{} and $dp$ can
improve the percentage of bugs detected, at the cost of greater
runtime overheads. Note the a change of \udp{} only changes the 
number of points being sampled for detection. The \pw{} corresponding
to this \udp{} only decides the relative distance between  
the sampled points.


\vspace{.5ex}
\noindent{\bf Soft error detection:\/}
We performed fault injection experiments to validate the coverage
guarantees provided by \FPD. Single bit flips were injected in
the kernel source code by corrupting array locations
involved in the stencil computation. 
Such a fault model abstracts the cause of the fault to 
an anomalous behaviour in data evolution. \FPD{} is designed
to detect an anomalous behaviour in data evolution 
with a precision guarantee irrespective of where it originates
as long as it impacts the computation.
The time steps and bit locations subject to injection were selected randomly with
equal probability across the iteration space, array index space,
and bit locations.  The bit flips
were injected with equal probability on all array locations, including
the unprotected boundaries. Each benchmark was exercised through a
fault injection campaign comprising of $10,000$ executions
to determine the detection rate.  In the reported results, \FPD was evaluated 
for detector configurations guaranteeing 15 bits of mantissa precision
(total of most significant 27 bits including 11 exponent bits and 1 sign bit) with a coverage of
60--90\%. 

%
%

We observe that, across all benchmarks and including the boundary region,
errors injected within the most significant \udp{} bits
(the guaranteed protection) were detected 92--99\% of the cases for the
90\% coverage, and 78--90\% of the cases for the 60\% coverage configuration.
We observed similar trends when multiple errors are injected, at multiple
and randomly selected locations in the iteration space.
Figure \ref{fig:detection-rate} 
 shows the detection
rates for the two coverage scenarios when only considering error
affecting the protected region (the most significant 15 bits, labeled
``protected'') and only those affecting the unprotected region (labeled
``unprotected'').
Furthermore, we observe that flips in bits beyond the unprotected region
(not guaranteed to be detected) were detected on average 26\% (13\%) of the cases
for 90\% (60\%) coverage scenario. 
Hence, even with a conservative configuration for guaranteeing user-defined bits of protection,
in practice, our approach {\em empirically} protects a larger fraction of the computation space and error scenarios.

Recent studies~\cite{multibit} show that single-bit errors yield a
higher percentage of SDCs in most cases, compared to multi-bit errors.
In cases of data corruption, two or more bit flips in the opposite direction,
can reduce the overall error magnitude thus filtering it out of the detectable
range. We performed a small scale multi bit-flip experiment to check \FPD's
detection ability in such scenarios over a selected number of representative benchmarks
for a coverage guarantee of 90\%.
Furthermore, the space of multi bit-flip errors is seemingly large.
We restrict ourselves to double bit-flip errors. Each injection
campaign executed at a random time step, splits the data array into
16 byte sections and selects one section randomly within which
2 randoms bits are flipped.

For multi bit-flip campaigns including atleast one bit flipped inside the protected range,
errors were detected in 88-99\% of the cases. Furthermore, multi bit-flips encountered outside the 
protection range were detected on average for 36\% of the cases. Hence, even though our
guarantees are restricted to single bit errors, empirically the detectors can handle
multi bit errors.

\vspace{.5ex}
\noindent{\bf{Comparative study}:\/}
Our detector synthesis strategy provides a variety of tuning knobs 
for flexibility in terms of minimum precision and protection coverage 
with precise bounds. To ascertain the effectiveness of our tool,
we compare them with two state-of-the-art soft error detectors,
AID~\cite{DBLP:journals/tpds/DiC16} and
SSD~\cite{DBLP:conf/ccgrid/SubasiDBBULCC16},
which build data value centric model for soft error detection.

SSD builds an epsilon-insensitive support
vector machine regression model to detect SDCs. 
As spatial features, it includes values of a given point's neighboring data points
as the training data. It runs into scalability issues by requiring the
generation of data traces of every point in the simulation over all time steps
before being fed to the learning model. We had to scale down the problem
to a small size of $1k$ points per dimension. It learns the
classifier first, triggering false positives for early stages of the simulation
trace. In particular, we encountered false positives in initial one to five
iterations. Hence, we devised the fault injection mechanism after atleast 10 iterations
have passed. SSD did not trigger a detection in any of our error-injected
runs past the first 10 iterations. We believe, in these benchmarks, the variations observed in
the first few iterations make SSD consider the error-injected behavior appear normal.

AID is an adaptive SDC detector wherein a best-fit prediction model gets selected 
adaptively based on local online data. AID faired better with our benchmarks than SSD. However,
the detection rates and overheads were extremely conservative.
The comparative analysis by \citet{comparative-analysis} reports significant slowdown at a 
maximum of 1.4$\times$ for AID when snapshots were taken for every time step. We observed similar overheads when
taking snapshots at every step. To reduce the overhead, we performed
two experiments with snapshot every 3 steps and every 10 steps. In the former case, AID 
reported similar overheads averaging around $80\%$ with a maximum detection
rate of $54\%$. In the latter case, the overhead reduced to the $20--30\%$ range,
while the average detection rate was around $25\%$ with a maximum and minimum
of $37\%$ and $21\%$, respectively.


\ignore{==>
SSD builds on an epsilon-insensitive support vector machine regression to detect silent data corruptions.
As spatial features, it includes values of neighbouring data points as the training data to learn the regression model.
We configured SSD for each type of our benchmarks but had to scale down the problem size to $1k\times 1k$, since 
it required to generate data traces of every point in the simulation over all time steps before being fed
to the learning model. Since it requires to learn the classifer first, it may trigger few false postives for the early stages 
of the simulation trace. We encountered false postive triggers in first one to five iterations, beyond which no more false postives were
triggered. Hence, we devised the fault injection mechanism after atleast 10 iterations have passed. In none of our benchmarks,
SSD triggered an error detection event for bit flip errors in the 
any significand bits, although it triggerred a few errors when flipping 
exponent bits, thus requiring a larger error threshold 
for detection.

AID is an adaptve SDC detector wherein a best-fit prediction model gets selected adaptively based on local online data.
AID faired better with our benchmarks than SSD, however the detection rates and overheads were extremely conservative.
The comparative analysis work in~\cite{comparative-analysis} reports significant slowdown at a maximum of 1.4x for AID when snapshots are taken
for every time-step. We encounteted similar overheads when taking snapshot every step.
To reduce the overhead we performed two experiments with snapshot every 3-steps and snapshots every 10-steps.
Even with every 3-steps snapshots the results were similar to every step snapshot wherein
AID reported a high overheads averaging around $80\%$ with a maximum detection rate of $54\%$.
Reducing the snapshot frequency to every 10 time steps, the overhead reduced significantly in
the range of $20 - 30$\%, while the average detection rate was around $25\%$ at a maximum
of $37\%$ and a minimum of $21\%$, for faults injected in the higher range of the mantissa, that is, 
within $15$ bits of the MSB.

<==}

\begin{figure*}
\centering
\begin{minipage}{0.68\textwidth}
\includegraphics[width=1.0\columnwidth, height=0.5\columnwidth]{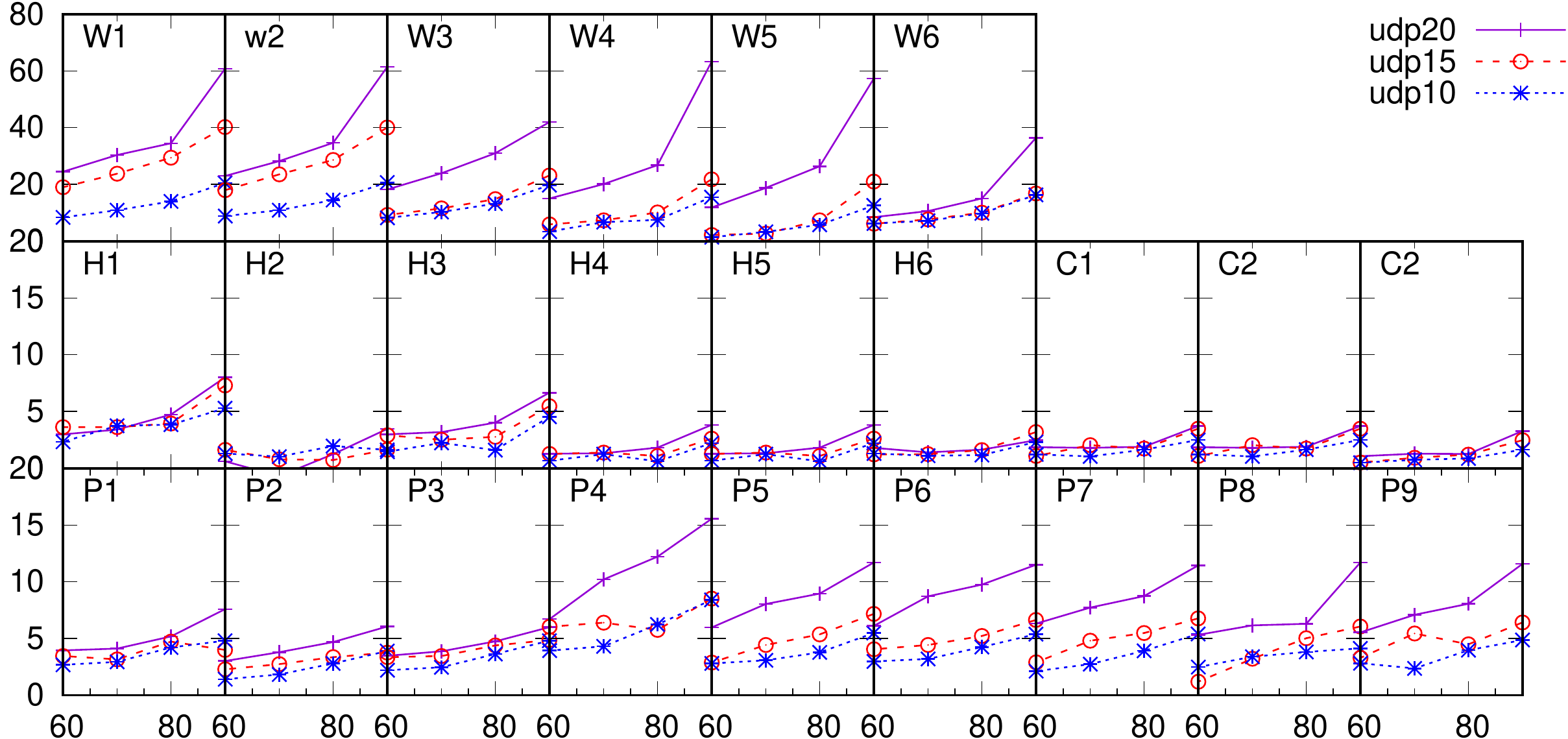}\vspace{0.5ex}
\caption{\label{fig:seq-ovh}
Sequential benchmark execution overheads for three different user-defined precision (\udp{})
values (20, 15, and 10) and varying. X-axis: coverage; y-axis: execution time overhead in percentages.
}
\end{minipage}
\hfill\hfill
\begin{minipage}{0.30\textwidth}
\includegraphics[width=0.8\columnwidth,height=0.6\columnwidth]{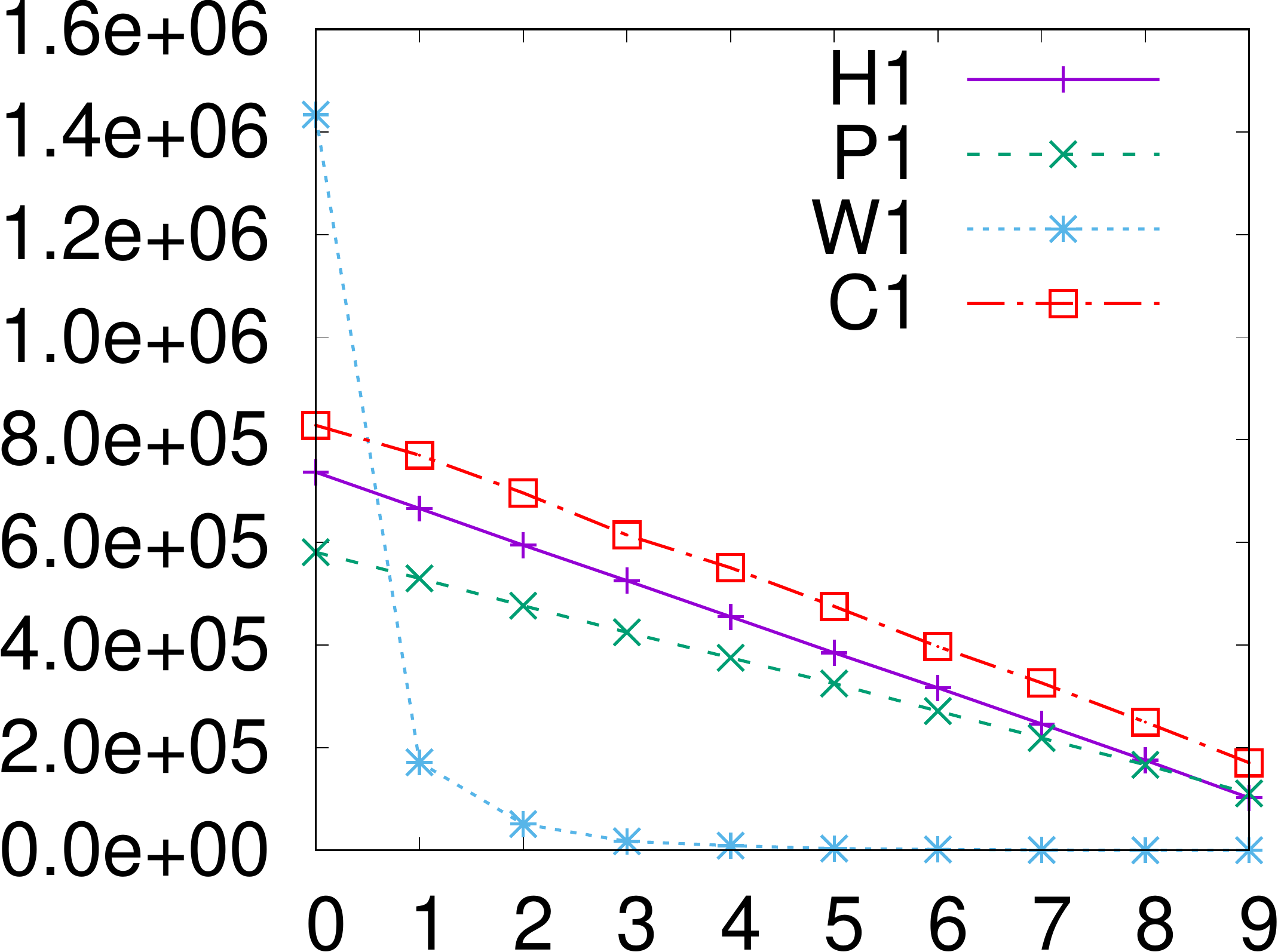}\\[.5em]
\caption{\label{fig:pwvol}
Protected region volume per detector (optimal configuration) with input exponent range for four benchmarks, \udp{}\!=\!15 and 80\% coverage. x-axis: exponent range ($2^0, 2^1,..$); y-axis: number of iteration points in the protected region per detector.}
\end{minipage}
\end{figure*}

\vspace{.5ex}
\noindent{\bf Sequential overheads:\/}
Figure~\ref{fig:seq-ovh} shows the sequential execution overheads of
our detection approach for three \udp{}-coverage configurations: 10, 15, and 20 bits.
We
observe that the overhead depends on the user-required coverage guarantee,
with
overheads, in general, below 10\% for \udp{}=15 and 80\% coverage, for heat, Poisson, and convection-diffusion benchmarks.
Overall, we observe that protecting
additional bits or providing greater coverage increases the overhead.
All wave benchmarks (and some Poisson benchmarks for 90\% coverage) incur far greater overheads,
reaching over 60\% for \udp{} of 20 bits and 90\% coverage. This is due to the nature of
the stencils that, as discuss below, leads to a sharp reduction in the protected region
(Figure~\ref{fig:pwvol}). Despite this increase, these results demonstrate the
approach's flexibility
in supporting low detection guarantees when a reduced overhead is desired.

\begin{figure*}
    \centering
    \begin{minipage}{0.46\textwidth}
\includegraphics[width=0.9\columnwidth, height=0.6\columnwidth]{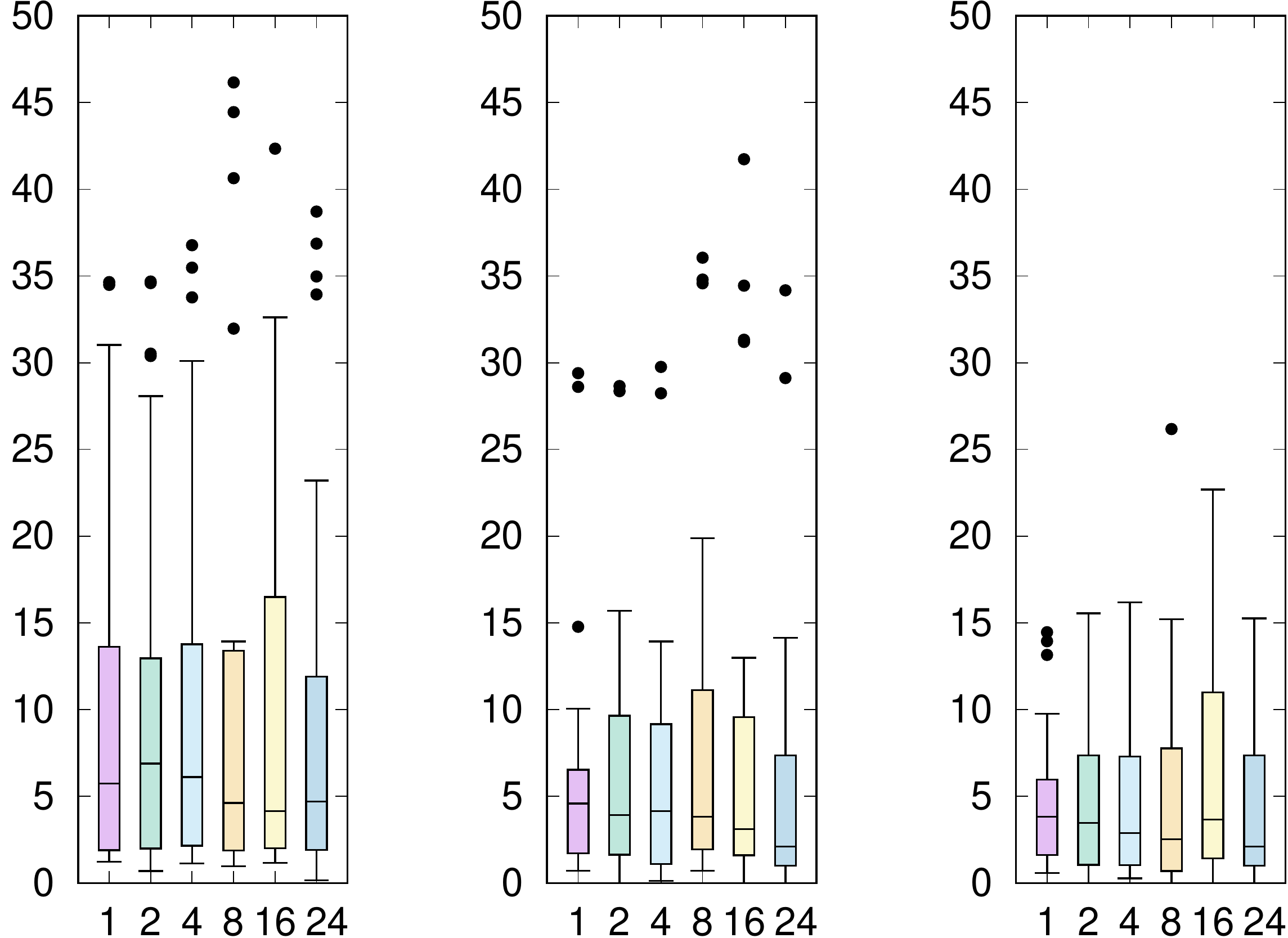}\\
\vspace{-1.5ex}
\caption{\label{fig:par_c90_var_udp_20_15_10}
Detection overheads across all
benchmarks for 80\% coverage for \udp{} values 20, 15 and 10. x-axis: thread count; y-axis: execution overhead in percentage.
}
    \end{minipage}
    \hfill
    \begin{minipage}{0.46\textwidth}
\includegraphics[width=0.9\columnwidth, height=0.6\columnwidth]{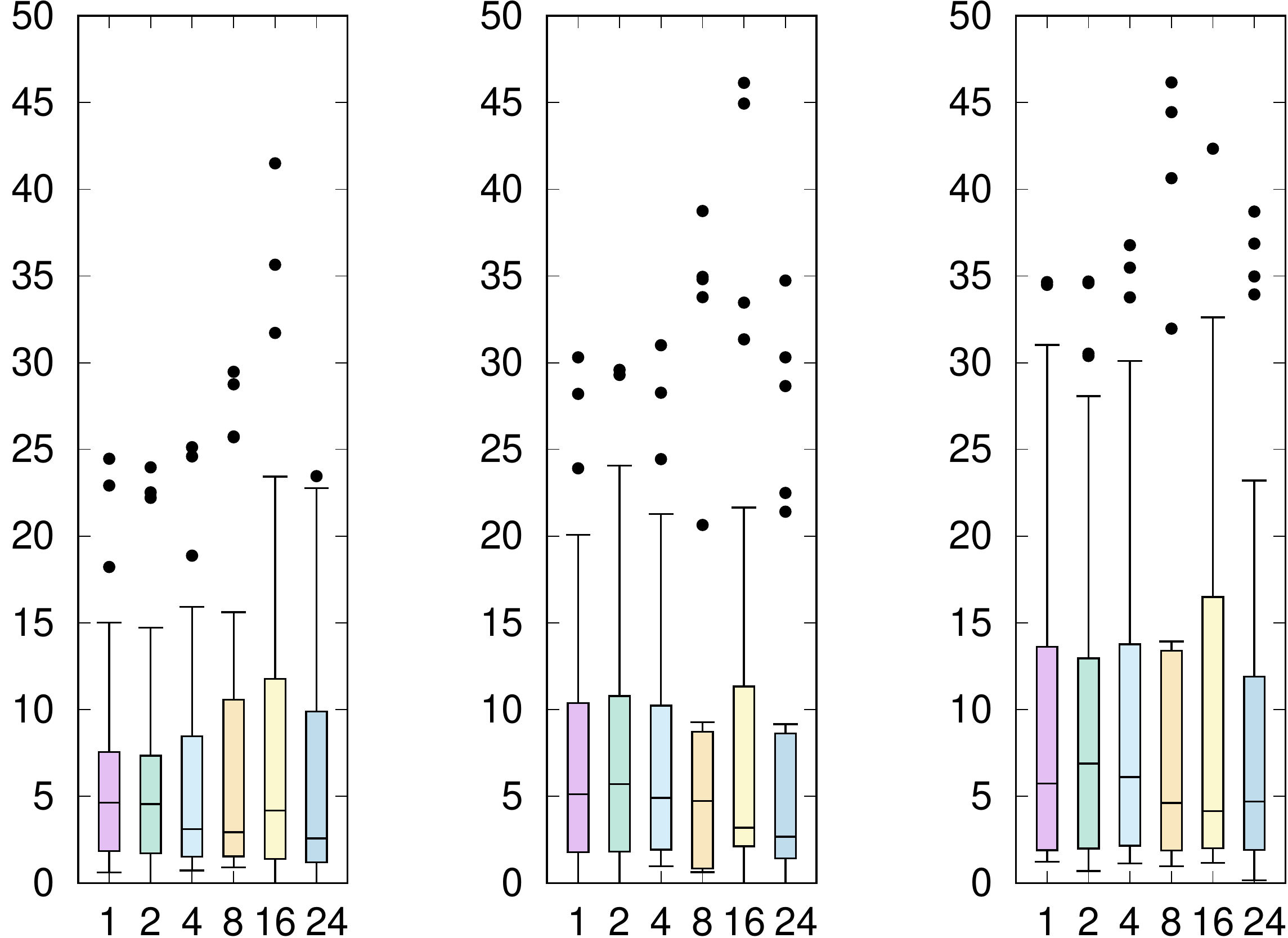}
\vspace{1.5ex}
\caption{\label{fig:par_u20_var_cov_60_80_90}
Detection overheads across all
benchmarks for \udp{} 20 and coverage of 60\%, 70\% and 80\%. x-axis: thread count; y-axis: execution time overheads in percentage.
}
    \end{minipage}
\vspace{1.5ex}
\end{figure*}

\begin{figure}
\centering
\begin{minipage}{0.65\textwidth}
\includegraphics[width=1.0\columnwidth, height=0.5\columnwidth]{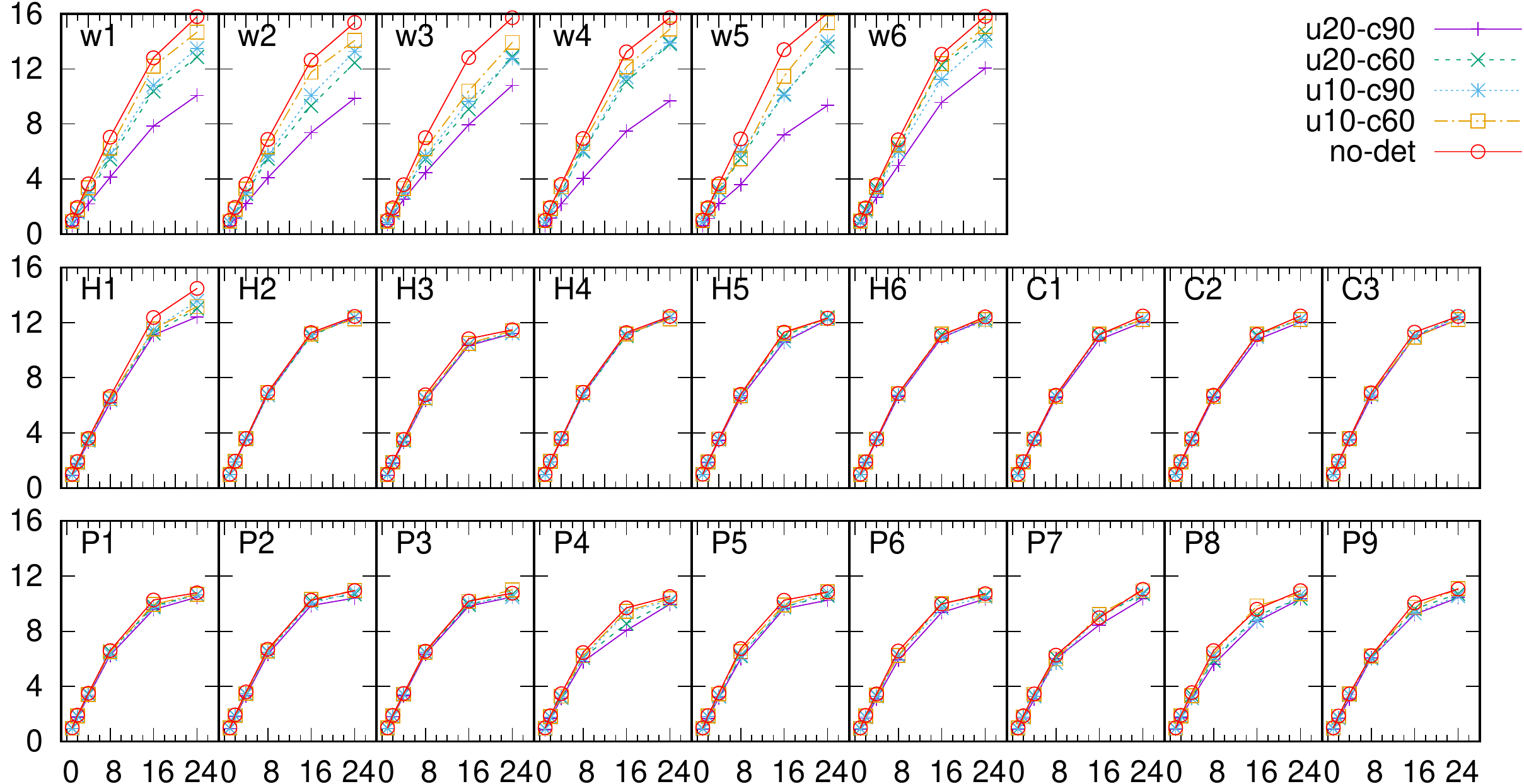}
\vspace{0.0ex}
\caption{\label{fig:par-detailed}
Scaling on up to 24 threads for the baseline
without detectors and with detectors over three combinations of \udp{} and coverage [(u=\udp{},c=cov)=(20,90), (10,90), (20,60), (10,60)]. x-axis: thread count;
y-axis: speedup over single-threaded execution with no
detectors.  
}
\end{minipage}\hfill%
\begin{minipage}{0.30\textwidth}
\includegraphics[width=\columnwidth,height=0.8\columnwidth]{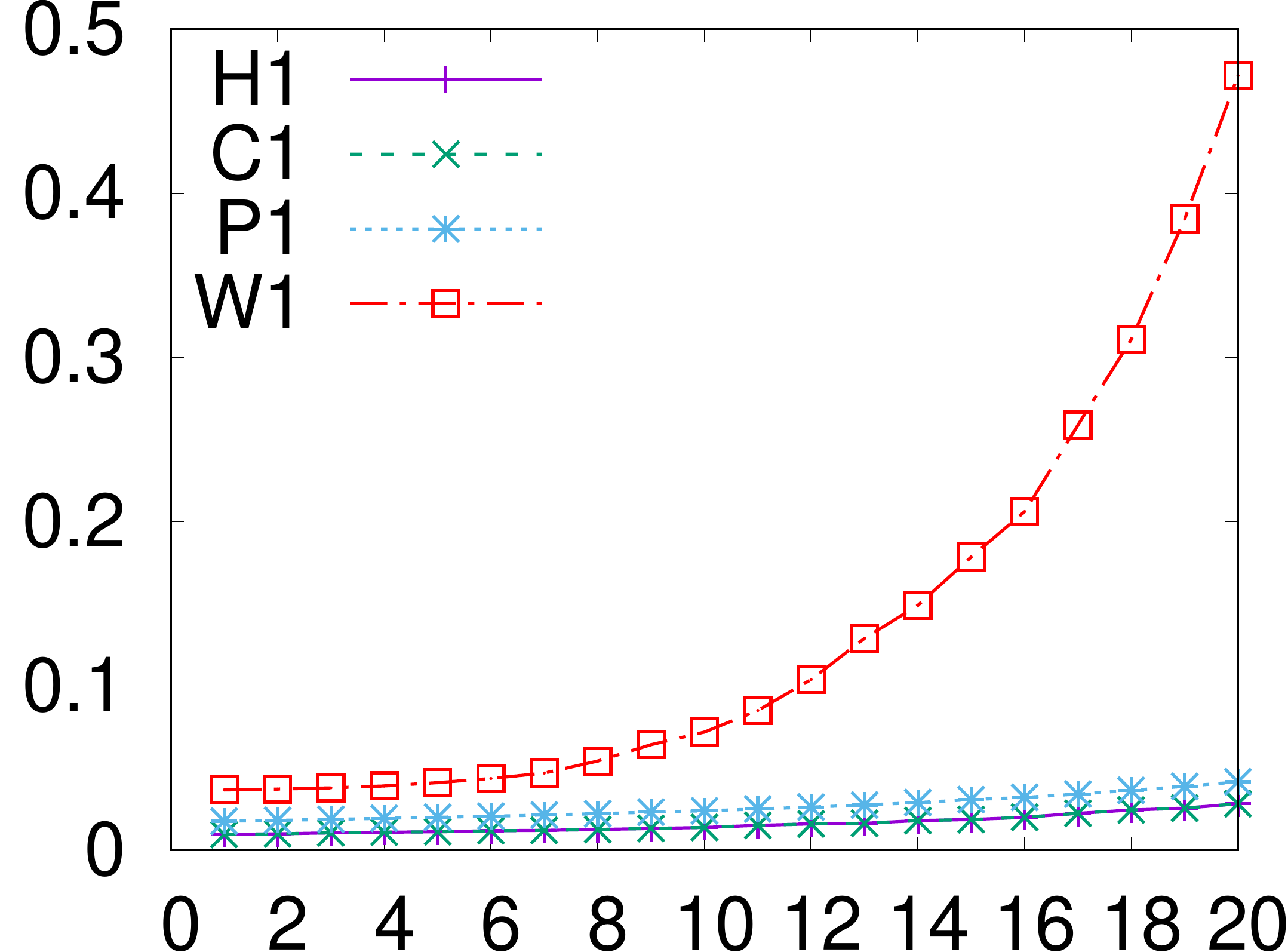}
\vspace{1.5ex}
\caption{\label{fig:cost-vs-udp}
Optimal cost from Equation~\eqref{eq:cost-function} (y-axis) vs \udp{} (x-axis) for four benchmarks.
}
\end{minipage}
\vspace{1em}
\end{figure}

\vspace{.5ex}
\noindent{\bf Scalability:\/}
Figure~\ref{fig:par-detailed} shows the scalability of the baseline and
detector-embedded versions on up to 24 threads using OpenMP. Across
all benchmarks, both variants achieve similar
speedups, demonstrating that the detectors do not interfere with
efficient parallel execution. 
While not shown here, we observed similar trends (in
terms of low overheads) with various thread counts 
for other user-defined precision and coverage values.

For the wave(``W'') benchmarks which model hyperbolic
PDE equations we encounter reduced scalability for stricter 
configurations of high \udp{} and coverage. 
We believe this is due to higher \udp{} values requiring 
more detectors. This coupled with the larger \ew{} results
in potentially increased cache contention and hence the associated
reduced scalability.

Summarizing the overheads, Figure~\ref{fig:par_c90_var_udp_20_15_10}
shows the distribution of overheads for 80\% coverage and three \udp{}
values. Figure~\ref{fig:par_u20_var_cov_60_80_90} shows the
distribution of overheads for a fixed \udp{} (20) and three different
coverage percentages. In both cases, we observe greater coverage
percentage or \udp{} requirement can increase overheads. However, the
median overhead remains close to 5\% and does not increase with scale.

\begin{figure*}
\centering
\begin{minipage}{0.45\textwidth}
\includegraphics[width=0.9\columnwidth, height=0.50\columnwidth]{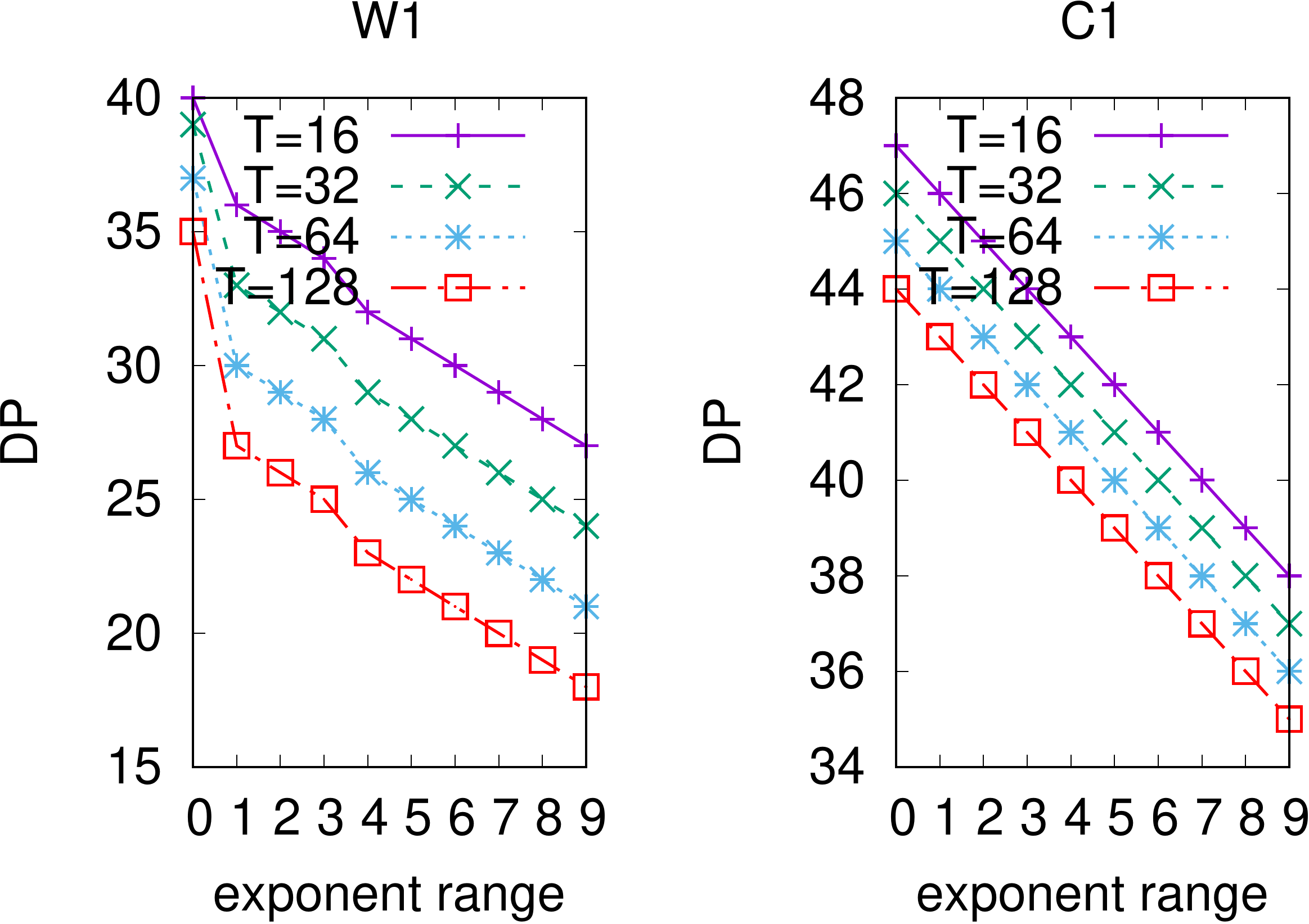}
\caption{\label{fig:dp-exp}
Maximum bits preserved by iterative stencil (y-axis) as function of 
input exponent range (x-axis), determined by our analysis, for two benchmarks (w1 and c1) for different maximum T-step choices.}
\end{minipage}\hfill
\begin{minipage}{0.25\textwidth}
\centering
\includegraphics[width=0.9\textwidth, height=0.55\columnwidth]{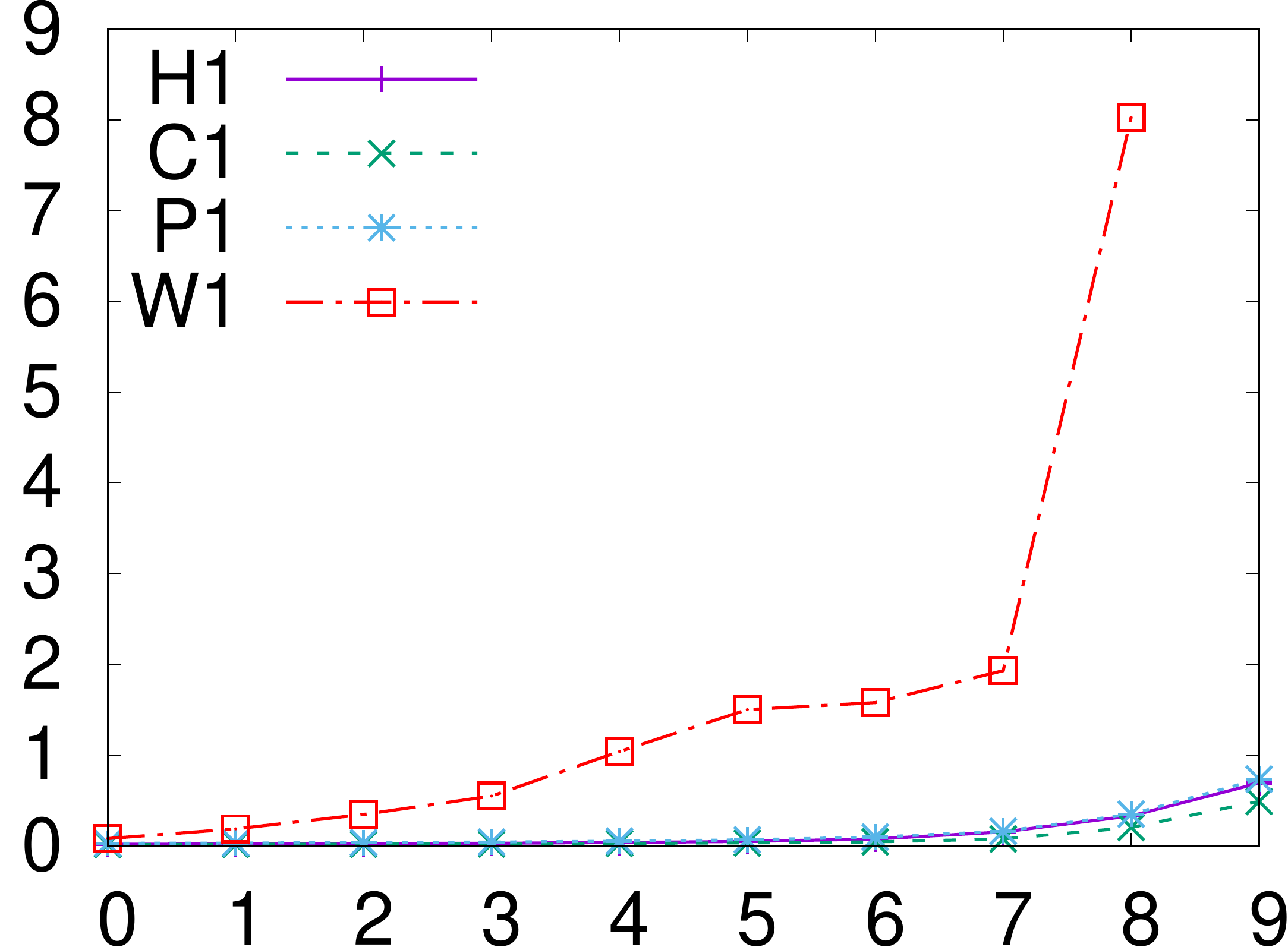}
\caption{\label{fig:cost-vs-input}
Optimal configuration's cost for different input exponent ranges for
a \udp{} of 20 and 60\% coverage. x-axis: input exponent range ($2^0,2^1,..$); y-axis:
cost for optimal configuration (from Equation~\eqref{eq:cost-function}).
}
\end{minipage}\hfill
\begin{minipage}{0.25\textwidth}
\centering
\includegraphics[width=0.9\textwidth, height=0.55\columnwidth]{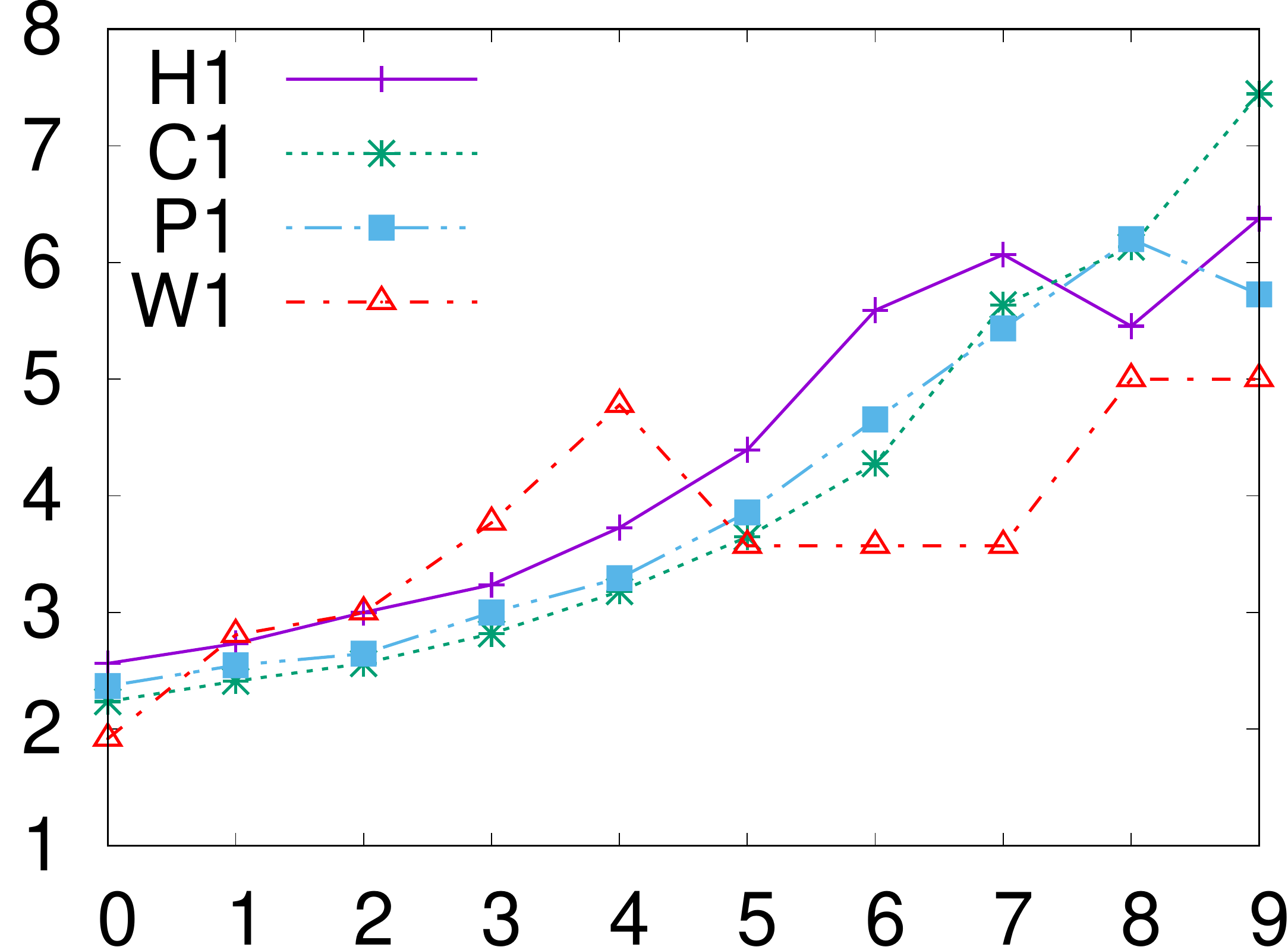}
\caption{\label{fig:ew-pw-ratio-vs-input}
Optimal configuration's \ew{} to \pw{} ratio for different input exponent ranges for
a \udp{} of 20 and 90\% coverage. x-axis: input exponent range ($2^0,2^1,..$); y-axis:
\ew{}/\pw{} for optimal configuration (from Equation~\eqref{eq:cost-function}).
}
\end{minipage}
\end{figure*}

Table~\ref{tab:summary} summarizes the optimal configuration, detection rates and overhead factors 
for user defined protection goals requiring $udp=20$ and 90\% coverage. We select the summary for a subset of
the example benchmarks that have shown the largest variations in their configuration and detection rates
in their class of numerical kernels.
\begingroup
\small
\begin{table*}[htbp]
\centering
\begin{tabular}{ccccccccc}
\toprule
\multirow{2}{*}{Benchmark} & \multicolumn{2}{c}{optimal config} & \multicolumn{2}{c}{\% det} & \multirow{2}{*}{\% Seq ovh} & \multicolumn{3}{c}{Scalability factor} \\
\cmidrule(lr){2-3} \cmidrule(lr){4-5} \cmidrule(lr){7-9}
      &  {(T,$dp_T$)} & {(ew,pw)}   & {single bit} & {Multi bit} & & {k=2}  & {k=8}  &  {k=24} \\
\midrule
H1 & (254,39) & (63,14) & 96.14 & 94.66 & 8.0   & 1.76 & 6.18 & 12.41 \\
H4 & (256,38) & (63,14) & 90.47 & 88.54 & 3.81  & 1.84 & 6.54 & 12.03 \\
P1 & (160,40) & (69,24) & 94.12 & 93.88 & 7.59  & 1.78 & 6.27 & 10.41 \\ 
P4 & (160,37) & (69,17) & 93.65 & 94.92 & 15.55 & 1.67 & 5.78 & 9.96 \\
P7 & (160,37) & (69,17) & 93.65 & 95.05 & 11.43 & 1.65 & 6.02 & 10.35 \\
W1 & (10,36)  & (20,3)  & 99.25 & 99.48 & 60.78 & 1.20 & 4.13 & 10.07 \\
W4 & (10,36)  & (20,6)  & 99.25 & 99.74 & 41.91 & 1.15 & 4.03 & 9.67  \\
C1 & (250,38) & (66,15) & 92.10 & 90.76 & 3.72  & 1.82 & 6.54 & 12.03 \\

\end{tabular}
\caption{\label{tab:summary}Summary table for optimal configurations, detection rate, sequential overhead overhead and scalability factors for udp of 20 and coverage 90\%}
\end{table*}
\endgroup


\subsection{Analysis of the Overhead Cost Function}

Because \FPD's detection strategy is sensitive to the specific stencil
and the characteristics of the initial/boundary conditions, we have
chosen multiple benchmarks and configurations, including different initial 
and mixed boundary conditions. For a broader analysis
of the space of choices evaluated in the offline phase, we
examine four candidate benchmarks---h1, c1, w1, and p1--in detail.


\vspace{.5ex}
\noindent{\bf Variation of essential and protected widths:\/}
Figure ~\ref{fig:pwvol} shows the variation in optimal \pw{} volume with input exponent range. W1, a second-order wave equation representing hyperbolic PDEs, exhibits the highest overhead and cost since its \pw{} volume rapidly diminishes with increasing exponent sizes. Benchmarks h1 and c1 exhibit a slower change in the optimal \pw{}.

Figure~\ref{fig:ew-pw-ratio-vs-input} plots the ratio of \ew{} to \pw{}
for their optimal configuration with varying input exponent range for a fixed {\em udp} and {\em cov}.
Larger binade differences in the input values result in increasing separation between the \ew{} and \pw{}
values, often increasing the
\ew{} to \pw{} ratio.
Together with $\rho$(equation$~\eqref{eq:cost-function}$) this can potentially lead to an increase in detector overhead.


\vspace{.5ex}
\noindent{\bf Impact of input range and \udp{}:\/}
Figure~\ref{fig:cost-vs-input} shows the cost function
values for four benchmarks (h1, c1, p1 and w1) for hypothetical range of
input values from $2^0$ to $2^9$ and user-defined
precision of 20 bits. We observe that cost increases with 
input range, especially for w1. 
Thus, accurate yet
efficient evaluation of the input range can help reduce the detector
overheads.
%

Figure~\ref{fig:cost-vs-udp} shows the evaluated
cost function value as \udp{} is varied.
Similar to the input exponent range, we observe that some benchmarks
(especially w1) are more sensitive than others to increase in \udp{}.
Thus, careful choice of \udp{}
can maximize coverage while
minimizing overhead.

\section{Additional Related Work}
\label{sec:rel-work}
Floating-point error analysis 
was the central driving concept in our work.
Boldo~\cite{DBLP:conf/arith/BoldoF07, boldo},
Darulova~\cite{Comp-for-reals},
Magron~\cite{Magron:2017:CRE:3034774.3015465},
and
Solovyev~\cite{DBLP:journals/toplas/SolovyevBBJRG19}
are three recent pieces of work that conduct
rigorous error analysis.
Zhang~\cite{zhang2015reduced} uses {\em reduced precision check}
to detect errors in the floating point units as a 
hardware solution.
Daumas et al. reason about floating-point 
operations using interval
arithmetic~\cite{DBLP:conf/arith/DaumasMM05}. 
These tools focus on programs operating on fixed
and small number of inputs.
They are unable to handle the kinds of
complexity presented by stencil expressions unfolded
in time.
They also do not steer their analysis toward
the synthesis of online error detectors like in
this work.

We exploit the structure of stencil operations to
simplify analysis of 
{\em parametrically} sized programs. 
Kramer established worst-case bounds for interval 
arithmetic~\cite{DBLP:conf/arith/Kramer97}.
We build on the guarantees for individual operations
(IEEE 754~\cite{kahan1996ieee})
to analyze worst-case bounds for 
stencil programs.

Chisel~\cite{DBLP:conf/oopsla/MisailovicCAQR14} 
and Rely~\cite{DBLP:conf/oopsla/CarbinMR13} consider 
potentially erroneous execution of portions of a program by
analyzing the probability of the output being erroneous. 
They track the probability of an erroneous output 
rather than its magnitude.
Soft error analysis has been performed for specific algorithm classes 
(e.g., linear algebra~\cite{DBLP:conf/hpdc/WuC14,DBLP:conf/ppopp/WuDGBCTLOC17}
and iterative solvers~\cite{DBLP:conf/hpdc/ElliottHM15,DBLP:conf/hpdc/TaoSKWLZKC16}).
Huang and Abraham introduced algorithm-based fault tolerance (ABFT)~\cite{DBLP:journals/tc/HuangA84}
to detect errors in matrix multiplication related operations.
Elliott et al.~\cite{DBLP:journals/corr/ElliottHM14} present selective reliability to provide numerical bounds on
anticipated behavior and
use this analysis in the design of resilient algorithms~\cite{DBLP:journals/jocs/ElliottHM16,DBLP:conf/ipps/ElliottHM14}. 
Our work focuses on soft error detection for stencil programs. 
Application-independent approaches for iterative programs rely on
observing the evolution of a value over time to detect anomalies (e.g., AID~\cite{DBLP:journals/tpds/DiC16} uses curve fitting,
SSD~\cite{DBLP:conf/ccgrid/SubasiDBBULCC16} uses support vector machines (SVM) regression, ~\cite{Sharma2015DetectingSE} uses a 
machine learning based approach to build regression models for synthesizing low cost detectors).
Gomez and Capello exploits multivariate interpolation in order to detect and correct
corruption in stencil application~\cite{capello-interpolate}.
Xiaoguang~\cite{gs-dmr} presents a {\em grid sampling} based DMR scheme which
determines the sampling points based on the error propagation
pattern in the grid. 
%

Gamell considers local recovery from fail-stop errors affecting
stencil programs~\cite{DBLP:conf/sc/GamellTHMKCP15}.
Fang et al.~\cite{DBLP:conf/icpp/FangCRC17} analytically model application overhead of 
recovery from detected soft errors via localized recovery. 
Our approach can complement such recovery algorithms
via efficient error detection.


%

\section{Conclusions and Future Work}
\label{sec:conclusions}
\label{sec:conclusions}

In this paper, we present \FPD, an approach to provide
error detection that covers both logical bugs and soft
errors in the data space of stencil computations.
Schemes comparable to \FPD have been observed to
generate false positives, incur higher 
overheads, or not provide similar rigorous guarantees.
%
%
We report \FPD's overheads
for different thread counts as well as its performance in conjunction with
polyhedral optimizations of the code over
for various user-defined precision values.
We believe that \FPD
can be used as part of a holistic
error detection system (e.g., involving
cross-layer concerns~\cite{DBLP:journals/tcad/ChengMSCCSSLABM18})
in which the most impactful of errors affecting
stencil programs can be protected.
%
%

As future work, we will investigate the use of \FPD for
runtime precision profiling, given that
developers often use high precision as a safety net for 
floating point errors, which may be wasteful in many cases.
\FPD's evaluation units are uniquely designed to allow
it to serve as a very
close proxy to real
values at runtime. Thus it can help profile runtime precision requirements.
%
%
%
In addition to re-instating confidence in the evolving results, this
approach may enable
the user to dynamically tune the working precision
based on the stability of the evolving results.
To this end, we have prototyped a machine
learning model built on profiled simulation data that attempts to predict the 
minimum precision around a rectangular region centered around a point for which
\FPD's evaluation unit was instantiated.
Current results show encouraging trends with
prediction accuracies within 2 precision bits.
Some of our ongoing work aims to
leverage this information in tuning simulation parameters for improved performance.


\bibliography{refs}

\clearpage

\end{document}


\title{\FPD: Supporting Material}
\author{}



\newcommand{\etc}{etc.\@\xspace}
\newcommand{\ie}{i.e.\@\xspace}
\newcommand{\eg}{e.g.\@\xspace}
\newcommand\ignore[1]{}
\newcommand\ggcmt[1]{\todo[inline, size=\small, color=green!40]{GG: #1}}
\newcommand\ggcmtside[1]{\todo[size=\scriptsize, color=green!40]{GG: #1}}

\newcommand\adcmt[1]{\todo[inline, size=\small, color=yellow!60]{AD: #1}}
\newcommand\adcmtside[1]{\todo[size=\scriptsize, color=yellow!40]{AD: #1}}

\newcommand\ibcmt[1]{\todo[inline, size=\small, color=blue!60]{IB: #1}}
\newcommand\ibcmtside[1]{\todo[size=\scriptsize, color=blue!40]{IB: #1}}

\newcommand\skcmt[1]{\todo[inline, size=\small, color=orange!60]{SK: #1}}
\newcommand\skcmtside[1]{\todo[size=\scriptsize, color=orange!40]{SK: #1}}


\maketitle

\tableofcontents
\newpage

In this material we provide the detailed proof
discussions and the analytical choices made in
our tool design. We have highlighted the key
results from our work inside the main paper,
and present the complete derivation of the 
concepts here for completeness.
The interested reader is encouraged to 
have a detailed look into the derivations.

We discuss broadly the four major sections.
In {\em Error Analysis} we detail the complete
derivation of our affine arithmetic based round-off error
analysis to obtain the detector precision. 
{\em Precision Driven Optimizations} describes the
methodology behind obtaining the \textbf{Essential Width} and
\textbf{Protected Width}.
{\em Overall Algorithm} provides the exact algorithms used for the offline/online split phases.
In {\em software Bug Detection} provide examples of the types of bugs
introduced and summary of the Pluto generated code and it's results.

\appendix
\section{Error Analysis}
\subsection{Floating point basics}
\label{sec:sfp-basics}
A floating point number system corresponding 
to a certain radix is a subset of the set of 
real numbers, and can be expressed as a 
5 tuple $(\beta, s,m,e,p)$.
Here $\beta$ is the radix or the base of the
floating point system. Our work concern with
$\beta=2$, that is the binary floating point
system in double precision with $p=53$.
$s \in \{-1,1\}$ is the sign bit, $m$ is 
called the significand or mantissa and 
$e$ is the exponent.
There are two types of floating point
numbers, the {\em normal} and the 
{\em denormals}, other than the special cases 
of NANs and infinities.
For the {\em normal}
numbers $m$ lies in the range $1 \leq m < 2$,
while for {\em denormals} the range is 
$0 \leq m \leq 1$ . 
The denormal range begins
after the lowest {\em normal} number to make
a gradual transition towards zero.
$e$ is the exponent in the range 
$-1022 \leq e \leq 1023$.
Any such number $x_f$ belonging to the 
floating point system $\mathbb{F}$ has 
the value $s\cdot m \cdot 2^e$ .
If $x \in \mathbb{R}$, then 
$x_f \in \mathbb{F}$ denotes an element 
in $\mathbb{F}$ closest to $x$ obtained
by applying the rounding operator 
$(\circ)$ to $x$. 

IEEE754~\cite{ieee754} defines four 
rounding modes for elementary 
floating point operations namely rounding towards positive infinity $(\circ_{+\infty})$, towards
negative infinity $(\circ_{-\infty})$, towards zero $(\circ_0)$ and to the nearest $(\circ_\backsim)$.
Every real number $x$ lying in the range of $\mathbb{F}$ can be approximated by an element  $x_f \in \mathbb{F}$
with a relative error no larger than the {\em unit round-off} $\boldsymbol{u} = \dfrac{1}{2}\beta^{1-p}$. 
Here $\beta^{1-p}$ corresponds to the {\em unit of least precision (ulp)} for exponent value of 1.
We use $\mu$ to denote ulp (1), such that $\mu = 2\boldsymbol{u}$.
In our case $\boldsymbol{u} = 2^{-53}$, and hence $\mu = 2^{-52}$. 
Hence, for all $\circ \in \{\circ_{+\infty},\circ_{-\infty},\circ_0, \circ_\backsim \}$, 
we have the following result
\begin{theorem}
	If $x \in \mathbb{R}$ lies in the range of $\mathbb{F}$, then
	\begin{equation}
	\label{eq:srounding}
	\circ(x) = x(1 + \delta),\quad |\delta| \leq \boldsymbol{u} = \mu/2
	\end{equation}
\end{theorem}

Given two exactly represented floating point numbers $x_f$ and $y_f$,
arithmetic operators $\diamond \in \{+,-,\times\}$ have the following guarantees across for any rounding
modes
\begin{equation}
\label{eq:sopbound}
x_f \diamond_f y_f = (x_f \diamond y_f)(1 + \delta),\quad |\delta| \leq 2\boldsymbol{u} = \mu
\end{equation}

Thus the computed floating point value post rounding represents an interval 
$[x_f\diamond y_f \pm |x_f\diamond y_f|2\boldsymbol{u}]$ of real numbers, any of which
could have produced the computed value. Multiplication operators have a stricter bound
of $|\delta| \leq \boldsymbol{u}$

%
%
%

\subsection{Affine Analysis}
\label{sec:saffine}
In Affine Arithmetica $(AA)$~\cite{stolfi}, a computed quantity $(\hat{x})$
is expressed in an affine form over its true value and 
its accumulated error quantities as shown in equation $~\eqref{eq:sbasic-affine}$.

\begin{equation}
\label{eq:sbasic-affine}
\begin{split}
	\hat{x} = x_0 + \sum_{i=1}^n x_i\epsilon_i
\end{split}
\end{equation}

Here, $x_0$ is the central value tracking the true value of $\hat{x}$.
The coefficients $x_i$ are finite floating point numbers and the 
$\epsilon_i$ are {\em formal noise variables} which are unknown 
until concretized but assumed to lie in the interval $[-1,+1]$.
To interprete an affine variable using an equivalent interval notation, 
the noise variables are concretized to their maximum range obtaining
the interval $I_x$.

\begin{equation}
\label{eq:saffine_interval}
  I_x = [x_0 - rad(\hat{x}), x_0 + rad(\hat{x})];\quad rad(\hat{x}) = \sum_{i=1}^n |x_i|
\end{equation}

The real advantage of AA can be seen by deferring the concretization process 
as far down the computation depth as possible since once concretized, it looses 
any variable correlation that might be present. Before concretizing, incoming 
contributions from the same or related noise variables are superposed, hence they
can accumulate or cancel based on the sign of the coefficients $x_i$. For example,
when evaluating $(x-x)$, affine arithmetic yields 0, unlike interval analysis where it
doubles the interval.

\subsection{Arithmetic Operations Using Affine Arithmetic}

We define the following three operators to deal with retrieval 
of noise variables and its coefficients.
\begin{compactitem}
\item {\em $\sigma$:} Defines the mapping from an affine variable
					  to its set of noise variables. Thus
							$\sigma(\hat{x}) = \{\epsilon_i\}_{i=1}^n$.
\item {\em $\sigma^{(0)}$:} A special $\sigma$ to retrieve {\em freshly 
							generated noise	variables} at the last 
							stage of compute of $\hat{x}$. Its purpose is to
							distinguish between freshly generated noise variables
							versus the existing ones.
\item {\em $\gamma$:} Defines the mapping from a 2-tuple of 
					  {\em (affine variable, noise variable)} to its 
					  corresponding noise coefficient. Thus \\ [-1em]
					\[
						\gamma(\hat{x},\epsilon_i) = \begin{cases}
												x_i;\quad \epsilon_i \in \sigma(\hat{x}) \\
												0 ; \quad otherwise
											 \end{cases}
					\]
\end{compactitem}

Based on the definitions for $\sigma$ and $\gamma$,
we now define the two affine variables
$\hat{x}$ and $\hat{y}$ as
\begin{equation}
\label{eq:saffine-representation}
\begin{split}
	\hat{x} &= x_0 + \sum_{\epsilon_i \in \sigma(\hat{x})} \gamma(\hat{x}, \epsilon_i)\epsilon_i ;\quad
	\hat{y} = y_0 + \sum_{\epsilon_i \in \sigma(\hat{y})} \gamma(\hat{y}, \epsilon_i)\epsilon_i \\
\end{split}
\end{equation}

We consider only, $(a)\ $ a scalar multiplication and $(b\ )$ an 
affine addition for dealing with dot products pertinent
to stencil computations.\\ [-1em]

\paragraph{\textbf{Scalar Multiplication}} Let $\alpha$ denote an exactly 
represented scalar quantity, then\\ [-1em]
\begin{equation}
	\alpha \hat{x} = \alpha x_0 + \sum_{\epsilon_i \in \sigma(\hat{x})} 
					 \alpha \gamma(\hat{x},\epsilon_i)\epsilon_i
\end{equation}

\paragraph{\textbf{Affine Addition}} 
The coefficients of the common noise variables are linearly superposed
preserving their correlation, hence coefficients of a common noise 
variable of equal but opposing sign can cancel out eliminating that 
noise variable.
\begin{equation}
\begin{split}
	\hat{x}+\hat{y} &= (x_0 + \sum_{\epsilon_i \in \sigma(\hat{x})} \gamma(\hat{x}, \epsilon_i)\epsilon_i) + 
	                   (y_0 + \sum_{\epsilon_i \in \sigma(\hat{y})} \gamma(\hat{y}, \epsilon_i)\epsilon_i) \\
					&= (x_0 + y_0) + \sum_{\underset{\sigma(\hat{x}) \cup \sigma(\hat{y})}{\epsilon_i \in} }
					   (\gamma(\hat{x},\epsilon_i) + \gamma(\hat{y},\epsilon_i)) \epsilon_i +
					   \{\hbox{higher order terms}\}
\end{split}
\end{equation}

Note that the $x_i$ coefficients obtained using $\gamma$ are finite
floating point numbers. Hence during the scaling or superposition
operations they incur rounding error as well leading to higher order
rounding terms.

\subsection{Affine model for Floating Point}
Equation $~\eqref{eq:srounding}$ shows the relation between a real number
$x \in \mathbb{R}$ and its rounded floating point representation,
$x_f = \circ(x) \in \mathbb{F}$.
 We can represent the floating point
 variable by associating the uncertainty radius
 with a noise variable as shown in equation $\eqref{eq:saffine-float-main}$
 \begin{equation}
 \label{eq:saffine-float-main}
 \begin{split}
 	x_f = \circ(x) &= x(1 + \delta);\quad |\delta| \leq \boldsymbol{u} \\
 				   &= x + (x\boldsymbol{u})\epsilon;\quad \epsilon \in [-1,+1] \\
 \end{split}
 \end{equation}

  Let $x_f$ and $y_f$ be two floating point variables 
  represented as in equation $\eqref{eq:saffine-representation}$
 such that $x_f=\hat{x}$ and $y_f=\hat{y}$. 
 Instead of the central values being $x_0$ 
 and $y_0$, we replace them by the real number 
 $x$ and $y$ respectively as per the defintion
 of our model .
 \begin{equation}
 \label{eq:saffine-float-representation}
 \begin{split}
 	x_f &= x + \sum_{\epsilon_i \in \sigma(\hat{x})} \gamma(\hat{x}, \epsilon_i)\epsilon_i \ ;\quad
 	y_f = y + \sum_{\epsilon_i \in \sigma(\hat{y})} \gamma(\hat{y}, \epsilon_i)\epsilon_i \\
 \end{split}
 \end{equation}

 Also, let $\alpha$ and $\beta$ be two exactly 
 represented scalar quantities as before. 
 In this model every floating point operation,
 $\diamond_f \in \{\circ(+), \circ(-), \circ(\times) \}$ 
 introduces a {\em fresh noise variable} with 
 coefficients equal to the uncertainty of maximum round-off 
 error introduced by that operation as
 in $\eqref{eq:sopbound}$.
  We consider two operations specific to the stencil inner product:
 (1) scaling $x_f$ by an exactly represented scalar $\alpha$ shown in 
 $\eqref{eq:saffine-float-mult}$, and
 (2) add or subtract two error-laden FP numbers $x_f$ and $y_f$ as shown in 
 $\eqref{eq:saffine-float-add}$
 \begin{equation}
 \label{eq:saffine-float-mult}
 \begin{split}
   \circ(\alpha{\!} \times{\!} x_f) &=\! \bigg (\alpha x\! +{\rs\rs\rs}\sum_{\epsilon_i \in \sigma(x_f)}{\rs\rs} \alpha \gamma(x_f,\epsilon_i)\epsilon_i
   							  \bigg )\! 
 							   +\! \boldsymbol{((\alpha x)\mu)}\epsilon_{new}\! +\! \{\mbox{he}\} 
 \end{split}
 \end{equation}
 
 \begin{equation}
 \label{eq:saffine-float-add}
 \begin{split}
 	\circ(x_f{\!} +{\!} y_f) &= (x+y) +{\rs\rs\rs} \sum_{\epsilon_i \in \sigma(x_f)\cup\sigma(y_f) } {\rs\rs\rs}\{\gamma(x_f,\epsilon_i) + \gamma(y_f,\epsilon_i) \}\epsilon_i  \\
 			  &\quad +\boldsymbol{((x+y)\mu)}\epsilon_{new} + \{\mbox{he}\}
 \end{split}
 \end{equation}

  Observations from equation $~\eqref{eq:saffine-float-mult}$ and $\eqref{eq:saffine-float-add}$
 \begin{compactitem}
     \item A new formal noise variable gets introduced, 
     denoted here as $\epsilon_{new}$ for every new floating point operation. 
     \item Furthermore, the $\alpha$ scaling operations in
	 		$\eqref{eq:saffine-float-mult}$
     		or the	linear superposition of 
			the coefficients in $\eqref{eq:saffine-float-add}$
			over the set of union of the noise variables show 
			how the error from previous stages get propagated 
			to future computations. 
			However, these operations 
			are also to be performed in finite precision and 
			hence introduces second order noise variable which
			we have bracketed out as {\em higher order error terms}.
			We choose to ignore second and higher order error 
			terms(that is $\boldsymbol{u}^2$ or higher) 
 			since they do not substantially affect the
 			analysis until the problem size approaches $2^{53}$. 
			Considering higher order noise variables leads to an 
 			explosion of new noise variables being produced 
			while not adding significant benefit to the
 			accuracy analysis of the system, we ignore the second
			order terms.
 	\item   If $\epsilon_i \in \sigma(x_f)\cap\sigma(y_f)$, 
			then in the linear superposition considering the affine model we get
			$(\gamma(x_f,\epsilon_i)+\gamma(y_f,\epsilon_i))$ while
			when using interval analysis would result in 
			$(|\gamma(x_f,\epsilon_i)| + |\gamma(y_f,\epsilon_i)|)$ 
			which would have been a large over-approximation when there is a mix of 
			addition and cancellation terms.
 \end{compactitem}

  \noindent{\textbf{Generation and Propagation of Errors}:\/}
  Darulova~\cite{Comp-for-reals} introduces the notion 
  of separating the {\em generation} and {\em propagation} of errors.
 This helps to analyze the effect of each individual 
 formal noise variable separately instead of over-approximating 
 all error terms together as done in interval analysis.
 We extend its application to stencil codes.
 Consider a real function $f$ and its floating point 
 version $\tilde{f}$. 
 The real set of variables is denoted by 
 $\mathbf{x}$ while the rounded off variable 
 set(with error laden terms intrinsically) are denoted as
 $\mathbf{\tilde{x}}$. 
 Then the total error can be written and decomposed in the form
 \begin{equation}
 \label{eq:sdarulova-separation}
 \begin{split}
 	|f(\mathbf{x}) - \tilde{f}(\mathbf{\tilde{x}})| &\leq \overbrace{|f(\mathbf{x}) - f(\mathbf{\tilde{x}})|}^\text{propagation of errors} + 
 											\overbrace{|f(\mathbf{\tilde{x}}) - \tilde{f}(\mathbf{\tilde{x}})|}^\text{Freshly generated errors}
 \end{split}
 \end{equation}

 The first term essentially represents the contribution 
 of the initial error in the input to the output. 
 Thus each input variable associates with itself a 
 propagation factor by which error on the input gets 
 propagated to the output. 
 The propagation factor essentially characterizes the rate
 at which the output changes for some delta change in the input.
 Here the delta change in the input is the round-off term
 associated with the error laden input variable.
 Thus the propagation factor essentially depends on the
 derivative of the function with respect to the input points.

\begin{figure}[H]
    \centering
\includegraphics[width=0.5\columnwidth]{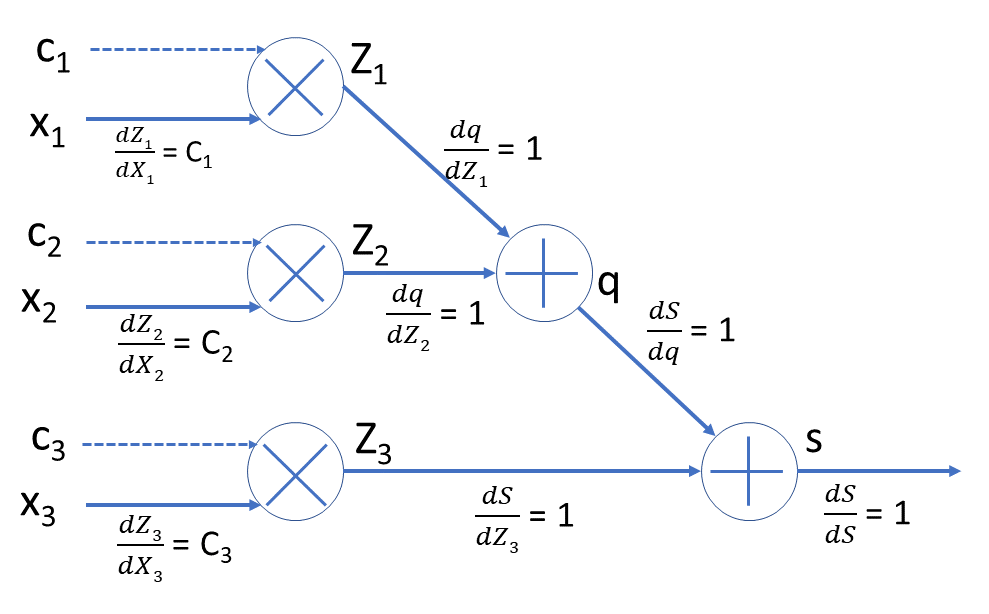}\\
\caption{\label{fig:scfg}
Computational graph with highlighted derivatives for 1 step of a 1-d 3-point stencil
}
    \hfill
\end{figure}

 Let $K_i$ denote the propagation factor for variable $x_i$
 and $\sigma(x_i)$ denotes the set of noise variables existing 
 at $x_i$ as part of incoming error. 
 Then
 \begin{equation}
 \label{eq:sslope-propagate}
 \begin{split}
   |f(\mathbf{x}) - f(\mathbf{\tilde{x}})| &\leq \sum_{i=1}^m K_i \bigg ( \sum_{\epsilon_i \in \sigma(x_i)} \gamma(x_i,\epsilon_i)\epsilon_i \bigg )\ ;\quad
   \mbox{where, } K_i = \underset{w_i \in\ \mathbf{\tilde{x}}}{sup} \bigg | \dfrac{\partial f}{\partial w_i}\bigg | \\
 \end{split}
 \vspace*{-1em}
 \end{equation}
 Evaluating $K_i$ factors is easy to visualize 
 once the computational graph is available as shown in Figure ~\ref{fig:scfg}.

  As a simple ilustration, consider a single step iteration
  of a 1-d 3-point stencil with stencil coefficients
  $\{c_1, c_2, c_3 \}$ centered around $x_2$ to evaluate
  $x_2$ for the next time step, denoted here as S in Figure$~\ref{fig:scfg}$.
  Thus the computational graph evaluates 
  $S=((c_1\times x_1)+(c_2\times x_2))+(c_3\times x_3)$. 
  If $\epsilon^\prime$
  is a noise variable belonging to $\sigma(x_1)$ , 
  then the error contribution's propagation at $S$ will be
  \begin{equation}
  \label{eq:schain-rule}
  \begin{split}
  	\gamma(x_1, \epsilon^\prime)\cdot K_{x_1} &=  \gamma(x_1, \epsilon^\prime)\cdot \bigg| \dfrac{dS}{dx_1}\bigg| 
  	                                     =  \gamma(x_1, \epsilon^\prime)\cdot \bigg| \dfrac{dS}{dS}\cdot
 										 	\dfrac{dS}{dq}\cdot
 											\dfrac{dq}{dz_1}\cdot
 											\dfrac{dz_1}{dx_1}\bigg| \\
 										&= \gamma(x_1, \epsilon^\prime)\cdot 1 \cdot 1 \cdot 1 \cdot c_1 
 										 = \gamma(x_1, \epsilon^\prime) \cdot c_1
  \end{split}
  \end{equation}
 
  Thus for all such $\epsilon^\prime \in \sigma(x_1)$,
  the $\gamma(x_i, \epsilon^\prime)$ gets propagated by $c_1$.
  Similarly for $x_2$ and $x_3$.

\subsection{Modeling rounding error in Stencils}
\label{sec:stencil-error}

Stencil applications modeling real world partial differential equations
in fluids, combustion, cosmology and other computing areas are required
to deal with multiple physical quantities. 
For example, when modeling
electromagnetic fields with finite difference time domain methods,
there are atleast two arrays corresponding to the electric and magnetic 
field whose updates are intrinsically dependent on both arrays.
To address more complicated structures,
we introduce the notion of multiple array updates.
 Suppose, every contribution depends on contributions
 from $N$ participating arrays.
We denote the iterative stencil update rule by $f_s^m$ which
associates the stencil update function to the domain of grid points,
$\vec{x}$, and the array being updated.
Equation $\eqref{eq:siterative}$
shows the {\em single step update rule} for updating location
$A_u[\vec{x},{\!}t]$,that is
location $\vec{x}$ of array $A_u$ for time step $t$ \\ [-2em]

\begin{equation}
\label{eq:siterative}
\begin{split}
 f_s^1 (\vec{x},t,A_u) = \sum_{v=1}^N\sum_{\vec{i}=-\vec{w}}^{\vec{w}}c_{(u,v,\vec{i},1)} \times A_v[\vec{x}+\vec{i},t-1] \\
\end{split}
\end{equation}

where, $c_{(u,v,\vec{i},1)}$ denotes the contributing coefficient
from array $A_v$ to array $A_u$ from the location at an offset 
$\vec{i}$ from $\vec{x}$ over a single step. 
The iterator $\vec{i}$
loops over the stencil-width $\vec{w}$ in the neighborhood of the point.
If we unroll $f_s^1$ over $k$ iterations symbolically
and reassociate the coefficients, we obtain a
{\em real-space equivalent $(=_{real})$ update rule 
to compute directly over $k$ iterations}. 
We define the {\em direct update rule by $f_d^k$} as:
\begin{equation}
\label{eq:sdirect}
\begin{split}
	f_d^k (\vec{x},t,A_u) &= \sum_{v=1}^N\sum_{\vec{i}=-k\vec{w}}^{k\vec{w}}c_{(u,v,\vec{i},k)} \times A_v[\vec{x}+\vec{i},t-k] \\
\end{split}
\end{equation}

such that $A_u[\vec{x},\!t] =_{real} f_d^k (\vec{x},t,A_u)$.
Note that as we unroll over $k$ steps, the stencil width proportionally
extends over $k\vec{w}$ neighborhood of the location $\vec{x}$.

{\em We use the multi-step update rule as our alternative evaluation
strategy for a detector} located at a point $A_u[\vec{x},\!t_0+T]$ 
using values from a baseline at $t=t_0$ as modeled by
$f_d^T(\vec{x},t_0+T,A_u)$ and refer to it as \textbf{direct evaluation}.
First, we determine the effective path contribution from each point 
on the baseline $t{\rs}={\rs}t_0$ towards $A_u[\vec{x},\!t_0\!+\!T]$ 
that fall inside its dependence cone and then perform a dot-product
of these coefficients with the corresponding data points from the baseline. \\ 

\begin{figure}[H]
\includegraphics[width=0.5\columnwidth]{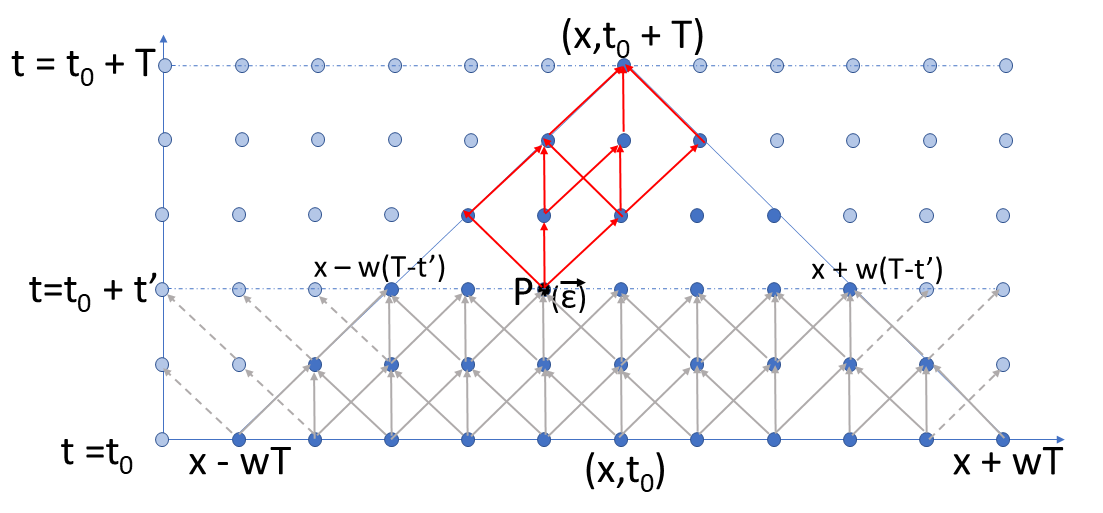}
\caption{\label{fig:serror-propagation}
Error propagation from point (x-1,t0+t') to (x,t0+T) (1-d stencil)
}
\vspace*{-1em}
\end{figure}

Figure ~\ref{fig:serror-propagation} shows the dependence 
cone and the data value flow paths when evaluating
a stencil point at $(t_0\!+\!T,\vec{x})$ from a baseline 
at $t{\rs}={\rs}t_0$ for a single array 1-d 3-point stencil.
Evaluation of each node of this stencil comprise of
three multiplications and two additions, same as the
computational flow graph shown in fig ~\ref{fig:scfg}.
In total five operations per node results in 5 fresh 
noise variable generated per node. Thus the space complexity
is linear in the number of data points.
For a d-dimensional problem with N-arrays and T-step detector model, the space complexity
for the noise variables will be $O\ (NT^d)$.
For notation ,we always consider the update towards array $A_u$, but the same 
holds for all participating arrays.
Consider a point $\vec{P}$ located within the dependence cone,
$t^\prime$ steps in the future from the baseline  and equivalently 
$(T-t^\prime)$ in the past from the point $(t_0+T,\vec{x})$. 
The value at $\vec{P}$ is obtained after
executing the iterative stencil over $t^\prime$ steps, that is 
$f_s^{t^\prime}(\vec{P},t_0+t^\prime,A_u)$. 
$\sigma(A_u[\vec{P},\!t_0\!+\!T])$ denotes the set of noise variables
at $\vec{P}$ for array $A_u$. It comprises of
the set of incoming noise variables already 
generated ({\em propagation of errors}) before reaching the stage
of evaluation of $f_s^1(\vec{P},t_0\!+\!t^\prime\!-\!1,A_u)$ and the new 
noise variables generated due to round-off from its execution.

Thus, we have \\ [-1em]

 \begin{equation}
 \label{eq:snoiseVar-union}
 	\begin{split}
		\sigma(A_u[\vec{P},\!t_0\!+\!t^\prime]) &=\!
					\overbrace{\bigcup_{v=1}^N{\rs}\bigg \{{\rs} \bigcup_{\vec{i}=-\vec{w}}^{\vec{w}}{\rs} 
										\sigma(A_v[\vec{P}\!+\!\vec{i},\!t_0\!+\!t^\prime\!-\!1] \bigg \}}^\text{propagation of incoming errors generated before $t+t^\prime$} 
										\cup \underbrace{\sigma^{(0)}(A_u[\vec{P},t_0+t^\prime])}_\text{freshly generated errors at $t+t^\prime$}
	\end{split}
 \end{equation} 

 In equation $~\eqref{eq:snoiseVar-union}$ , the first term on the rhs indicate the
 propagation of existing errors before reaching the compute stage of $\vec{P}$.
 The second term of the union retrieves the freshly generated noise variables 
 at $\vec{P}$ using the operator $\sigma^{(0)}$.

 Once a noise variable is freshly generated at $\vec{P}$, its magnitude is scaled
by the effective path contribution from $\vec{P}$ to the destination $A_u[\vec{x},\!t_0\!+\!T]$
which in the derivative form can be summarized as 
\begin{equation}
\label{eq:scoeff-deriv-form}
  \bigg | \dfrac{dA_u[\vec{x},\!t_0\!+\!T]}{dA_v[\vec{P},t_0+t^\prime]} \bigg | =
											c_{(u,v,\vec{x}-\vec{P},T-t^\prime)}
\end{equation}

depending on which participating array $\vec{P}$ belongs to for a multi-array update.
Thus, instead of dynamically propagating the noise variables at every step,
we associate a set of fresh noise variables with each compute node of
the CFG, and then scale them directly with the pre-computed coefficients 
of equation$~\eqref{eq:scoeff-deriv-form}$.
For an intermediate time tile at $t=t_0+t^\prime$, let
$E_{u,v}^{t^\prime}$ denote the total error contribution
for the new noise variables at $(t_0+t^\prime)$ due to $A_v$ 
contributing towards $A_u$. Then we have \\ [-2em]

\begin{equation}
\label{eq:sperArray-error-per-tile}
E_{u,v}^{t^\prime}{\rs} ={\rs} \sum_{\underset{-\vec{w}(T-t^\prime)}{i=}}^{\vec{w}(T-t^\prime)}{\rs\rs} \bigg (
							\sum_{\underset{\sigma^{(0)}\!\left(\!\vec{P}_{(t^\prime,\vec{i})}\! \right)}{\epsilon_i \in }}{\rs\rs\rs}
							\bigg |
								\overbrace{\gamma(\vec{P}_{(t^\prime,\vec{i})} , \epsilon_i)}^\text{noise coefficient}
									\times {\rs}
								\underbrace{c_{(u,v,\vec{i}, T-t^\prime)}}_\text{Propagating coefficient}
							\bigg |
							\bigg ) 
\end{equation}
$\text{where, } \vec{P}_{(t^\prime,\vec{i})} = A_v[\vec{x}+\vec{i},t_0+t^\prime] $.
$E_{u,v}^{t^\prime}$ needs to be summed over all the participating arrays that
contribute to $A_u$ to obtain the total error term at $t_0+{t^\prime}$.
To find the total error generated over $T$ steps, 
we need to sum for all the intermediate
$t^\prime $ tiles.
Then the total error ($T_E$) and the relative error ($R_E$)
in the evaluation of $A_u[\vec{x},\!t_0\!+\!T]$ are: 
\begin{equation}
\label{eq:serror-equation}
\begin{split}
	\boldsymbol{T_E}(A_u[\vec{x},t_0+T]) &= \sum_{t^\prime=0}^T \left ( \sum_{v=1}^N  E_{u,v}^{t^\prime} \right )\\
	\boldsymbol{R_E}(A_u[\vec{x},t_0+T]) &= \bigg | \dfrac{T_E(A_u[\vec{x},t_0+T]}{f_d^T(\vec{x}, t_0+T, A_u)} \bigg |
\end{split}
\end{equation} 
The relative error measure normalizes the total error by
the actual stencil value at that point(ideally the exact real value)
hence providing a quantification about how severe the encountered 
loss is with repect to the actual operating 
data. Note that when normalizing, the denomiator used 
detector evaluation, $f_d^T$, and not the stencil's iterated value.
In theory, we must normalize by the ideal $\textbf{real}$ value
but in practice we have to compute in finite precision.
However, our direct evaluation is designed to be a very close proxy to
real evaluation since parts of it can be exercised at very high precision.

\paragraph{Iterated vs direct loss.}
	Evaluating equation$~\eqref{eq:serror-equation}$ for all $(\vec{x}+\vec{i},t_0)$ 
	belonging to a bounded input interval $[I_l, I_u]$ derives the relative error 
	for the iterated evaluation of the stencil denoted by $\boldsymbol{R_s}$. 
	To evaluate the error loss for the direct evaluation, we start from $(\vec{x}, t_0+T)$ 
	and do a back propagation to compute the coefficient set for $f_d^T$, followed by 
	a $f_d^T$ evaluation on the input data set bounded in the interval $[I_l,I_u]$.
	Since, $f_d^T$ computes a large dot product that can introduce 
	significant round off error, we deploy Kahan's algorithm ~\cite{kahan1996ieee} 
	for the additive portion of $f_d^T$, which is computationally more expensive
	but more precision-preserving summation algorithm. 
	Let the relative error term evaluated for the direct evaluation be $\boldsymbol{R_d}$. 
	The precision of value computed	using direct evaluation need not be any greater than 
	that produced by the iterative stencil. 
	\begin{equation} \label{eq:sdet-prec}
	\begin{split}
		\hbox{bits preserved}_T  &= (dp)_T = p - \log_2(\max(\boldsymbol{R_s}, \boldsymbol{R_d})) \\
	\end{split}
	\end{equation} 

Equation$~\eqref{eq:sdet-prec}$ evaluates the minimum matching precision between the
iterated and direct evaluation schemes over T-steps that survives the loss due to 
 round-off error. This is denoted by \textbf{detector precision (dp)},
and represents the maximum precision that can potentially be guaranteed by our detection
strategy. Note that our error analysis is conservative in its treatment of
input values, and might not capture the maximum precision preserved by a specific
set of input values. On the other hand, it provides a rigorous detection guarantee and attempts to
minimize total cost in providing that guarantee.

\clearpage
\section{Precision Driven Optimizations}
\label{sec:soptimization}

In the main paper we discuss the precision driven optimization 
techniques employed to reduce the detector computation cost
while still guaranteeing the necessary quality detection.
The concepts of \textbf{Essential Width} and \textbf{Protected Width}
were introduced briefly with discussions on how they help the 
optimization process. Here, we delve into further details 
of these concepts begining with an intuitive description 
followed by formalization of the concepts.

\subsection{Overview}
To simplify the intricacies of the methodologies of \FPD, we begin with a simple
one dimensionsal heat equaation description represented in the continuous domain 
as

\begin{equation}
    \label{eq:sheat-cont}
    \begin{split}
        \dfrac{\partial u}{\partial t} &= \nabla^2 u(x),\quad x \in \Omega \times (0,T] \\
        u(x) &= u_D,\quad x \in \partial \Omega \times (0,T] \quad\hbox{(Dirichlet Boundary Conditions)}\\
        u &= u_0(x),\quad t=0\ ;\quad \hbox{(Initial Conditions)} \\
    \end{split}
\end{equation}

where $u(x)$ corresponds to the exact solution, $\Omega$ is the solution domain
and $\partial \Omega$ corresponds to the boundary points of the domain.
An explicit finite difference solution of $\eqref{eq:sheat-cont}$ is expressed as 
\begin{equation}
\label{eq:sheat-discrete}
    \begin{split}
        u_i^{n+1} 
                 &= \alpha\dfrac{\Delta t}{\Delta x}u_{i-1}^n + ( 1 - 2\alpha\dfrac{\Delta t}{\Delta x})u_i^n + \alpha\dfrac{\Delta t}{\Delta x}u_{i+1}^n 
    \end{split}
\end{equation}

Here $u_i^n$ is corresponds to the $i'th$ discretized point 
during the time iteration  $n$. 
Thus neighbouring points from the $n'th$ iteration are used
to update a point during its next iteration.
The correct choice of the discretization parameters constitute 
the amount of contribution received from the
neighbouring elements in the evolution of a stencil. 
We consider the following coefficients only for
demonstrating the methodologies. 
Let $A$ denote the one dimensional grid on which $u(x)$ 
is discretized into (N+1) points including the boundaries. 
With a set of chosen coefficients conforming to the 
discretization and stability criterions, 
a common 1d heat equation form is written as
\begin{equation}
\label{eq:sheat-unit}
    A[x,t+1] = 0.25*A[x-1,t] + 0.5*A[x,t] + 0.25*A[x+1,t]
\end{equation}

On the rhs of $\eqref{eq:sheat-unit}$ one can further expand by replacing them using
points from step $(t-1)$ and so on. Notice that the offset dependence around $i$ for one iteration in this
case is $\{-1,0,+1\}$, while when unrolled for 2 steps it will be $\{-2,-1,0,1,2\}$ and so on.
Let $Set_k$ denote the offset set corresponding to the $k'th$ unroll. For our example, if we unroll this way, we can see the evolution pattern in the corresponding coefficients as shown in Table ~\ref{table:sunroll-table}.

\begin{table}[h]
  \caption{Evolution of Coefficient values as we unroll the stencil} 
  \centering
\begin{tabular}{cccccccc}
  \toprule
  \bfseries Set Number & \multicolumn{7}{c}{Coefficients at Offsets} \\\cmidrule(lr){2-8}
	  offset index $\rightarrow$ & -3 & -2 & -1 & 0 & 1 & 2 & 3 \\
  \midrule
    $Set_1$ & 0 & 0 & 0.25 & 0.5 & 0.25 & 0 & 0 \\ 
    $Set_2$ & 0 & 0.0625 & 0.25 & 0.375 & 0.25 & 0.0625 & 0 \\ 
    $Set_3$ & 0.015625 & 0.09375 & 0.234375 & 0.3125 & 0.234375 & 0.09375 & 0.015625 \\
    \bottomrule
  \end{tabular}
  \label{table:sunroll-table}
\end{table}


While our direct evaluation method by leap-frogging Tsteps is equivalent
in the real space, it is still subject to round off error accumulation
due to the non compositional nature of floating point numbers.
\FPD tries to find the upper bound on the minimum matching precision
between the stencil and the direct evaluation executions for 
a given interval range. using the methods described in section ~\ref{sec:stencil-error}.
We call this the detector precision $(dp_T)$.

\begin{figure*}
\centering 
	\begin{minipage}{0.32\textwidth}
	\includegraphics[width=1.0\columnwidth]{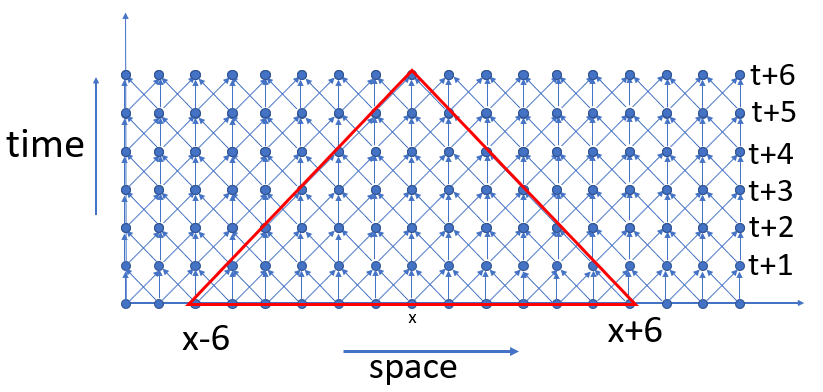}
	\caption{\label{fig:siteratedStencil} Simplified 1D stencil over 6 time steps}
	\end{minipage}\hfill
\begin{minipage}{0.32\textwidth}
\includegraphics[width=1.0\columnwidth]{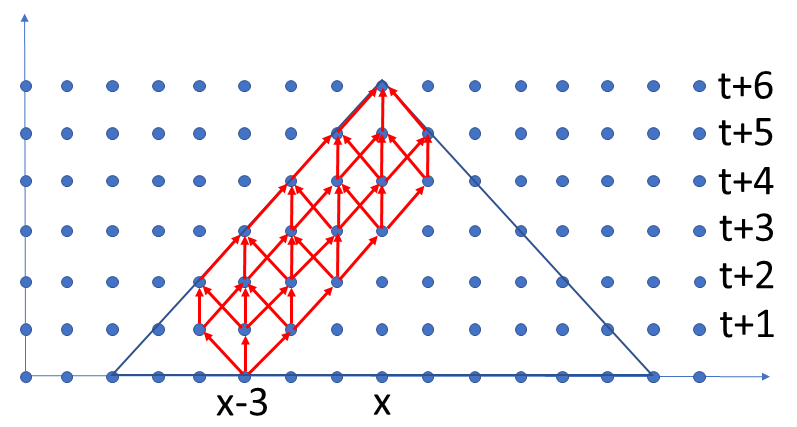}
\caption{\label{fig:sPathDominance} Illustration of path dominance}
\end{minipage}\hfill
	\begin{minipage}{0.32\textwidth}
	\centering
	\includegraphics[width=1.0\columnwidth]{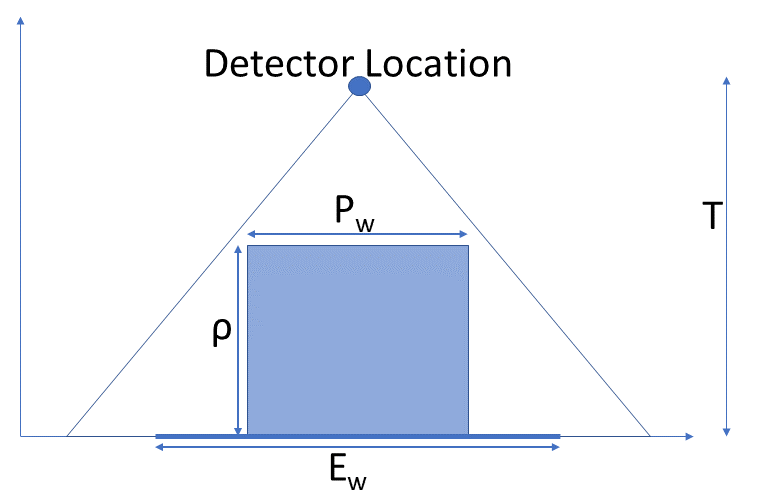}
	\caption{\label{fig:sew_and_pw} Essential (\ew{}) and protected widths (\pw{})}
	\end{minipage}
\end{figure*}

Given we obtain this bound on the precision to check at the detector, it
opens up the possibility of driving optimizations constrained by $dp_T$.
The $Set_k$ values show that as we increase $k$, the set of points required 
in the offset to compute the detector increases.
Furthermore, the points at the edges of these dependence sets
have reduced contributions to the compute than the interior 
points.
This is due to a combination of associated primary coefficients
due to the discretization process and the phenomenon of
{\em path dominance}. 
We do not consider discretization effects
and assume the discretized numerics are designed to be stable
inlcuding the method error.
However, we consider the effect of {\em path dominance} as shown in 
fig ~\ref{fig:sPathDominance} by the highlighted path contribution.

Essentially, points at the boundary of the cone have lesser path contributions 
than the interior ones. 
Notice that when computing in infinite precision
we require computing over the entire dependence 
set to obtain the exact real value.
However in finite precision, we have already trimmed our expectation
of the preserved accuracy at the detector location by $dp_T$. 
For example, when computing in double precision with $p=53$ bits,
\FPD says a detector precision of \textbf{45} bits can be matched
between the stencil and the detector over 64 iterations for the 
one dimensional heat example. 
Hence, points in the dependence set 
that do not possess sufficiently strong contributing coefficients
to affect the detector within 45 bits are good candidates to be
\textbf{trimmed away} from the detector's dependence set
for the evaluation.
Thus, instead of the offset being $\{-64,-63,\dots, 0, \dots, 63, 64\}$
will become say $\{-40, -39, \dots, 0, \dots, 39, 40\}$ based on the 
analysis presented later. 
We carve out an \textbf{essential width} (\ew{})
out of the full dependence set.
This process introduces \textbf{controlled approximations}
to the detector compute process while still retaining the
detector precision bound.
Fig ~\ref{fig:sew_and_pw} shows a representative \ew{} for a detector.

Furthermore, each point in a detectors dependence cone, 
contribute to the detector a varying amount of precision.
We provide the user with a parametric knob called the
\textbf{user defined precision (udp)} that lets the user choose 
how many bits they want to be preserved for fractions of the 
computation space. 
For example, suppose the user selects
$udp=20$ bits. 
In our example, with $dp_{64}=45$ at the detector,
\FPD finds the \textbf{minimal set of points which contribute atleast
20 bits of precision to reatin 45 bits of precision at the 
detector}. 
Thus any mismatch beyond $dp_{64}$ at the detector implies
atleast one or more of the points in this minimal set of
points been affected.
This forms the basis of \FPD's protection mechanism
and this minimal set of points conforming to the $udp$
protection guarantee is called \textbf{Protected Width} \pw{}.
Fig ~\ref{fig:sew_and_pw} shows a representative \pw{} tile
for the given detector.
Next we discuss how the \ew{} and \pw{} are obtained by
detailing the interactions between operating floating
point numbers.

\subsection{Precision reasoning on operand exponents}
In section ~\ref{sec:fp-basics}, we denote a floating 
point number as a 5-tuple $(\beta, s,m,e,p)$.
For a given radix,$\beta\ (=2)$ and a fixed working precision
$p(=53)$ we can represent a number in this system using
the 3-tuple of $(s,m,e)$.
Let $a=(s_a, m_a, e_a)$ and $b=(s_b, m_b, e_b)$ be two 
such floating point numbers under a fixed $\beta$ and $p$
floating point system.
We define a \textbf{distance metric (d)}, that quantifies the 
relative distance of the binades
\footnote{A binade is a set of binary floating-point 
	values that all have the same sign and exponent. 
	The distance between successive members belonging to the 
	same binade are equal to the ulp of that binade}
in which the two
floating point numbers are represented.
\begin{equation}
  \label{eq:sdist-metric}
	d_{ab} = |e_a - e_b|
\end{equation}
Let $S_{ab}$ denote the summation of these two floating point
numbers
\[
	S_{ab} = \circ(a + b) = \circ((s_a, m_a, e_a) + (s_b, m_b, e_b))
\]

For floating point addition/subtraction, the two operands are adjusted 
to bring them in the same binade and then binary
addition/subtraction takes effect between the mantissa
components followed by rounding.
If $e_a > e_b$, then to evaluate $S_{ab}$, $m_b$ must be shifted
by $d_{ab}$ bits to its lsb implying $d_{ab}$ bits of $m_b$ are 
lost.
In other words, in the system with working precision $p$, only
$(p-d_{ab})$ bits of $m_b$ take part in the binary addition to determine
the final outcome of $S_{ab}$. Likewise, if $e_b > e_a$, then
$m_a$ shifts by $d_{ab}$ bits towards its lsb.

\subsection{Modeling Protected Width (\pw{})}

Suppose $a$ and $b$, instead of being two floating point numbers,
belongs to an interval such that
\[
	a \in [a_l, a_u];\quad b \in [b_l, b_u]
\]

where the subscripts $l$ and $u$ denote the lower and upper bounds
respectively. We can further break down the floating point representations
of the intervals bounds in terms of the 3-tuple expressions as

\[
	[a_l, a_u] = [(\underline{s_a}, \underline{m_a}, \underline{e_a}),
	              (\overline{s_a}, \overline{m_a}, \overline{e_a})]; \quad
	[b_l, b_u] = [(\underline{s_b}, \underline{m_b}, \underline{e_b}),
	              (\overline{s_b}, \overline{m_b}, \overline{e_b})]; 
\]

To quantify the \textbf{minimal possible influence} $a$ has on the summation
$S_{ab}$, we need to find the distance of the exponents between the maxima 
of $b$ and the minima of $a$. Hence, the \textbf{minimal number of precision bits}
($\mathcal{P}_a$)
of $a$ that part take in the summation for the resulting $p$ precision
bits can be quantified as in equation $~\eqref{eq:sminPrec-a}$ where $p$
is the working precision of the system.

\begin{equation}
\label{eq:sminPrec-a}
\begin{split}
	\mathcal{P}_a &= \begin{cases}
						p - (\overline{e_b} - \underline{e_a}) &:   (\overline{e_b} - \underline{e_a}) > 0 \\
						p &: (\overline{e_b} - \underline{e_a}) \leq 0
					\end{cases} \\
	\mathcal{P}_a &= p - \max(0, (\overline{e_b} - \underline{e_a}))
\end{split}
\end{equation}

Similarly, we can quantify the \textbf{minimal possible influence} $b$ has
on the $S_{ab}$ as

\begin{equation}
\label{eq:sminPrec-b}
	\mathcal{P}_b = p - \max(0, (\overline{e_a} - \underline{e_b}))
\end{equation}

We will next show how to extend the analysis from two points to a set of points.
\FPD{} works by synthesizing detectors for the expected ranges of binades
that the stencil computation may begin with. A $T-step$ detector's
support comprises of $(2*T*\vec{w}+1)$ points centered around a 
grid location $\vec{x}$, that is the points in the set
$X_D = [\vec{x} - \vec{w}T, \vec{x} + \vec{w}T]$, where
$\vec{w}$ represents the footprint of one iteration of 
the stencil expressed as a vector to include higher
dimensional spaces.
Each point $x_i$ in $X_D$ belongs to the tighest encompassing
interval $I$. Upon multiplying each $x_i$ by their respective $T-step$
coefficients, $c_i$, generates separate intervals for each individual product term
denotes as
\[
[ \underline{y_i}, \overline{y_i}] = [\min(c_i\underline{x_i}, c_i\overline{x_i}),
						\max(c_i\underline{x_i}, c_i\overline{x_i}) ]
\]
This results in a new set of $N$ points, $Y = [y_1, y_2, \dots, y_N]$,
where $N = (2T\vec{w} +1)$. Thus , we are now extending our {\em precision enfluence}
analysis from two points to a larger set of $N$ points each belonging to separate intervals.
%
Let $S_Y$ denote the summation of the points in $Y$
over the respective intervalss, then
\begin{equation}
\label{eq:sintv-sum}
\begin{split}
	S_Y &= \sum_{j=1}^N y_j = \sum_{j=1}^N [\underline{y_j}, \overline{y_j}] \\
	    &= [\underline{S_Y}, \overline{S_Y}]
\end{split}
\end{equation}

To analyze the minimal contribution of point $y_i$, we need to compare
its contribution with respect to the maximal contribution from
all the remaining points over that interval.
Let $S_{Y\setminus i}$ denote the partial sum over all points in $Y$
excluding $y_i$.
\begin{equation}
\label{eq:spartial-sum}
\begin{split}
	S_{Y\setminus i} &= \sum_{\underset{i \neq j}{j=1}}^N y_j \\
	      &= [ \underline{S_{Y\setminus i}}, \overline{S_{Y\setminus i}} ]
\end{split}
\end{equation}

We redefine the distance metric from equation $~\eqref{eq:sdist-metric}$
corresponding to point $y_i$ as $d_{\max}(y_i)$ to be distance between
the exponent of the upper bound of $S_{Y\setminus i}$ and the exponent of the 
lower bound of $y_i$ such that

\begin{equation}
	d_{\max}(y_i) = \max(0, exp(\overline(S_{Y\setminus i}) - \exp(\underline{y_i})))
\end{equation}

Then the minimal precision contribution of point $y_i$ to the 
final sum, $S_Y$, is computed upto $p$ precision bits will 
be expressed as
\[
	minContrib_p(y_i, S_Y) = p - d_{\max}(y_i)
\]

Our bound of correctness of $p$ precision bits holds upto 
the detector precision, $dp_T$, for  the error analysis 
of a T-step model of the detector. Hence, we have
\begin{equation}
	\label{eq:smincontrib}
		minContrib_{dp_T}(y_i, S_Y) = dp_T - d_{\max}(y_i)
\end{equation}

Equation $~\eqref{eq:smincontrib}$ evaluates the minimal precision
of each point in $Y$ that affects $S_Y$ within $dp_T$ bits of precision.
This now enables us to partition the set $Y$ based on a 
\textbf{user defined precision(udp)} that the user selects
for protection guarantees. 
Let $Y_{udp}$ denote the set of points in $Y$ whose
contribution to $S_Y$ is atleast $udp$ precision. We
can define such a set as
\begin{equation}
	\label{eq:spw-carve}
		Y_{udp} = [y_i : minContrib_{dp_T}(y_i, S_Y) \geq udp,\quad y_i \in Y]
\end{equation}

The most significant $udp$ bits of precision of each point inside $Y_{udp}$
is then protected if $S_Y$ is evaluated correctly to $dp_T$ bits of
precision. The set $Y_{udp}$ forms the building block for
obtaining the the \textbf{Protected Width} (\pw) parameter.

In our stencil model, for a detector placed at $A_u[\vec{x}, t_0+T]$,
we define \textbf{Protected Width} $(\pw{}(T,\rho, udp))$ as the width
of the multidimensional rectangular region centerd around $\vec{x}$ 
such that with respect to a detector placed at time $t_0+T$ and 
spatial position $\vec{x}$, for each 
$\vec{p} \in \vec{x} \pm \pw{}(T,\rho, udp)$ at time $t_0+\rho$,
the protection guarantee holds.

\begin{lemma}
	\label{sPW-guarantee}
	An error affecting any of the MSB \udp{} bits of a 
	point inside the protected width is detected.
\end{lemma}

For a point $y_i$ in $Y$, if $y_i$ belongs to $Y_{udp}$
and its minimal precision contribution is $p_i$
at the final output, then $p_i \geq udp$. 
If $y_i \in [\underline{y_i}, \overline{y_i}]$, then an error,
$err_{y_i}$ affecting within the MSB $udp$ bits will be
bounded by $err_{y_i} \geq 2^{exp(\underline{y_i}) - p_i +1}$.
Since, we are matching $dp$ bits at the detector, 
hence the threshold of detectable error is bounded by 
$2^{exp(\overline{S_Y})-dp_T+1}$. 
For our guarantee to hold, the following must be true \\ [-0.8em]
\begin{equation}
\begin{split}
	\hbox{generated error} &\geq \hbox{error threshold} \\
	2^{exp(\underline{y_i}) - p_i +1} &\geq 2^{exp(\overline{S_Y})-dp_T+1} \\
\end{split}
\end{equation}

Taking logarithms and simplifying the above terms leads
to the following relation that must hold true 
for all points inside the $P_w$ region.

\begin{equation} \label{eq:spw-limit}
\begin{split}
	exp(\overline{S_Y}) - exp(\underline{y_i}) &\leq dp_T - p_i \\
	udp \leq p_i &\leq dp_T - \{exp(\overline{S_Y}) - exp(\underline{y_i})\}
\end{split}
\end{equation}

In figure ~\ref{fig:sew_and_pw} , the enclosed region ($\ty \times \pw{}$)
for the given detector location T-steps ahead forms the protected region
for that detector.
In determining the protected region,
where a detector is used to protect iteration points in multiple
time steps, we choose \ty{} such that the protected width chosen
at \ty{} is also valid for all time points $t_0 \leq t \le t_0 + \ty{}$:
\begin{equation}
  \forall\ 0\le t\le \ty{}: P_w(T,\ty{},udp) \le P_w(T,t,udp)
\end{equation}

\subsection{Modeling Essential Width (\ew{})}
Essential width models the maximum set of points
in $X$ absolutely necessary to obtain $dp_T$ precision
at the detector location. 
If the maximal contribution of $y_i$ to the
output falls below ulp of the desired precision, 
then $y_i$ can be safely discarded. 
Ofcourse, all
discarded points collectively could still affect within
the required precision. 
Hence, there is an
iterative process to carve out a set of excluded points.
To this end, we compare $y_i's$ maximal contribution
with respect to the minimal contribution from all the 
remaining points.
%
%
We define $d_{\min}(y_i)$ as the distance between the exponent of the lower bound of $S_{Y\setminus i}$ and the exponent 
of the upper bound for $y_i$ such that \\ [-0.7em]
\begin{equation} \label{eq:smin-distance}
	d_{\min}(y_i) = \max(0,exp(\underline{S_Y^i}) - exp(\overline{y_i})) 
\end{equation}
The maximal precision contribution of point $y_i$ to 
the final sum $S_Y$ when $S_Y$ is computed
up to $p$ precision bits is
\\ [-0.7em]
\begin{equation} \label{eq:smaxContrb}
	maxContrib_p(y_i, S_Y) = p - d_{\min}(y_i)
\end{equation}
In our stencil model for detector placed at 
$A_u[\vec{x}, t_0+T]$, we define \textbf{Essential Width} ($E_w$)
as the width of the multidimensional rectangular region around 
$\vec{x}$ such that the direct evaluation over $E_w$
is sufficient to guarantee $dp_T$ precision at the detector.
Specifically, the region is defined by extents $E_{wl}$ and $E_{wr}$ 
such that $E_w = E_{wl}+ E_{wr} +1$.
Figure~\ref{fig:sew_and_pw} 
illustrates $E_w$ for evaluation of a 1-D stencil over T-steps.

\begin{lemma}
	\label{sFalse-positive}
	 In the absence of an error, direct evaluation over the 
	 essential width ($E_w$) ensures the correctness of the computation
	to at least $dp_T$ bits of precision at the detector location.
\end{lemma}

For all points in the region excluded from $E_w$ 
$\forall \tilde{X}: \tilde{X}\subset ([\vec{x}\pm \vec{w}T] - E_w)$, the following holds \\ [-0.7em]
\begin{equation}
	\forall x_i \in \tilde{X}: maxContrib_{dp_T} \left(\sum (y_i = c_i\times x_i), S_Y \right ) \leq 0
\end{equation}

One form of false positives involves detection of soft errors when no
error affects any iteration point on which the detector depends (aka
the detector's dependence cone). The preceding lemma guarantees that,
despite the reduced evaluation cost, in the absence of errors
affecting the dependence code, the direct evaluation is equivalent to
the iterative evaluation within $dp$ bits and no soft error
notification(false alarms) is triggered.

In essence, Essential Width (\ew{}) provides a method to reduce the 
detector evaluation cost while retaining the quality of the detector
precision ($dp_T$). The Protected Width (\pw{}) on the other hand characterizes 
a detectors maximum reach to provide rigorous protection for an enclosed
region of space per detector. Hence, obtaining an optimal combination of
$(\ew{}, \pw{}, \ty)$ allows \FPD to situate detectors sparsely and effectively
over the entire compute space.

\clearpage
\section{Overall Algorithm}
\label{sec:soffline-online}

To efficiently deploy \FPD's detectors during online execution,
we build a look-up table of detector configurations using an
offline profiling pass without actually looking at the input data.
This allows the detector configurations to be valid for a large 
range of initialization and boundary constraints when being
instantiated online. For each anticipated choice of 
$udp$, probability of detection, and anticipated input range, the offline analysis
determines the detector configuration to be used during online execution.
A detector configuration consists of: $T$, the distance at which the single-direct detector
is evaluated: $\rho$, the number of iterations between certified baselines: \ew{}, the essential width to
evaluate the detector; Detector coefficients for the values in the essential width;
and \pw{}, the spatial separation between detectors in the same time step. These parameters completely specify
the detector configuration at runtime.

\subsection{Offline algorithm.}
We consider all T and \ty{} such that
$ T \leq T_{\max}, \ty{} < T $. To
constrain the search space of possible configurations, we use a
customizable upper bound on the Tstep (say, $T_{\max}=256$) that one
may use for the detector evaluation. Furthermore, having
a very large $T$ impacts coverage around the boundary region.
%
%
For a d-dimensional stencil, \pw{} is
	d-dimensional vector, corresponding to protected region of iteration points centered at the detector.
    \ew{} is represented by a rectangular region with a computed number of points along left and right of
    the detector along each direction. These extents along dimension $i$ are denoted by
    $ewl_i$ and $ewr_i$ with $ew_i$ = $ewl_i + ewr_i + 1$.
    The cost function for a 5-tuple
	detector configuration $(exp,udp,T,\ty{},Coeffs)$ is given as\\[-1em]
\begin{equation}
  Cost = \dfrac{1}{\rho}\prod_{i=1}^d \left(\dfrac{ew_i}{pw_i}\right) \label{eq:scost-function}
\end{equation}
Even though the tuple elements $exp$, $T$, and $Coeffs$ do
	not explicitly appear in the cost function, they implicitly influence
	the cost through \ew{} and \pw{}. 
    

\begin{figure}[h]
\begin{lstlisting}
struct Config {
       double volume, cost;
       int dt, [@\ty{}@], pwvec, ewlvec, ewrvec;
};
auto offline_analysis(Tmax, exp_set, udp_set, cov_set) {
  for exp in exp_set:
    //compute maxdp[t] for 1<=t<=Tmax
    //compute ewcost[t] as product of elements of vector ew[exp,t,maxdp[t]]
    for b in udp_set:
      COV[nt, nt, wvec] = //number of points in region [dt:dt+[@\ty{}@]][-i:i][-j:j] with b bits of coverage
      for dt = nt-1 .. 0 and [@\ty{}@] = nt-dt .. dt+1:
        forall ivec:
          cost = VOLUME(dt, [@\ty{}@], ivec) / ewcost[dt+[@\ty{}@]]);
          for c in cov_set:
            if COV(dt, [@\ty{}@], ivec) >= c && configs[exp,b,c].cost < cost
              configs[exp,b,c] = {vol, cost, dt, [@\ty{}@], ivec, ewl[dt+[@\ty{}@]], ewr[dt+[@\ty{}@]]}
  return configs;
}
\end{lstlisting}
\caption{Offline profiler for optimal detector configurations}\label{alg:sprofiler}
\end{figure}

Figure~\ref{alg:sprofiler} outlines the steps involved in the offline
analysis. The algorithm takes as input parameters for which offline
profiles need to be determined: the maximum number of time steps
(Tmax), the set of input ranges (as exponents in exp\_set), set
of \udp{} values (as udp\_set), and probabilistic coverage values
(cov\_set). The algorithm determines the configurations one input
range choice at a time.  Using floating-point round-off analysis, the
maximum number of bits that will be preserved for each possible time
step $t$ is computed (as maxdp). For each $t$, the cost of evaluating
each detector is computed as the product of the $\vec{ew}$ dimensions
to guarantee $maxdp[t]$ bits of precision of the direct evaluation $t$
time steps away. Then, for each candidate protected region, the
fraction of points with guaranteed coverage of $b$ bits (where $b$ is
a candidate \udp{} in the input parameter udp\_set) is computed. If
this fraction is greater than a desired coverage and if the associated
cost is lower than that of any configuration seen thus far, this
configuration is chosen. After evaluating all feasible solutions, the
algorithm returns the last chosen configurations.


\begin{algorithm}
	\caption{Online Detection} \label{alg:sonline}

\begin{algorithmic}[1]
\footnotesize
\Function{OnlineDetection}{A=Stencil,$udp$, $cov$}

    \State I = ScanInput(A)

    \State exp = MapInterval(I)

	\State in\_cov = AdjustCoverage(cov)

    \State \{$T,\rho,dp,\vec{E_w}, \vec{P_w}$\} = lookup\_configuration[(exp, udp,in\_cov)]

    \State Coeffs = get\_coefficient(Stencil, T, $\vec{E_w}$)

    \State
    configure\_detectors(\{$T,\rho,dp,\vec{E_w}, \vec{P_w}$\})

\State
    cur = 0; $t_{\delta}$ = T - $\rho$

\State D[0] =
    eval\_detector(A, Coeffs)

\State execute\_stencil(min(ITER,
    $\rho$),A)

\For {$t$ = $\rho$; $t$ < ITER; $t$ +=
    $\rho$}

    \State $cur = 1 - cur$

    \State D[$cur$] = eval\_detector(A, Coeffs)

    \State
      execute\_stencil(min(ITER-t, $t_{\delta}$), A)

    \If {t + $t_{\delta}$ $\leq$ ITER}

        \State detector\_check(A,  D[$1 - cur$])

    \EndIf

    \If {t + $t_{\delta}$ < ITER}

        \State
          execute\_stencil(min(ITER-(t + T - $\rho$), $\rho -
          t_{\delta}$))

     \EndIf

\EndFor

\If {ITER > $\rho$ $\mbox{and}$ ITER\%$\rho$ < $t_{\delta}$}

    \State
      trailing\_detector(D[$1-cur$], A)

\EndIf{}

\If {ITER\% $\rho$ !=   0}

    \State trailing\_detector(D[$cur$],  A)

\EndIf

\EndFunction

\end{algorithmic}
\end{algorithm}

\subsection{Online Detector-Embedded Execution}
\label{sec:sonline}

Algorithm~\ref{alg:sonline} outlines the online detector-embedded
execution of a given stencil kernel. The input values are scanned to
compute the input exponent range to be handled in the
floating-point space (lines 2--3). To account for the unprotected boundary regions,
user's input cov value is mapped to an equivalent internal
coverage value (in\_cov) that corresponds to the detection probability
on the grid excluding the boundaries. The adjustment is done with the guarantee 
that the fraction of points covered due to (in\_cov) is at least as many required
by cov over the entire computation space. If an equivalent mapping is not found by
(\texttt{AdjustedCoverage}), the program exits conveying 
an error message for unsupported coverage.
The input range, the user-specified \udp{}, and the modified detection probability (in\_cov)
are used to lookup the detector configuration
(lines 5--7). 

After an initial evaluation of the detector and the
stencil for \ty{} iterations (lines 9--10), the execution of the
stencil (\texttt{execute\_stencil}) is split into two segments:
iterations till the next detector evaluation (\texttt{eval\_detector}
and iterations till the next detector check
(\texttt{detector\_check}). Thus, the stencil is executed with
interleaved detector evaluation and checking (lines 11--18). Note that
the algorithm assumes \ty{} is greater than half the detector evaluation
T-step. This scenario requires at most two ``live'', i.e., unchecked
detectors per spatial position in the iteration space, at any time.
If not, at each spatial position,
multiple detectors need to be evaluated and retained until they are
checked.

When all iterations of the stencil have been executed, the iterations
past the last certified baseline need to be checked. 
To protect any leftover iterations, the computation since the last
certified baseline is checked using the
trailing detection strategy (lines 19--22).

\clearpage
\section{Software Bug Detection}
\label{sec:ssoftware-bug-detection}

\FPD's evaluation units can be efficiently utilized to trap logical and compiler transformation
(polyhedral transforms exploiting data locality and parallelism) induced bugs.
The design of these evaluation units model the actual stencil in real equivalent
space and hence are logically equivalent to the actual stencil without any modeling bias.
Thus any transformation bugs introduced into the stencil codes later trhough
optimization will result in mismatch with our evaluation units.
Hypothesizing that such bugs lead to systematic errors impacting floating point
values non-trivially, we intend to trap bugs in the floating point space within
a precision threshold.
As such, there remains possibility that bugs escape beyond the threshold
we check for (controlled by the detector precision).
%

\paragraph{\bf{Summary of Pluto generated code:\/}}
\begin{table}[htbp]
	\caption{Summary of Pluto generated code.}
	\centering
	\begin{tabular}{ccccc}
	\toprule
	Benchmarks & SLOC spec & SLOC Pluto & Num Loops & Nesting Level	\\
	\midrule
	H1	&	27	&	905		&	374		&	7 	\\
	H2	&	52  &   1164	&	370		&	7 	\\
	H3	&	65	&	905		&	374		&	11	\\
	H4	&	54	&	2314	&	486		&	6 	\\
	H5	&	54	&	2314	&	486		&	6 	\\
	H6	&	54	&	2314	&	486		&	6 	\\
	P1	&	23	&	31		&	12		&	6 	\\
	P2	&	23	&	31		&	12		&	6 	\\
	P3	&	19	&	31		&	12		&	6 	\\
	P4	&	63	&	1163	&	370		&	6 	\\
	P5	&	39	&	1163	&	370		&	6 	\\
	P6	&	39	&	1163	&	370		&	6 	\\
	P7	&	47	&	1165	&	414		&	6 	\\
	P8	&	47	&	956		&	416		&	6 	\\
	P9	&	47	&	956		&	416		&	6 	\\
	W1	&	70	&	1157	&	360		&	6 	\\
	W2	&	70	&	1157	&	360		&	6 	\\
	W3	&	70	&	1157	&	360		&	6 	\\
	W4	&	70	&	1157	&	360		&	6 	\\
	W5	&	70	&	1157	&	360		&	6 	\\
	W6	&	70	&	1157	&	360		&	6 	\\
	C1	&	53	&	1157	&	360		&	6 	\\
	C2	&	53	&	1157	&	360		&	6 	\\
	C3	&	53	&	1157	&	360		&	6 	\\
   \bottomrule
\end{tabular}
  \label{table:sPluto-summary}
\end{table}

We studied the effectiveness of \FPD's evaluation units in bugs introduced to evaluate
polycheck~\cite{DBLP:conf/popl/BaoKPRS16} on Pluto~\cite{uday08cc} 
optimized stencil codes. 
To generate optimized versions using Pluto, we implemented all the
stencils in the form of affine loop nests enclosed in \texttt{scop}
pragmas. This code input to Pluto has consists of median 53 source
lines of code (SLOC)\footnote{Source lines of code are measured using
sloccount tool} per benchmark. Pluto takes
these \texttt{scop}-annotated benchmarks as input to generate
optimized versions with median 1157 SLOC. The median of the number of
loops in the transformed code was 365 across all the benchmarks with a
median nesting depth of 6. In checking such large and complex
generated codes, tools such as \FPD{} are essential.
Table ~\ref{table:sPluto-summary} provides a summary
of the Pluto generated code.
The `spec' refers the original stencil code prior to
feeding into Pluto. We report the source lines of code (SLOC) for both the `spec' and
the Pluto optimized version alongwith the number of loops and nesting levels in the
transformed code. The number of loops and nesting levels drive the complexity of the 
tests being performed since each `for loop' entry point becomes a potential
bug-injection site.

The bugs introduced in the source potentially affect multiple runtime operations
making them easier to detect. These tests fall under three categories based on the generated errors impacting:
(1){\em Loop bounds}, resulting in an incorrect stencil pattern at the tile boundaries,
(2){\em array access}, affecting the stencil footprint used, and
(3){\em Loop reordering}, affecting the data dependencies.
We discuss each of them with an example as discussed below.

\paragraph{\bf{Loop Bounds:\/}} To inject errors impacting `loop'
bounds, we first identified all the `for' loop sites present in the
transformed code. Error injection was performed by either adding one to a lower
bound of a for loop or subtracting one from an upper bound.
This was automated by replacing each bound, $x$, for an identifed 'for loop' 
with a call to a function {\bf{\em bound}}$(c,x)$
where $c$ is a constant integer number identifying  each unique callsite 
and $x$ is the actual bound in the original `for loop' of the transformed code
as shown in figure \ref{fig:sfig-loop-bound}. 
\lstset{escapeinside={(*@}{@*)}}
\begin{figure}[htp]
\begin{lstlisting}
for (t1=-938)t1<=floord(3*steps,32);t1++) {
  for (t2=lbp;t2<=ubp;t2++) {
\end{lstlisting}
\begin{lstlisting}
for (t1=(*@  {\bf bound}  @*)(0, -938);t1<=(*@ {\bf bound} @*)(1, floord(3*steps,32));t1++) {
  for (t2=(*@ {\bf bound} @*)(2, lbp);t2<=(*@ {\bf bound} @*)(3, ubp);t2++) {
\end{lstlisting}
\caption{Source code before and after bound replacement}
\label{fig:sfig-loop-bound}
\end{figure}

A callsite can be selected by command line argument and the resulting execution
will have the indicated bound changed.
The {\bf{\em bound}} function returns the $x$ value (the original bound) if the callsite does not
match the selected callsite for fault injection. IF it is the selected call-site, then
it return $x+1$ if $c$ is even or $x-1$ if $c$ is odd.

\paragraph{\bf{Array Access:\/}} Similarly the array access errors were injected using a function utilizing a
unique integer. Each `for' loop site was replace by the function {\bf{acs}} with a unique integer ID as shown in
the figure \ref{fig:sfig-access}. On encountering an {\bf{acs}} function call, it returns the original access value if 
its unique ID number do not match the intended injection call site ID. for injection. When it matches, it returns the modiefied the access value by dividing it by two.

\begin{figure}[htp]
\begin{lstlisting}
A[( ((32*t1+15056)/3) - 1) % 2][10000][(-32*t1+3*t6-15056)/3)]
\end{lstlisting}
\begin{lstlisting}
A[(*@ {\bf acs} @*)(51, ( ((32*t1+15056)/3) - 1) % 2)][(*@ {\bf acs} @*)(52, 10000)][(*@ {\bf acs} @*)(53,  ((-32*t1+3*t6-15056)/3))]
\end{lstlisting}
\caption{Source code before and after access replacement}
\label{fig:sfig-access}
\end{figure}

For both these methods of automatic injection there are three counts that come
about.
One is the number of source code locations modified.
Another is the number of locations reached by the running code.
The final number is the number of locations reached which actually changed the
computation.
%
Not all source locations where an error can be injected are reachable at run time.
Table ~\ref{table:ssoftware-bugs} identifies the source locations (\#$SL$) identified and the number of runtime 
locations (\#$RL$) actually reached for injecting a bug. 
For example, in benchmark H4, there are 486 locations for injecting a bug that
can potentially impact a loop-bound, however only 106 of them are reached at
runtime.

\paragraph{False positives}
Some bugs do not result in any difference in our bitwise comparison of
the result with the non-buggy version. Therefore, our approach incurs
no false positives in our evaluation. This is because some
source-level bugs do not manifest at runtime.  For example, a loop
iterator might be constrained by multiple loop bounds
(e.g., \texttt{min} or \texttt{max} of multiple expressions), with the
loop bound never reaching the error-injected expressions.For example, 
in benchmark H4, there are 486 locations for injecting a bug that
can potentially impact a loop-bound, however only 106 of them are reached at
runtime. Figure ~\ref{fig:sfig-Pluto-if} lists an example where the `for' loop iterator
passes trhough two nested `if' checks that can potentially mask an error in the
iterator due to a bug impacting the `for loop' bound value.

\begin{figure}[htp]
\begin{lstlisting}
for (t1=-938; t1<value; t1++) {
 if (2*t1%3 == 0) {
  if ((t1+3)%6 == 0) {
    <body>
  }
 }
}
\end{lstlisting}
\caption{An example of nested if statements in Pluto transformed code}
\label{fig:sfig-Pluto-if}
\end{figure}

Table~\ref{table:ssoftware-bugs} summarizes \FPD{}'s bug detection
effectivness for the above two categories of bugs. The table only
lists detection percentages when the bitwise comparison with the
non-buggy version flags an error-injected version as being in error.

\begin{table}[htbp]
	\caption{Software bug detection results. }
	\centering
	\begin{tabular}{ccccccc||ccccccc}
	\toprule
	\multirow{2}{*}{} & \multicolumn{3}{c}{Loop bound} & \multicolumn{3}{c}{Array access} & \multirow{2}{*}{} & \multicolumn{3}{c}{Loop bound} & \multicolumn{3}{c}{Array access} \\
	\cmidrule(lr){2-4} \cmidrule(lr){5-7} \cmidrule(lr){9-11} \cmidrule(lr){12-14}
	& {\#SL}  & {\#RL} & {\%det}  & {\#Sl}  & {\#Rl} & {\%det} & & {\#SL}  & {\#RL} & {\%det}  & {\#Sl}  & {\#Rl} & {\%det}\\
	\midrule
	H1  &  374  & 374 & 100    & 300 & 267   & 99.67 & H2	&  370  & 370 & 100    & 300 & 258   & 100	\\ 
	H3  &  374  & 374 & 100    & 300 & 267   & 98  & H4  &  486  & 106 & 99.79  & 300 & 26    & 100  	\\ 
	H5  &  486  & 106 & 100    & 300 & 26    & 100  & H6  &  486  & 106 & 100    & 300 & 26    & 100  \\ 
	P1  &  10   & 10  & 100    & 18  & 17    & 100  & P2  &  12   & 12  & 100    & 18  & 17    & 100  \\ 
	P3  &  12   & 12  & 100    & 18  & 17    & 100  & P4  &  370  & 310 & 100    & 300 & 223   & 87.67  \\ 
	P5  &  370  & 310 & 96.22  & 300 & 223   & 82.67  & P6  &  370  & 310 & 96.22  & 300 & 223   & 82.33 \\ 
	P7  &  414  & 104 & 98.55  & 300 & 48    & 97     & P8  &  416  & 104 & 100    & 300 & 44    & 98.67   \\ 
	P9  &  416  & 104 & 98.56  & 300 & 44    & 98     & W1  &  360  & 280 & 100    & 300 & 215   & 55.33   \\ 
	W2  &  360  & 280 & 44.72  & 11  & 6     & 54.67  & W3  &  360  & 350 & 31.11  & 300 & 266   & 45.33 \\ 
	W4  &  260  & 280 & 45.28  & 300 & 215   & 60    & W5  &  360  & 280 & 53.61  & 300 & 215   & 62.67 \\
	W6  &  360  & 350 & 100    & 300 & 266   & 100  & C1	&  360  & 360 & 100    & 300 & 262   & 100  \\ 
	C2	&  360  & 360 & 100    & 300 & 262   & 100   & C3	&  360  & 360 & 100    & 300 & 262   & 100 \\ 
        \bottomrule
\end{tabular}
  \label{table:ssoftware-bugs}
\end{table}

%

\paragraph{\bf{Loop ordering:\/}}
Loop reordering was performed by hand.
Figure \ref{sfig-loop-reorder} contains an example of loop reordering where the
first two loops are swapped.
Changing the order of updates to $A$ substantially changed the result of the
computation.

Unlike changes to array accesses and loop bounds, changing loop orders
required non-local changes to the Pluto-generated code. Therefore, we
handcrafted the error injected versions. The handcrafted scenarios
included swapping nested loop pairs and re-ordering of non-nested loop
blocks.  The optimized codes included a maximum of 360
nested \texttt{for} loops with a maximum loop nesting depth of 11.
Testing all possible combinations of loop reorderings will be
prohibitively expensive. We limited our testing to 15 loop order bug
injections per benchmark.
Figure \ref{sfig-loop-reorder} contains an example of loop reordering where the
first two loops are swapped.

\begin{figure}[htp]
\begin{lstlisting}
for (t4=max(1,ceild(16*t1+48*t2+48*t3,3));
     t4<=min(min(min(min(floord(16*t1+48*t2+48*t3+141,3),steps),64*t2-1),64*t3-1),16*t1-16*t2-16*t3+9999);
     t4++) {
 (*@ {\bf for (t5=max(64*t2,-16*t1+16*t2-48*t3+3*t4-78);t5<=min(64*t2+63,-16*t1+16*t2-48*t3+3*t4);t5++) }@*) {
  lbv=max(64*t3,-16*t1+16*t2+16*t3+3*t4-t5-15);
  ubv=min(64*t3+63,-16*t1+16*t2+16*t3+3*t4-t5);
  for (t6=lbv;t6<=ubv;t6++) {
   A[( t4 - 1) % 2][ (-t4+t5)][ (-t4+t6)] = (((((0.5 * A[ t4 % 2][ (-t4+t5)][ (-t4+t6)]) + ...
  }
 }
}
\end{lstlisting}
\begin{lstlisting}
(*@ {\bf for (t5=max(64*t2,-16*t1+16*t2-48*t3+3*t4-78);t5<=min(64*t2+63,-16*t1+16*t2-48*t3+3*t4);t5++) }@*) {
 for (t4=max(1,ceild(16*t1+48*t2+48*t3,3));
      t4<=min(min(min(min(floord(16*t1+48*t2+48*t3+141,3),steps),64*t2-1),64*t3-1),16*t1-16*t2-16*t3+9999);
      t4++) {
  lbv=max(64*t3,-16*t1+16*t2+16*t3+3*t4-t5-15);
  ubv=min(64*t3+63,-16*t1+16*t2+16*t3+3*t4-t5);
  for (t6=lbv;t6<=ubv;t6++) {
   A[( t4 - 1) % 2][ (-t4+t5)][ (-t4+t6)] = (((((0.5 * A[ t4 % 2][ (-t4+t5)][ (-t4+t6)]) + ...
  }
 }
}
\end{lstlisting}
\caption{Source code before and after manual loop reordering}
\label{sfig-loop-reorder}
\end{figure}

\clearpage

\bibliography{refs}

\clearpage